\documentclass[%
 reprint,
 superscriptaddress,
 showkeys,
 amsmath,amssymb,
 aps,
 prl,
]{revtex4-2}

\usepackage{physics}
\usepackage{graphicx}
\usepackage{dcolumn}
\usepackage{bm}
\usepackage{booktabs} 

\usepackage{xcolor}

\begin{document}

\title{Revealing Hydroxide Ion Transport Mechanisms in Commercial Anion-Exchange Membranes at Nano-Scale from Machine-learned Interatomic Potential Simulations}

\author{Jonas H{\"a}nseroth}
\affiliation{Theoretical Solid State Physics, Institute of Physics, Technische Universität Ilmenau, 98693 Ilmenau, Germany}

\author{Muhammad Nawaz Qaisrani}
\affiliation{Theoretical Solid State Physics, Institute of Physics, Technische Universität Ilmenau, 98693 Ilmenau, Germany}

\author{Mostafa Moradi}
\affiliation{Fraunhofer-Institut für Keramische Technologien und Systeme IKTS, 
 Hydrogen Technologies, 99310 Arnstadt, Germany}

\author{Karl Skadell}
\affiliation{Fraunhofer-Institut für Keramische Technologien und Systeme IKTS, 
 Hydrogen Technologies, 99310 Arnstadt, Germany}

\author{Christian Dre{\ss}ler}
\email{christian.dressler@tu-ilmenau.de}
\affiliation{Theoretical Solid State Physics, Institute of Physics, Technische Universität Ilmenau, 98693 Ilmenau, Germany}

\date{\today}

\begin{abstract}

Hydroxide ion transport in anion-exchange membranes fundamentally limits the efficiency of alkaline water electrolysis for green hydrogen production, yet the atomic-scale transport mechanisms remain poorly understood due to the computational challenges associated with modeling ion dynamics. 
Given that anion-exchange membranes enable alkaline electrolysis with abundant catalysts while avoiding perfluoroalkyl and polyfluoroalkyl materials, a deeper mechanistic understanding of hydroxide transport in these systems is essential for advancing sustainable hydrogen production.
Here, we show that large-scale molecular dynamics simulations with fine-tuned machine-learned interatomic potentials provide atomistic insight into hydroxide mobility in a commercial membrane over tens of nanoseconds and over ten nanometer.
We find that increasing water content transforms isolated water clusters into a connected hydrogen-bond network that enables long-range proton transfer.
Under dry conditions hydroxide ions are trapped near positively charged groups and transport is strongly hindered, whereas well-hydrated membranes exhibit extended proton migration and diffusion coefficients approaching those of dilute aqueous solutions.
The simulations reproduce experimental trends in diffusion and activation energies.
Our results establish a direct link between nano-scale structure and macroscopic transport.
Beyond mechanistic insight, the presented simulation framework enables predictive, simulation-guided optimization of membrane chemistry and architecture, opening a pathway toward the rational design of more efficient anion-exchange membranes for green hydrogen technologies.
\end{abstract}


\maketitle

Green hydrogen production through water electrolysis has emerged as a cornerstone technology for carbon-neutral industrial processes and the establishment of a sustainable energy infrastructure.
Among the available technologies, anion-exchange membrane (AEM) water electrolysis enables efficient hydrogen production using platinum-group-metal-free catalysts based on earth-abundant elements such as iron and nickel \cite{dekel2013alkaline,leng2012,pan2012designing,park2019,henkensmeier2020,hren2021,muhyuddin2025anion}.
In contrast, proton exchange membrane (PEM) electrolysis relies on harsh acidic environments, perfluoroalkyl and polyfluoroalkyl (PFAS) materials, and scarce noble metals such as platinum and iridium \cite{zou2015,schalenbach2018,alia2021}.
Recent advances in AEM technology include alkaline-free water feeding strategies for hydrogen production and the extension of these materials toward high-temperature fuel cell applications \cite{douglin2020high,douglin2021high,muhyuddin2025anion}.
Despite this progress, AEM materials still exhibit lower hydroxide ion conductivity compared to the proton conductivity achieved in PEM systems \cite{henkensmeier2020,wijaya2024}.
A detailed molecular-level understanding of hydroxide transport is therefore essential for the rational design of next-generation membranes.

Computational predictions of hydroxide mobility provide a powerful route to elucidate ion transfer mechanisms and guide AEM optimization \cite{han2014,dekel2018_sim,wang2018,karibayev2022}.
Predictive simulations can accelerate membrane material development even prior to experimental synthesis.
However, simulating solvated nanochannels in polymeric AEMs requires computational resources that exceed the capabilities of conventional \textit{ab initio} molecular dynamics (AIMD).
Previous theoretical studies have therefore relied on simplified membrane representations in which functional groups and polymer backbones were mimicked by small molecular fragments \cite{park2017,takaba2017,zelovich2019hydration,zelovich2019mimics,zelovich2020,zelovich21_ohvsh3o,zelovich2024}.
Although these AIMD investigations provided valuable mechanistic insight, the resulting diffusion coefficients remained poorly converged, with hydroxide ions diffusing less than $4$~{\AA} on average \cite{zelovich23}.
Classical force fields offer substantially improved computational efficiency for simulations of water molecules, hydroxide ions, and nano-structured materials, yet they fail to reproduce hydroxide mobility qualitatively \cite{chakraborti2024unravelling,delucas2024,sharma2025understanding,lagerweij2026}.
This limitation arises from the bond breaking and bond forming events inherent to hydroxide diffusion, which cannot be described within conventional non-reactive force field approaches \cite{tuckerman2006acs,ouma2022,zelovich2024}.

Several strategies have been proposed to overcome these methodological limitations.
The multiscale Lattice-Monte-Carlo approach extends ion dynamics to millisecond timescales by employing an analytical expression for proton transfer (PT) probabilities that depends solely on the oxygen--oxygen distance \cite{dressler2016,kabbe2016,kabbe2017,dutta2024,qaisrani2025bridging,haenseroth2025lmc}.
Machine-learned interatomic potentials (MLIPs) provide an alternative route by combining near-\textit{ab initio} accuracy with computational efficiency comparable to classical force field molecular dynamics \cite{hellstrom2018,jinnouchi23_proton,karibayev2022,grunert2025,haenseroth2025mace,flototto2026large,haenseroth2026htscreening}.

In the present work, we employ a carefully constructed MLIP to obtain atomistic insight into hydroxide dynamics in nanochannels of a commercially available AEM on timescales of several tens of nanoseconds.

Early machine-learning approaches such as Gaussian Approximation Potentials demonstrated the feasibility of data-driven force fields for molecular dynamics simulations \cite{bartok2010,unke2021,friederich2021,reiser2022,batzner2022}.
Substantial progress has been achieved through the transition from traditional descriptors, including atom-centered symmetry functions and smooth overlap of atomic positions, toward the atomic cluster expansion (ACE), which provides a flexible and efficient representation of local atomic environments \cite{behler2007,bartok2010,bartok2013representing,drautz2019}.
The introduction of equivariant graph neural networks has further enhanced predictive accuracy by incorporating symmetry constraints directly into the model architecture \cite{unke2021,reiser2022}.
A major breakthrough has been the development of MLIPs trained on extensive materials databases such as the Materials Project, giving rise to foundation models with broad transferability \cite{mp_1,mp_2,pymatgen,mptrj}.
These foundation models enable efficient fine-tuning for specific chemical systems while retaining general applicability across diverse materials classes \cite{mace_mp,grace_2}.
As a result, highly accurate and versatile MLIP frameworks have emerged that substantially expand the scope of molecular dynamics simulations at near-\textit{ab initio} accuracy \cite{mace_1,mace_2,mace_mp,batzner2022,grace_2}.
In this study, we focus on the \textsc{MACE} framework, one of the most promising publicly available MLIP approaches for describing many-body atomic interactions \cite{mace_1,mace_2,mace_mp}.

Understanding with the help of such MLIPs the hydroxide ion transport mechanism under realistic operating conditions remains a central challenge for AEM development.
In general both OH$^-$ and H$_3$O$^+$ exhibit anomalously high mobilities in water compared to other ions due to the Grotthuss mechanism \cite{grotthuss, tuckerman1995jpc}.
This mechanism involves proton hopping along hydrogen-bonded water networks followed by structural reorganization of the solvent \cite{grotthuss,tuckerman1995jpc,marx2006,tuckerman2006acs}.
It enables rapid charge transport without requiring long-range mass diffusion of individual ions \cite{tuckerman2010rev}.
Within AEM materials, hydroxide transport is strongly influenced by hydration and temperature \cite{henkensmeier2020}.
Water content and thermal conditions directly affect ionic conductivity, reaction kinetics, and membrane stability \cite{franck1965,marino2014hydroxide,henkensmeier2020}.
In this work, we investigate hydroxide mobility in fumasep{\texttrademark} FAA-3 from Fumatech BWT GmbH \cite{henkensmeier2020}.
We consider a broad range of membrane hydration levels ($\lambda=3$--$50$) and temperatures between $298$--$370$~K to establish a comprehensive picture of hydroxide transport mechanisms inside the AEM nanochannels \cite{henkensmeier2020,miller2020,busacca22019}.
The hydration parameter $\lambda$ denotes the number of water molecules per functional group, with low $\lambda$ corresponding to dry membrane conditions and high $\lambda$ to well-hydrated states.

\begin{figure}[ht]
    \centering
    \includegraphics[width=0.48\textwidth]{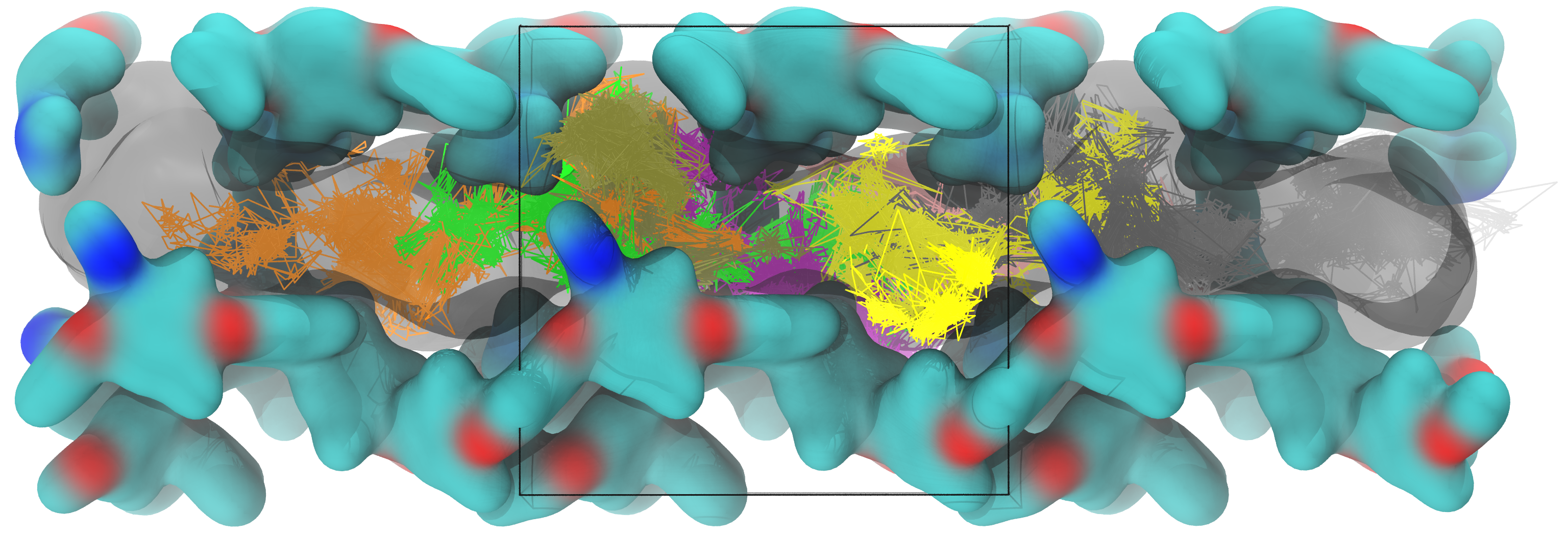}
    \caption{\textbf{Hydroxide ion diffusion within a single anion-exchange membrane channel.}
    Diffusion pathways of hydroxide ions over $0.5$~ns within the membrane channel at hydration level $\lambda = 10$ and temperature $350$~K, obtained from fine-tuned MLIP molecular dynamics simulations. 
    Each colored line represents the trajectory of one of the eight hydroxide ions. 
    Hydrogen, carbon, oxygen, and nitrogen atoms in the polymer backbone are shown in white, turquoise, red, and blue, respectively. 
    The black lines outline the simulation cell of one periodic image with a width of $20$~\AA. 
    The single-channel systems comprise up to 1,700~atoms, while larger systems containing more than 30,000~atoms were also investigated in this work. 
    For additional details, see Figure~\ref{fig:2}, Supplementary Note~7 and the Methods section.}
    \label{fig:0}
\end{figure}

The molecular structure of commercially available AEMs is often proprietary, which has historically limited the realism of theoretical models \cite{zelovich2019mimics}.
Recently, multinuclear solid-state NMR studies by Giovanelli et al.\ have provided detailed structural insight into the polymer architecture of the FAA-3 membrane \cite{giovanelli24_faa3}.
These investigations reveal that the ionomer is based on functionalised poly(\textit{p}-phenylene oxide) (PPO).
The phenyl rings of PPO are functionalised with \mbox{-CH$_2$N$^+$(CH$_3$)$_2$(CH$_2$CH$_3$)} groups that serve as cationic sites for anion exchange.
The degree of functionalisation is approximately $40$~\mbox{mol-\%} for PPO.
In the commercially available material, bromide ions act as counterions.
The FAA-3-PK variant additionally contains polyether ether ketone (PEEK) alongside functionalised PPO \cite{giovanelli24_faa3}.
PEEK does not interfere with anion conduction mechanisms and therefore preserves hydroxide transport pathways \cite{giovanelli24_faa3}.
Instead, it provides mechanical reinforcement and reduces excessive swelling upon water uptake \cite{giovanelli24_faa3,favero2024}.

Building on these structural insights and advances in MLIP methodology, the present work establishes a comprehensive molecular picture of hydroxide ion transport in realistic AEM nanochannels (see Figure \ref{fig:0}).
By systematically varying hydration and temperature, we bridge the gap between earlier theoretical studies based on artificial membrane models and experimentally relevant commercial systems.
The large-scale---with multi-channel systems with up to 32,000 atom---near-\textit{ab initio} accurate simulations presented here enable the calculation of converged diffusion coefficients under practical operating conditions.
These findings provide fundamental insight into hydroxide transport mechanisms and contribute to the targeted optimization of AEM materials for efficient and durable green hydrogen production and fuel cell technologies.

\section*{Results}

\subsection*{Structure of the membrane nanochannels}

\begin{figure*}[]
    \centering
    \includegraphics[width=\textwidth]{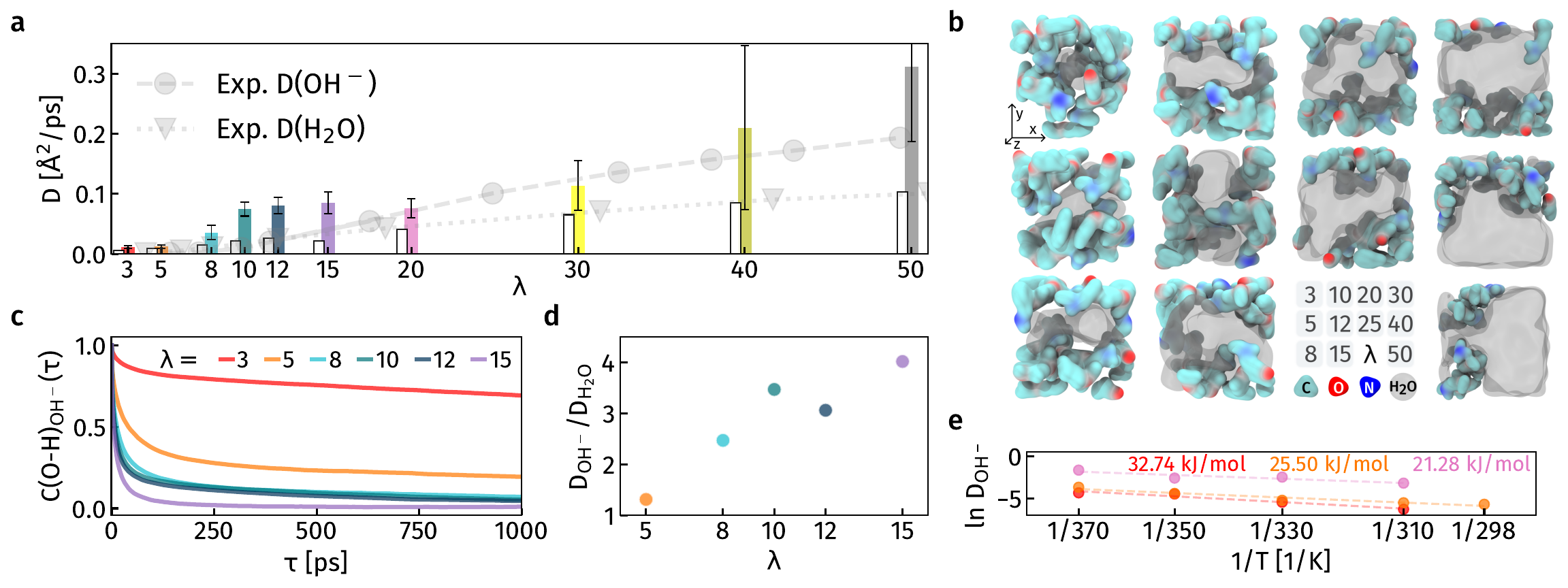}
    \caption{\textbf{Hydroxide diffusion in anion-exchange membrane channels as a function of hydration.}
    \textbf{a}~Diffusion coefficients of hydroxide ions (colored bars) and water molecules (white bars) as a function of membrane water content $\lambda$, defined as the number of water molecules per functional group.
    Hydroxide error bars represent the standard deviation obtained by evaluating the diffusion of each OH$^-$ ion individually within the same trajectory.
    Simulation results are derived from fine-tuned MLIP molecular dynamics simulations at $350$~K.
    Experimental diffusion coefficients from Marino et al. \cite{marino2014hydroxide}, measured at $298$~K, are shown for comparison.
    A enlarged representation of the low hydration regime is provided in Supplementary Note~3.
    \textbf{b}~Representative membrane channel morphologies viewed along the $z$ direction.
    The aqueous phase is rendered as a gray continuous medium.
    Carbon, oxygen, and nitrogen atoms in the polymer strands are colored turquoise, red, and blue, respectively.
    The legend summarizes the structural appearance at different $\lambda$ values.
    All structures are shown at identical visual scale for direct comparison of morphology.
    Consequently, the apparent channel dimensions do not reflect their absolute physical size at the respective hydration levels.
    \textbf{c}~Oxygen–hydrogen bond lifetime within hydroxide ions, expressed as the fractional (O--H)$_{\text{OH}^-}$ bond correlation function, at different hydration levels.
    \textbf{d}~Ratio of hydroxide to water diffusion coefficients, $D$(OH$^-$)/$D$(H$_2$O), highlighting the contribution of structural proton transfer to long-range transport.
    \textbf{e}~Arrhenius representation of hydroxide diffusion coefficients and corresponding activation energies assuming Arrhenius behavior.
    The underlying numerical data are compiled in Supplementary Note~2.
    }
    \label{fig:1}
\end{figure*}

The membrane channel models of fumasep{\texttrademark} FAA-3 were equilibrated by NPT molecular dynamics simulations for $10$~ns using the fine-tuned \textsc{MACE} MLIP as described in the Methods and Supplementary Note~1.
Each model consists of four polymer strands and a hydration-dependent number of water molecules, resulting in systems containing up to 1,700~atoms.
The polymer strands extend along the $z$-axis, as illustrated in Figure~\ref{fig:1}b.
A pronounced restructuring of the polymer–water morphology is observed with increasing hydration level $\lambda$.
At low $\lambda$ values the polymer strands form a confined environment that leaves only narrow aqueous regions in between, such that water diffusion occurs---if observed---predominantly along the channel direction.
The cationic functional groups \mbox{-CH$_2$N$^+$(CH$_3$)$_2$(CH$_2$CH$_3$)} are primarily oriented toward this central water-filled region.
With increasing water content the polymer strands are progressively pushed apart and distinct aqueous channels emerge.
At intermediate hydration levels water molecules frequently separate neighboring polymer strands, and the functional groups no longer exclusively point toward a common channel center but also toward adjacent polymer segments.
At very high $\lambda$ values extended water compartments dominate the morphology, while the polymer strands exhibit a tendency to cluster.
These structural transformations establish the microscopic framework for the hydration-dependent transport properties discussed below.

\subsection*{Hydration-dependent diffusion behavior} 

The diffusion of hydroxide ions and water molecules was investigated for all hydration conditions at $350$~K (for all temperatures see Supplementary Note~2).
Diffusion coefficients were obtained from NVT simulations of up to $25$~ns using the fine-tuned \textsc{MACE} potential as detailed in the Methods and Supplementary Note~1.
Figure~\ref{fig:1}a summarizes the predicted diffusion coefficients of hydroxide ions and water molecules in comparison with experimental data \cite{marino2014hydroxide,zhegur2020measuring}.
The error bars for hydroxide diffusion reflect the standard deviation derived from analyzing each OH$^-$ ion individually within the same trajectory.
The simulations reproduce the experimentally observed increase of both water and hydroxide diffusion coefficients with increasing water content.

For very low hydration levels experimental data are scarce, yet the simulations predict systematically reduced mobilities under these dry conditions.
Although the calculated diffusion coefficients at $\lambda \leq 15$ slightly exceed experimental values, the overall trend is captured accurately (see Supplementary Note~3).
At intermediate hydration the mobility increases dramatically, with $D$(OH$^-$) at $\lambda=20$ being approximately five times larger than at $\lambda=5$.
At $\lambda \geq 40$ the system approaches the limit of polymer strands embedded in a diluted aqueous hydroxide solution.
Under these conditions the water diffusion coefficient converges toward the experimental value of neat water, $0.19$~{\AA$^2$/ps} \cite{tuckerman2006acs}.
Similarly, hydroxide mobility becomes comparable to that in diluted potassium hydroxide solution with $0.31$~{\AA$^2$/ps} \cite{tuckerman2006acs}.

A substantial spread in hydroxide diffusivities is observed in well-hydrated systems.
At $\lambda=50$ the average diffusion coefficient is $0.31$~{\AA$^2$/ps}, while individual ions range from $0.16$ to $0.48$~{\AA$^2$/ps}.
The fastest ions spend significantly less time within $3.0$–$4.0$~{\AA} of the nearest nitrogen atom of a functional group compared to the slowest ions which is twice as often in this region.
This heterogeneous local environment explains the larger statistical variance in hydroxide mobility at high hydration.
In contrast, under dry conditions the diffusivities of individual hydroxide ions are more uniform, reflecting the uniformly confined morphology.

\subsection*{Proton transfer kinetics and free energy barriers}

The proton transfer probability as a function of oxygen–oxygen distance is reduced for $\lambda=3$ and $5$ compared to higher hydration levels, where the behavior resembles that of aqueous potassium hydroxide obtained from AIMD simulations (see Supplementary Note~4).
The absolute number of proton transfer events per hydroxide ion and per picosecond increases from less than one at $\lambda=3$ to approximately four to five at $\lambda=10$, $12$, and $15$.
At even higher hydration levels the frequency decreases to roughly two events per hydroxide ion and per picosecond.
When excluding non-productive back-and-forth transfers between the same oxygen atoms, the fraction of such wiggling events increases with $\lambda$, showing a higher ratio of productive PT (see Supplementary Note~5).
Free energy profiles for proton transfer provide further insight and can give an explanation for lower PT frequency at higher $\lambda$.
For $\lambda<20$ a barrier of approximately $0.14$~eV is obtained, whereas at high hydration the barrier increases slightly to values between $0.18$ and $0.23$~eV (see Supplementary Note~6).

\begin{figure*}[]
    \centering
    \includegraphics[width=\textwidth]{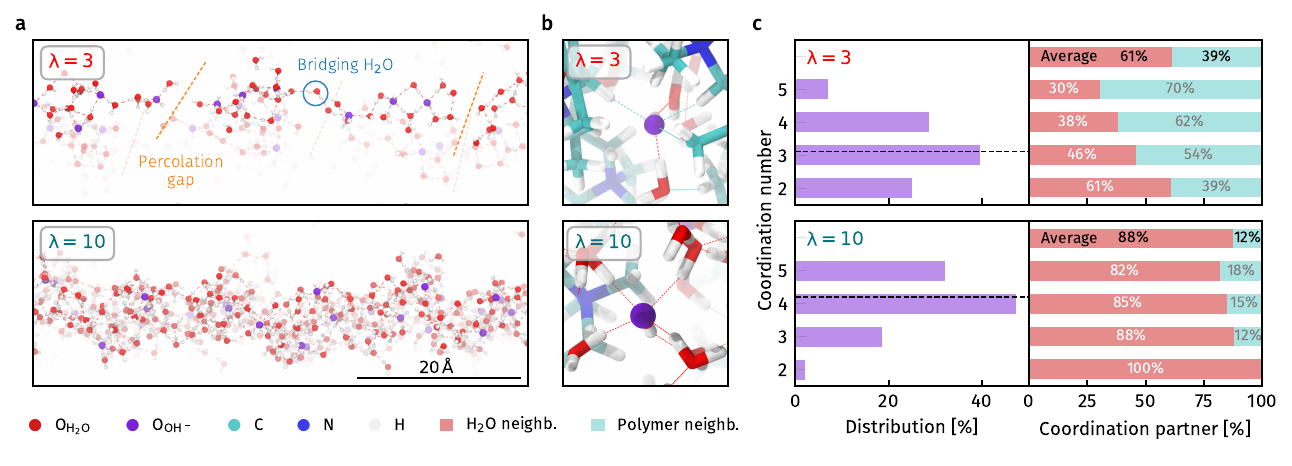}
    \caption{\textbf{Hydrogen-bond network topology and hydroxide coordination in membrane nanochannels.}
    \textbf{a}~Hydrogen-bond network within the aqueous phase of the membrane channel at $\lambda=3$ (upper panel) and $\lambda=10$ (lower panel).
    The polymer backbone and functional groups are omitted for clarity.
    Hydrogen atoms are shown in white, oxygen atoms of water molecules in red, and oxygen atoms of hydroxide ions in purple.
    Hydrogen bonds are indicated by red dashed thin lines.
    Orange dashed lines mark percolation gaps between disconnected water clusters.
    The blue circles highlights water molecules that transiently bridge two clusters and act as dynamic bottlenecks for proton transport.
    Snapshots are taken from fine-tuned MLIP molecular dynamics simulations of the large simulation cells, as described in Supplementary Note~7 and in Methods.
    \textbf{b}~Representative coordination geometries of a hydroxide ion in the membrane channel at $\lambda=3$ (upper panel) and $\lambda=10$ (lower panel), extracted from AIMD structures.
    Water oxygen atoms are colored red, hydroxide oxygen atoms purple, carbon atoms turquoise, nitrogen atoms blue, and hydrogen atoms white.
    Hydrogen bonds are shown as dashed lines, colored red when the coordination partner is a water molecule and turquoise when coordinated by the polymer.
    \textbf{c}~Distribution of hydroxide coordination numbers obtained from AIMD molecular dynamics trajectories at $350$~K for $\lambda=3$ (upper panel) and $\lambda=10$ (lower panel).
    The right-hand histograms display the average contribution of coordination partners as a function of coordination number, with water neighbors shown in red and polymer contributions in turquoise.
    The coordination number is determined by identifying the two nearest atoms for each hydrogen atom.
    }
    \label{fig:2}
\end{figure*}

\subsection*{Grotthuss contribution and coordination environment}

Trimethyl alkyl ammonium groups (TMA) are known to exhibit kosmotropic behavior, as their relatively small cationic head group strongly structures the surrounding water network and can substantially reduce water mobility at low $\lambda$ values \cite{zelovich23}.
The functional group in the present membrane is chemically closely related to TMA, which provides a structural-level explanation for the strongly reduced water diffusion coefficient observed in the low hydration regime. 
The ratio $D$(OH$^-$)/$D$(H$_2$O), referred to as the Grotthuss quota, exhibits a strong dependence on hydration (see Fig. \ref{fig:1}d).
Under dry conditions this ratio is close to unity, indicating that hydroxide mobility is dominated by vehicular diffusion similar to water.
With increasing $\lambda$ the ratio rises toward values of $3$–$4$, as known for diluted potassium hydroxide solutions \cite{tuckerman2006acs}.
This behavior correlates with the hydrogen bond lifetimes formed by the oxygen atoms of hydroxide ions and hydrogen atoms of water molecules shown in Figure~\ref{fig:1}c.
A rapidly decaying correlation function corresponds to short O--H bond lifetimes and thus frequent proton transfer events.
At $\lambda=3$ and $5$ the limited number of coordinating water molecules suppresses efficient proton hopping.
The dynamical hyper-coordination mechanism, which involves transitions between threefold (PT-active) and fourfold (PT-inactive) coordinated hydroxide states, becomes increasingly accessible with rising hydration \cite{tuckerman1995jpc,tuckerman2002nature,tuckerman2006acs,tuckerman2010rev}.
If the hydroxide hydrogen atom is donating an H-bond to a nearby water molecule coordination numbers of $5$ can be reached.
At $\lambda=3$, only three water molecules are available per hydroxide ion. 
From this, one could numerically estimate that the Grotthuss mechanism is hindered or even impossible to observe in the trajectory.
Because the hyper-coordination concept proposes hydroxide ion coordination geometries build up from $3$ and $4$ water molecules.

\subsection*{Hydration-dependent H-bond network and long-range proton transport}

We have also computed $70$~ps AIMD simulations of the membrane channel at $\lambda=3$ and $10$ (see Methods).
Evaluating the coordination environment of the hydroxide ions throughout these trajectories for $\lambda=3 $, leads to an average coordination number of $3.1$. 
The corresponding distribution is shown in Figure~\ref{fig:2}c in the upper panel.
Most hydroxide ions are coordinated by $2$, $3$, or $4$ neighbors under these dry conditions.
Because of the limited number of water molecules, hydrogen atoms corresponding to the functional groups donate H-bonds to OH$^-$, as illustrated in Figure~\ref{fig:2}b (upper panel).
This additional polymer contribution enables coordination numbers larger than three.
The coordination partner analysis in Figure~\ref{fig:2}c reveals that on average $61$~\% of the neighbors are water molecules and $39$~\% originate from the polymers.

At higher coordination numbers the fraction of polymer participation further increases.
Although threefold coordination can enable proton transfer events, efficient long-range proton migration requires a continuous wide network of water molecules or hydroxide ions that supports consecutive transfers. 
Inspection of the aqueous nano-structure at $\lambda=3$ demonstrates that such connectivity is not fully established.
As shown in Figure~\ref{fig:2}a, the dry membrane contains temporary and partially interconnected water clusters embedded between the polymer strands.
These clusters are intermittently separated by regions without hydrogen-bond connectivity.
We refer to these interruptions as percolation gaps because they impede long-range proton transport across the channel.
In some instances single water molecules transiently bridge two clusters and form the only hydrogen-bond connection between them.
This molecule acts as a dynamic connectivity bridge that temporarily enables long-range proton transfer, yet frequent rearrangements can rapidly disrupt this pathway.
As a consequence, the H-bond network remains spatially confined.

In contrast, at $\lambda=10$ the aqueous phase forms a continuous and robust hydrogen-bond network, as depicted in Figure~\ref{fig:2}a (lower panel).
The PT-inactive configuration of a hydroxide ion under these hydrated conditions is shown in Figure~\ref{fig:2}b (lower panel), where the ion is coordinated by four water molecules and donates a hydrogen bond to a neighboring water molecule.
The average coordination number increases to $4.2$, with $88$~\% water coordination and only $12$~\% polymer contribution.
Here the same trend as at $\lambda=3$ is observed: at higher coordination number a greater part of polymer-based coordination can be observed.
These structural differences provide a direct explanation for the hydration dependence of the activation energy displayed in Figure~\ref{fig:1}e.

The activation energy derived from hydroxide diffusion coefficients decreases from $33$~kJ/mol at $\lambda=3$ to $21$~kJ/mol at $\lambda=20$.
The experimentally determined value of $27$~kJ/mol obtained from conductivity measurements under hydrated conditions lies within this range, as summarized in Table~\ref{tab:tab1} (see Methods).

\begin{table}[ht]
    \centering
    \caption{
    \textbf{Experimental conductivity measurements and activation energy.}
    Conductivity measurements of fumasep{\texttrademark} FAA-3-PK-130 in aqueous $0.1$~mol~l$^{-1}$ KOH solution. 
    The corresponding Nyquist plots are shown in Supplementary Note~8. 
    T is absolute temperature, the resistance R, area-specific resistance (ASR), conductivity ($\sigma$).
    Leading to an activation energy of E$_\mathrm{a} = 27$ kJ/mol.}  
    \begin{tabular}{l@{\hspace{2em}}c@{\hspace{2em}}c@{\hspace{2em}}c}
        \toprule
        T [K] & R [$\Omega$] & ASR [$\Omega$ cm$^2$] & $\sigma$ [S cm$^{-1}$] \\
        \midrule
        323 & 17.65 & 31.59 & 0.32 \\
        333 & 11.04 & 19.77 & 0.51 \\
        343 & 9.85  & 17.64 & 0.57 \\
        \bottomrule
    \end{tabular}
    \label{tab:tab1}
\end{table}

\begin{figure*}[]
    \centering
    \includegraphics[width=\textwidth]{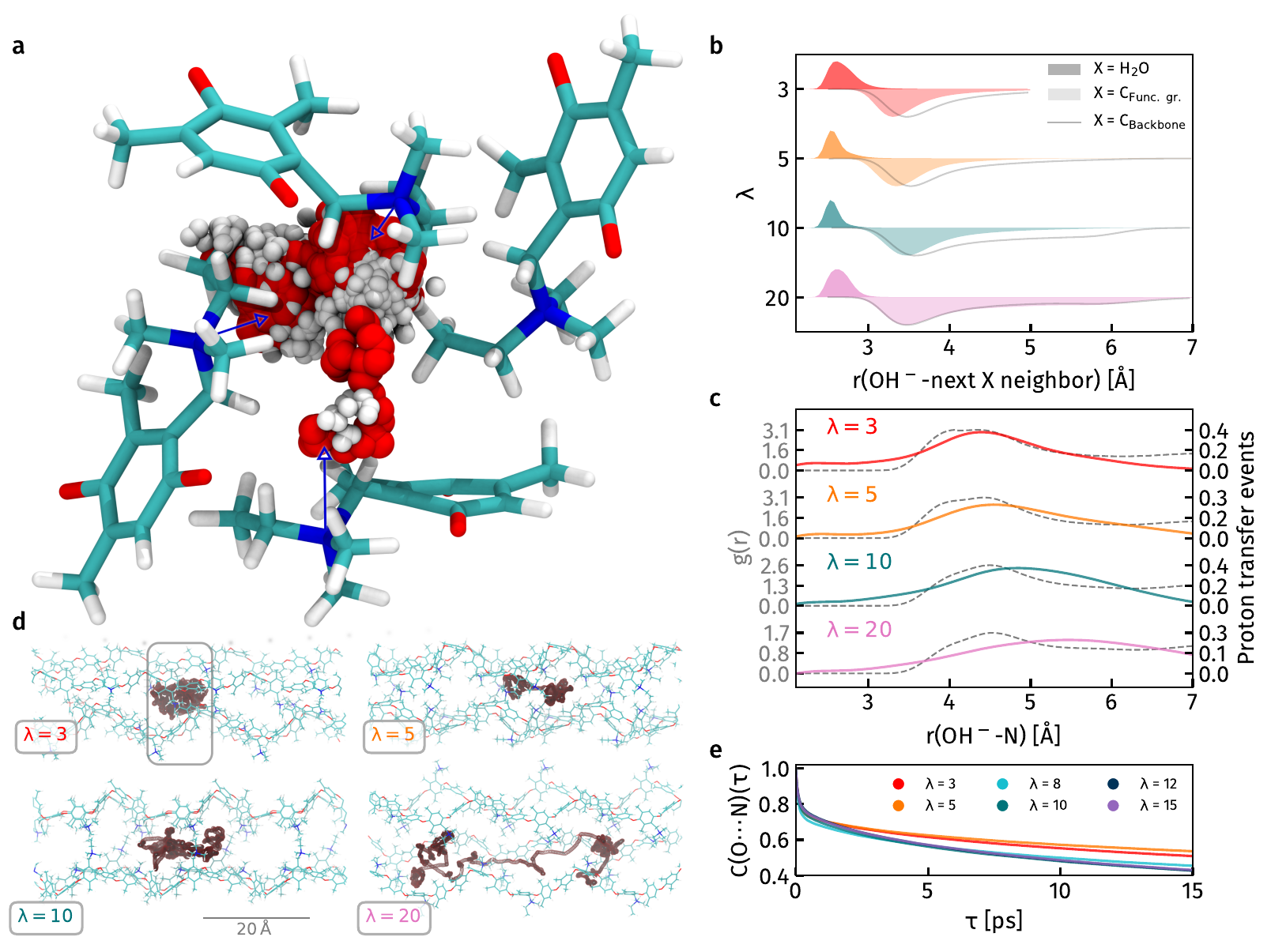}
    \caption{\textbf{Long-range hydroxide transport and interactions with functional groups.}
    \textbf{a}~Diffusion pathway of a single OH$^-$ ion over $60$~ps within a thin slice of the membrane channel at $\lambda=3$ and $350$~K.
    For clarity, only a confined section of the simulation cell along the $z$ direction is shown. 
    The four polymer strands are represented by one monomer unit each, and the backbone is truncated at the oxygen atoms for clarity.
    Oxygen, carbon, nitrogen, and hydrogen atoms are colored red, turquoise, blue, and white, respectively.
    The trajectory is extracted from the AIMD simulation at $350$~K and $\lambda=3$. 
    \textbf{b}~Distance distributions between hydroxide ions and their nearest neighboring species at different hydration levels.
    The intense colored filled curves represent the OH$^-$–water oxygen distance for each $\lambda$.
    For clarity, the distributions of distances to the nearest functional group carbon atom and to the nearest backbone carbon atom are plotted upside down.
    The OH$^-$–functional group carbon distribution is shown in semi-transparent color, whereas the OH$^-$–backbone carbon distribution is indicated by a gray line.
    \textbf{c}~Radial distribution functions between hydroxide oxygen atoms and nitrogen atoms of the functional groups for different hydration levels, are shown as dashed gray lines.
    The colored solid curves display the distribution of proton transfer events as a function of the instantaneous OH$^-$–nitrogen distance.
    \textbf{d}~Hydroxide diffusion pathways over $0.5$~ns within the membrane channel at $\lambda=3$, $5$, $10$, and $20$, shown in the upper left, upper right, lower left, and lower right panels, respectively.
    Trajectories are obtained from fine-tuned MLIP molecular dynamics simulations at $350$~K.
    \textbf{e}~Partnership correlation function between hydroxide ions and their nearest nitrogen neighbors, illustrating the residence time of hydroxide ions in proximity to functional groups at different hydration levels. 
    }
    \label{fig:3}
\end{figure*}

A direct visualization of hydroxide trajectories over $0.5$~ns, shown in Figure~\ref{fig:3}d, illustrates the progressive transition from confined to long-range transport.
At $\lambda=3$ the hydroxide ion remains trapped within a single water cluster.
At $\lambda=5$ transitions between neighboring clusters become possible but remain spatially limited.
At $\lambda=10$ no isolated clusters are visible and frequent proton transfer events are observed.
At $\lambda=20$ long-range proton migration spanning several tens of {\AA} occurs within the continuous aqueous channel.

Closer inspection of the membrane interior at $\lambda=3$ reveals that hydroxide ions preferentially reside near the cationic functional groups of the polymer strands, in agreement with previous reports \cite{zelovich21_ohvsh3o,zhuoran23cylindernano,zelovich23}.
The trajectory of one representative hydroxide ion over $100$~ps is shown in Figure~\ref{fig:3}a.
The oxygen atom of OH$^-$ is typically oriented toward the nitrogen atom of the positively charged group, while the hydrogen atom points into the aqueous region.
The visual evaluation of all hydroxide ions in this channel is shown in Supplementary Note~9.

A statistical analysis of nearest-neighbor distances supports this observation (see Fig. \ref{fig:3}b).
The closest neighbor is most often a water oxygen atom at all hydration levels, yet at low $\lambda$ the distance distributions of water oxygen and functional group carbon atoms overlap significantly more than in hydrated systems.
The overlap amounts to $11$~\% at $\lambda=3$ and decreases to $3$~\% at $\lambda=20$ as shown in Supplementary Note~10.
This indicates a substantially higher probability of close hydroxide–functional group contact in the dry regime.
With increasing hydration the average hydroxide--functional group distance shifts to larger values, and the distinction between backbone and functional group carbons---clearly observable at lower $\lambda$ values---diminishes. 
In the dry regime the carbon atoms from the functional group are closer to the hydroxide ions than the carbon atoms from the polymeric backbone.

The radial distribution function (RDF) between hydroxide oxygen and nitrogen atoms exhibits a corresponding shift of the first maximum from $3.8$ to $4.5$~{\AA} as $\lambda$ increases, as shown in Figure~\ref{fig:3}c.
Proton transfer events at low hydration occur predominantly at distances corresponding to this first RDF peak.
In well-hydrated membranes the distribution of proton transfer distances shifts toward larger separations, indicating that proton hopping increasingly occurs within the aqueous phase rather than in direct proximity to the cationic head groups.
At this hydration levels the cationic head groups are not directly involved in the Grotthuss mechanism, the hydroxide ion conducts this structural diffusion mechanism in the aqueous phase distant from the functional groups. 

The lifetime analysis of the closest hydroxide–nitrogen association further supports this interpretation.
The partnership correlation decays with a half-life of approximately $15$~ps at $\lambda=3$ and $7.5$~ps for $\lambda>5$, as shown in Figure~\ref{fig:3}e.
The shorter residence time at higher hydration reflects reduced trapping by functional groups and enhanced mobility of hydroxide ions.
This dynamic decoupling from the polymer matrix establishes the structural prerequisite for sustained long-range proton transport inside the membrane channel.

\section*{Discussion}

In this work we combine recent structural insight into a commercial anion-exchange membrane with large-scale molecular dynamics simulations based on a fine-tuned machine-learned interatomic potential to uncover the molecular origin of hydroxide transport.
The results demonstrate that hydration governs a structural transition from isolated water clusters confined by polymer strands to a percolating hydrogen-bond network that enables long-range proton migration.
Under dry conditions hydroxide ions remain trapped near cationic functional groups and transport is dominated by localized motion within disconnected aqueous domains.
With increasing water content the polymer–ion interaction weakens, the aqueous phase becomes continuous, and structural diffusion through consecutive proton transfer events emerges as the dominant transport mechanism.

Beyond the macroscopic transport trends, this study provides a level of molecular detail that, to our knowledge, has not previously been reported for a realistic membrane material at this time- and length-scale.
The coordination geometries of hydroxide ions within an operating membrane environment are resolved with explicit consideration of their interactions with the functional groups.
In addition, proton transfer rates, free energy profiles along the proton transfer coordinate, and correlation analyses linking hydroxide motion to its instantaneous molecular environment are obtained within the same consistent framework.
Importantly, these insights are derived for a chemically and structurally resolved membrane system rather than for a strongly simplified model representation.
This enables a direct connection between atomistic interactions, nano-scale morphology, and experimentally accessible transport coefficients.

The calculated diffusion coefficients and activation energies follow the experimentally observed trends, supporting the validity of the approach.
At high hydration levels the predicted mobilities approach those of dilute aqueous hydroxide solutions, indicating that the simulations capture the essential physics of the system.

Despite this overall agreement, several sources of uncertainty must be considered.
First, the machine-learned potential is obtained by fine-tuning on a limited set of reference data of a subsystem for a chemically complex and comparatively large system.
Second, the membrane channel is constructed manually based on available structural information therefore we equilibrated the systems for $10$~ns in the isothermal–isobaric ensemble.
While this equilibration ensures a stable nano-structure, the true morphology of the polymer network may exhibit additional heterogeneity that is not fully sampled within the accessible simulation time and system size.
Third, the underlying density functional approximation used to generate the reference data introduces systematic functional errors.
Only selected functionals such as BLYP reliably reproduce the correct hydroxide coordination geometry and proton transfer mechanism.
However, this functional is known to underestimate diffusion coefficients compared to experiment, which may partly explain the remaining quantitative deviations \cite{blyp_overstructure1,blyp_overstructure2}.

Furthermore, for the most confined systems at $\lambda=3$ and $5$, simulations in the canonical ensemble do not yield fully converged diffusion coefficients within the accessible trajectory length.
In these cases hydroxide and water mobility strongly depend on the instantaneous nano-structure of the disconnected water clusters.
Significantly longer simulations would be required to sample rare restructuring events that reconnect clusters and enable larger displacements.
An alternative strategy would be to perform multiple statistically independent replicas of the same hydration state to average over morphological variations.
These limitations highlight the intrinsic challenge of describing ion transport in heterogeneous polymer environments across multiple time and length scales.

The proposed workflow establishes a predictive framework for the rational design of anion-exchange membranes.
Because hydroxide transport can now be quantitatively described at the molecular level within realistic membrane nano-structures, key materials parameters such as the chemical identity of functional groups, the degree of functionalization, and the architecture of the polymer backbone can be systematically explored in silico.
This capability enables the direct prediction of hydroxide diffusion and transport mechanisms without requiring empirical parametrization or prior experimental input.
Consequently, membrane chemistries can be screened and optimized computationally before synthesis, allowing the identification of promising design strategies for maximizing ion mobility while maintaining structural stability.
From a broader perspective, this represents a long-sought objective in computational membrane science: the ability to move beyond post hoc interpretation of experimental observations toward truly predictive, simulation-guided materials design for next-generation anion-exchange membranes.

Future work should further analyze whether specific membrane pockets or nano-scale cavities exist that can selectively capture or transiently trap hydroxide ions and thereby modulate long-range mobility.
Another important direction concerns membrane stability.
The pronounced interactions between hydroxide ions and functional groups in the low hydration regime, revealed here by frequent contact events and prolonged residence times, may represent a molecular pre-stage of degradation pathways that are widely discussed in the literature.
The present molecular dynamics perspective extends this discussion to nanosecond time scales and provides a dynamic view of how local chemical environments evolve under operating conditions.

Nevertheless, the present study establishes a consistent microscopic picture that links polymer–ion interactions, hydrogen-bond network connectivity, and macroscopic transport properties.
The results demonstrate that machine-learned molecular dynamics can bridge the gap between first-principles accuracy and experimentally relevant time scales for complex membrane materials.
This approach opens a pathway toward predictive simulation-guided optimization of anion-exchange membranes for efficient and durable green hydrogen production.

\section*{Methods}

In this study, approximately $0.4$ microseconds of near-\textit{ab initio} accuracy MD simulations of anion exchange membrane channels were performed, corresponding to $15$~thousand GPU hours. 
The high-throughput MLIP simulation workflow (including input generation and post-processing) was automated using our toolkit \cite{haenseroth2025atk} (see Code Availability statement).
For the data supporting this work please refer to the Data Availability statement.

\subsection*{AIMD simulations}

AIMD simulations were performed on CPUs using CP2K (version 2025.1) with the BLYP exchange-correlation functional, GTH pseudopotentials, DZVP-MOLOPT basis set, and the Nos\'{e}-Hoover chain thermostats \cite{cp2k_1, cp2k_2, cp2k_3, cp2k_4, cp2k_5, cp2k_quickstep, cp2k_basis-set, cp2k_orb_trans, cp2k_gth-pseudopot1, cp2k_gth-pseudopot2, cp2k_gth-pseudopot3, blyp1, blyp2, nose1, nose2, nose3}.
The training dataset of 2,500 data points was generated through a first-principle MD simulations of a representative subsystem of the membrane channel (see Supplementary Note~11).
This subsystem comprised one functionalised PPO polymer chain constructed from five monomers (two functionalised), four hydroxide ions, two potassium ions, and $55$ water molecules ($296$ atoms total), yielding a $\lambda$ value of $27.5$. 
This system is at the boundary between medium and high $\lambda$ values with elevated hydroxide ion concentration, captures a comprehensive range of relevant intra- and intermolecular interactions occurring in the full membrane channel system, ensuring high accuracy for simulations of larger systems. 
Previous work has demonstrated that foundation models fine-tuned at medium hydroxide concentrations (equivalent to medium $\lambda$ conditions) exhibit broad concentration transferability \cite{haenseroth2025mace}.

For validation AIMD simulations of membrane channel systems consisting out of four polymer strands and the respective number of water molecules the systems at $\lambda=3$ and $10$ were conducted for $100$~ps and $70$~ps at $350$~K with the same settings.

\subsection*{\textsc{MACE} foundation model fine-tuning}

The \textsc{aMACEing\_toolkit} was employed to establish the fine-tuning protocol for refining the \textsc{MACE} (version 0.3.14) foundation model \textsc{MACE}-MP-0-small using the following hyperparameter: batch size $5$, $200$ epochs, learning rate $0.01$, and force-energy weight ratio $100$ on GPU \cite{mace_1, mace_2, mace_mp, haenseroth2025atk}. 
The root-mean-square errors of the forces and energies amount to $0.025$~eV/{\AA} and $0.001$~eV/atom for the training set, and $0.026$~eV/{\AA} and $0.001$~eV/atom for the validation set, respectively.
To further assess transferability, the fine-tuned foundation model was evaluated by recalculating frames from the membrane channel system (four polymer strands and water) MD trajectories using first-principles methods at two different $\lambda$ values, yielding force errors of $0.116$~eV/{\AA} ($\lambda = 3$) and $0.097$~eV/{\AA} ($\lambda = 10$). 
These accuracies are sufficient for molecular dynamics simulations and comparable to other methods such as Gaussian Approximation Potentials.
The resulting model was also evaluated by comparing the radial distribution functions at different lambda values in comparison with AIMD simulations (see Supplementary Note~11). 

\subsection*{\textsc{MACE} MD simulations}

MD simulations for property evaluation were performed using \textsc{MACE}-torch (version 0.3.14), \textsc{LAMMPS} and \textsc{ASE} with system-specific temperatures controlled by Nos\'{e}-Hoover chain thermostats \cite{mace_1,mace_2,mace_mp,lammps,ase, nose1, nose2, nose3}. 
Calculations were executed on the compute cluster of Technische Universität Ilmenau using GPUs for interference runs.

\subsection*{Membrane channel model construction}

Membrane structures within aqueous environments were constructed from scratch using information from the multinuclear solid-state NMR study, \textsc{Packmol}, and \textsc{Avogadro2} \cite{giovanelli24_faa3, martinez2009packmol, hanwell2012avogadro, avogadro2}. 
The constructed systems comprised four polymer chains, each built from five monomers ($40$~\% functionalised), totaling $484$ atoms, $8$ hydroxide ions, and the $\lambda$-specific number of water molecules. 
Structures for different values of $\lambda~\in~\{ 3, 5, 8, 10, 12, 15, 20, 25, 30, 40, 50\}$ at various temperatures: $298$~K, $310$~K, $330$~K, $350$~K, $370$~K, were prepared, resulting in membrane channels containing up to nearly 1,700 atoms (see Supplementary Note~1).

First-principles geometry optimization of the systems was performed using \textsc{CP2K}, followed by $10$~ns NPT MD simulations using the fine-tuned \textsc{MACE}-MP-0 foundation model for equilibration to ensure reasonable membrane channel structure design. 
Calculations for analyzing hydroxide ion mobility within membrane channels consisted of NVT MD simulations with the fine-tuned \textsc{MACE}-MP-0 foundation model for different trajectory length due to system size constraints, with up to $25$~ns (see Supplementary Note~1).

A second set of membrane models containing multiple channels was generated for the hydration levels $\lambda=3$ and $\lambda=10$ by constructing $3\times3\times5$ supercells of the equilibrated single-channel systems (see Supplementary Note~7).
The resulting simulation cells span more than $50$~{\AA}$\times50$~{\AA}$\times100$~{\AA} and contain up to 32,220~atoms.
These multichannel systems were subsequently equilibrated using $0.5$~ns NPT MD simulations with the fine-tuned \textsc{MACE}-MP-0 model.

\subsection*{Hydroxide ion diffusion and environment analyses}

The high mobility of hydroxide ions primarily results from the Grotthuss mechanism. 
In this process, the hydroxide ion is not defined by a single, fixed pair of hydrogen and oxygen atoms identifiable by their indices in the trajectory. 
Instead, the hydroxide ion comprises the oxygen atom in the trajectory that is covalently bonded to only one hydrogen atom. 
Proton transfer events cause the specific oxygen and hydrogen atoms constituting the hydroxide ion to change over time. 
To compute dynamical and static properties of hydroxide ions, the particle positions must be known at every time step. 
This is achieved by classifying oxygen atoms at each trajectory frame into water molecules (two covalently bonded hydrogen atoms) and hydroxide moieties (one covalently bonded hydrogen). 
The classification algorithm assigns each hydrogen atom to its nearest oxygen neighbor, resulting in oxygen atoms covalently bonded to either one or two hydrogen atoms, corresponding to hydroxide or water, respectively. 
Importantly, oxygen atoms in the membrane structure are not considered by this algorithm. 
For subsequent analyses, the hydroxide ion position is defined by the coordinates of the hydroxide oxygen atom. 
During molecular dynamics simulations, proton transfer manifests as a change in the nearest oxygen neighbor of a hydrogen atom, leading to interconversion between hydroxide and water molecules. 
This causes a discontinuous `jump' in the hydroxide ion position from one oxygen atom to another.

Dynamical properties of hydroxide ions were evaluated by calculating mean-squared displacements using Equation \ref{eq:msd}, with $\vec{r}_{i}(t)$ representing the position of particle $i$ at time $t$. 
The $\mathrm{MSD}(\tau)$ is calculated by averaging over all particles $i$ and all time origins $t$ satisfying $t + \tau < t_{\text{max}}$:

\begin{equation}
    \mathrm{MSD}(\tau) = \left\langle \left| \vec{r}_{i}(t+\tau) - \vec{r}_{i}(t) \right|^2 \right\rangle
    \label{eq:msd}
\end{equation}

From the mean-squared displacement, the diffusion coefficient $D$ is calculated using the relation \ref{eq:diffusion}

\begin{equation}
    \mathrm{D} = \lim_{\tau \to \infty} \frac{\mathrm{MSD}(\tau)}{6\tau}
    \label{eq:diffusion}
\end{equation}

The activation energy for hydroxide ion diffusion is determined using the Arrhenius equation given in Equation \ref{eq:arrhenius}, which describes the temperature dependence of the diffusion coefficient $D$. 
Here, $A$ is the pre-exponential factor, $T$ denotes the absolute temperature, $k_b$ is the Boltzmann constant, and $E_a$ represents the activation energy. 
By plotting the natural logarithm of the diffusion coefficient, $\ln(D)$, against the inverse temperature, $1/T$, the activation energy can be extracted from the slope of the resulting linear fit, which equals $-E_a/k_b$.

\begin{equation}
    \mathrm{D}(T) = A \ \exp\left(-\frac{E_a}{k_b T}\right)
    \label{eq:arrhenius}
\end{equation}

\subsection*{Experimental conductivity measurements}

The conductivity of the AEM-130-PK/KOH system with $c(\text{KOH})=0.1$~M was measured via electrochemical impedance spectroscopy (EIS) using a potentiostat (Zahner\textsuperscript{\textregistered} Zennium Pro equipped with a PP242 module) connected to an in-house built cell placed in a temperature-controlled oven. 
Two symmetrical round metal plates made of stainless steel with a high Ni and Mo content were used as electrodes in a two-electrode configuration. 
Prior to the measurements, the AEM-130-PK membrane was immersed in $0.1$~M KOH solution for $24$~h to ensure complete ion exchange and activation. 
The electrolyte was freshly prepared from KOH pellets ($\geq 85$~\%) using Millipore water ($18.2$~M$\Omega$cm) at $298$~K.

Due to the design of the cell, which contains a single electrolyte reservoir, the measurement was performed using two identical membranes assembled symmetrically in the cell and filled with the electrolyte $0.1$~M solution.
In this configuration, the measured impedance includes the resistance contributions of the electrolyte and two membranes connected in series.

EIS was carried out potentiostatic at $100$~mV with an amplitude of $0.01$~mV and a frequency range of $1$ to $300000$~Hz.
The impedance spectra were analyzed in the Nyquist representation, and the ohmic resistance R was determined from the high-frequency intercept on the real axis. 
To support this determination, the high-frequency region of the impedance spectra was fitted using RelaxIS~3 software with an equivalent circuit consisting of a series resistor (R1) and a parallel branch composed of a resistor (R2) in series with a Warburg element (W) and a constant phase element (CPE) (see Supplementary Note~8 and 12).
For the fitting procedure, only the first 15 data points in the high-frequency region of the impedance spectrum were considered.

First, a reference measurement was performed using the electrolyte alone under identical conditions of temperature and KOH concentration. 
The ohmic resistance associated with the membranes ($R_\text{mem,2}$) was then obtained by subtracting the electrolyte resistance ($R_\text{electrolyte}$) from the total measured resistance ($R_\text{total}$):

\begin{equation}
    R_\text{mem,2} = R_\text{total} - R_\text{electrolyte}.
    \label{eq:exp1}
\end{equation}

Since the cell configuration contains two membranes connected in series, the resistance of a single membrane was obtained by dividing this value by two.
To enable comparison independent of cell geometry, the resistance of a single membrane was converted to area-specific resistance ($\mathrm{ASR}$) according to:

\begin{equation}
    ASR = R \ \cdot \ A.
    \label{eq:exp2}
\end{equation}

where $A$ is the active area of the cell $1.79$~cm$^2$ (see Supplementary Note~13). 
The ionic conductivity $\sigma$ was calculated using a calibrated cell constant ($K_\text{cell}=5.587$~cm$^{-1}$) according to:

\begin{equation}
    \sigma = K_\text{cell} / R.
    \label{eq:exp3}
\end{equation}

Measurements were performed at $323$, $333$, and $343$~K after thermal equilibration. 

\section*{Data Availability}

The dataset containing equilibrated membrane model structures, \textsc{CP2K}, \textsc{LAMMPS} and \textsc{ASE} input files, the fine-tuning data and the resulting model are available at \url{doi.org/10.5281/zenodo.18889931}.

\section*{Code Availability}

The used third-party codes \textsc{CP2K}, \textsc{MACE}, \textsc{ASE}, \textsc{LAMMPS} are available at \url{cp2k.org}, \url{github.com/acesuit/mace}, \url{gitlab.com/ase/ase}  and \url{lammps.org}, respectively. 
The high-throughput input creator used in this study is available at \url{github.com/jhaens/amaceing_toolkit}.

\section*{Acknowledgments}

We thank the staff of the Compute Center of the Technische Universität Ilmenau, especially Mr.~Henning~Schwanbeck for providing an excellent research environment. 
This work is supported by the Carl-Zeiss-Stiftung (SustEnMat, funding code: P2023-02-008), the Thüringer Aufbaubank (TAB) (KapMemLyse, grant no.~2024 FGR 0081 / 0082) and the European Social Fund Plus (ESF+). 

We thank Dario R. Dekel and Alexander A. Baranov for the insightful discussions.

\section*{Competing interests}

The authors declare no competing interests.

\section*{Author contributions}

J.H. conceived the idea; J.H. constructed the membrane structures; J.H. performed all calculations; M.M. conducted the experimental conductivity measurements; J.H., M.Q. and C.D. analyzed the data; J.H. visualized all results and J.H. wrote the first draft of the manuscript. 
K.S., M.Q. and C.D. supervised the work; all authors revised and approved the manuscript.

\bibliography{bibliography.bib}

\begin{thebibliography}{96}%
\makeatletter
\providecommand \@ifxundefined [1]{%
 \@ifx{#1\undefined}
}%
\providecommand \@ifnum [1]{%
 \ifnum #1\expandafter \@firstoftwo
 \else \expandafter \@secondoftwo
 \fi
}%
\providecommand \@ifx [1]{%
 \ifx #1\expandafter \@firstoftwo
 \else \expandafter \@secondoftwo
 \fi
}%
\providecommand \natexlab [1]{#1}%
\providecommand \enquote  [1]{``#1''}%
\providecommand \bibnamefont  [1]{#1}%
\providecommand \bibfnamefont [1]{#1}%
\providecommand \citenamefont [1]{#1}%
\providecommand \href@noop [0]{\@secondoftwo}%
\providecommand \href [0]{\begingroup \@sanitize@url \@href}%
\providecommand \@href[1]{\@@startlink{#1}\@@href}%
\providecommand \@@href[1]{\endgroup#1\@@endlink}%
\providecommand \@sanitize@url [0]{\catcode `\\12\catcode `\$12\catcode `\&12\catcode `\#12\catcode `\^12\catcode `\_12\catcode `\%12\relax}%
\providecommand \@@startlink[1]{}%
\providecommand \@@endlink[0]{}%
\providecommand \url  [0]{\begingroup\@sanitize@url \@url }%
\providecommand \@url [1]{\endgroup\@href {#1}{\urlprefix }}%
\providecommand \urlprefix  [0]{URL }%
\providecommand \Eprint [0]{\href }%
\providecommand \doibase [0]{https://doi.org/}%
\providecommand \selectlanguage [0]{\@gobble}%
\providecommand \bibinfo  [0]{\@secondoftwo}%
\providecommand \bibfield  [0]{\@secondoftwo}%
\providecommand \translation [1]{[#1]}%
\providecommand \BibitemOpen [0]{}%
\providecommand \bibitemStop [0]{}%
\providecommand \bibitemNoStop [0]{.\EOS\space}%
\providecommand \EOS [0]{\spacefactor3000\relax}%
\providecommand \BibitemShut  [1]{\csname bibitem#1\endcsname}%
\let\auto@bib@innerbib\@empty
\bibitem [{\citenamefont {Dekel}(2013)}]{dekel2013alkaline}%
  \BibitemOpen
  \bibfield  {author} {\bibinfo {author} {\bibfnamefont {D.~R.}\ \bibnamefont {Dekel}},\ }\bibfield  {title} {\bibinfo {title} {Alkaline membrane fuel cell ({AMFC}) materials and system improvement-state-of-the-art},\ }\href {https://doi.org/10.1149/05002.2051ecst} {\bibfield  {journal} {\bibinfo  {journal} {ECS Trans.}\ }\textbf {\bibinfo {volume} {50}},\ \bibinfo {pages} {2051} (\bibinfo {year} {2013})}\BibitemShut {NoStop}%
\bibitem [{\citenamefont {Leng}\ \emph {et~al.}(2012)\citenamefont {Leng}, \citenamefont {Chen}, \citenamefont {Mendoza}, \citenamefont {Tighe}, \citenamefont {Hickner},\ and\ \citenamefont {Wang}}]{leng2012}%
  \BibitemOpen
  \bibfield  {author} {\bibinfo {author} {\bibfnamefont {Y.}~\bibnamefont {Leng}}, \bibinfo {author} {\bibfnamefont {G.}~\bibnamefont {Chen}}, \bibinfo {author} {\bibfnamefont {A.~J.}\ \bibnamefont {Mendoza}}, \bibinfo {author} {\bibfnamefont {T.~B.}\ \bibnamefont {Tighe}}, \bibinfo {author} {\bibfnamefont {M.~A.}\ \bibnamefont {Hickner}},\ and\ \bibinfo {author} {\bibfnamefont {C.-Y.}\ \bibnamefont {Wang}},\ }\bibfield  {title} {\bibinfo {title} {Solid-state water electrolysis with an alkaline membrane},\ }\href {https://doi.org/10.1021/ja302439z} {\bibfield  {journal} {\bibinfo  {journal} {J. Am. Chem. Soc.}\ }\textbf {\bibinfo {volume} {134}},\ \bibinfo {pages} {9054} (\bibinfo {year} {2012})}\BibitemShut {NoStop}%
\bibitem [{\citenamefont {Pan}\ \emph {et~al.}(2012)\citenamefont {Pan}, \citenamefont {Chen}, \citenamefont {Zhuang},\ and\ \citenamefont {Lu}}]{pan2012designing}%
  \BibitemOpen
  \bibfield  {author} {\bibinfo {author} {\bibfnamefont {J.}~\bibnamefont {Pan}}, \bibinfo {author} {\bibfnamefont {C.}~\bibnamefont {Chen}}, \bibinfo {author} {\bibfnamefont {L.}~\bibnamefont {Zhuang}},\ and\ \bibinfo {author} {\bibfnamefont {J.}~\bibnamefont {Lu}},\ }\bibfield  {title} {\bibinfo {title} {Designing advanced alkaline polymer electrolytes for fuel cell applications},\ }\href {https://doi.org/10.1021/ar200201x} {\bibfield  {journal} {\bibinfo  {journal} {Acc. Chem. Res.}\ }\textbf {\bibinfo {volume} {45}},\ \bibinfo {pages} {473} (\bibinfo {year} {2012})}\BibitemShut {NoStop}%
\bibitem [{\citenamefont {Park}\ \emph {et~al.}(2019)\citenamefont {Park}, \citenamefont {Kang}, \citenamefont {Oh}, \citenamefont {Kim}, \citenamefont {Lim}, \citenamefont {Ahn}, \citenamefont {Cho},\ and\ \citenamefont {Sung}}]{park2019}%
  \BibitemOpen
  \bibfield  {author} {\bibinfo {author} {\bibfnamefont {J.~E.}\ \bibnamefont {Park}}, \bibinfo {author} {\bibfnamefont {S.~Y.}\ \bibnamefont {Kang}}, \bibinfo {author} {\bibfnamefont {S.-H.}\ \bibnamefont {Oh}}, \bibinfo {author} {\bibfnamefont {J.~K.}\ \bibnamefont {Kim}}, \bibinfo {author} {\bibfnamefont {M.~S.}\ \bibnamefont {Lim}}, \bibinfo {author} {\bibfnamefont {C.-Y.}\ \bibnamefont {Ahn}}, \bibinfo {author} {\bibfnamefont {Y.-H.}\ \bibnamefont {Cho}},\ and\ \bibinfo {author} {\bibfnamefont {Y.-E.}\ \bibnamefont {Sung}},\ }\bibfield  {title} {\bibinfo {title} {High-performance anion-exchange membrane water electrolysis},\ }\href {https://doi.org/10.1016/j.electacta.2018.10.143} {\bibfield  {journal} {\bibinfo  {journal} {Electrochim. Acta}\ }\textbf {\bibinfo {volume} {295}},\ \bibinfo {pages} {99} (\bibinfo {year} {2019})}\BibitemShut {NoStop}%
\bibitem [{\citenamefont {Henkensmeier}\ \emph {et~al.}(2020)\citenamefont {Henkensmeier}, \citenamefont {Najibah}, \citenamefont {Harms}, \citenamefont {Žitka}, \citenamefont {Hnát},\ and\ \citenamefont {Bouzek}}]{henkensmeier2020}%
  \BibitemOpen
  \bibfield  {author} {\bibinfo {author} {\bibfnamefont {D.}~\bibnamefont {Henkensmeier}}, \bibinfo {author} {\bibfnamefont {M.}~\bibnamefont {Najibah}}, \bibinfo {author} {\bibfnamefont {C.}~\bibnamefont {Harms}}, \bibinfo {author} {\bibfnamefont {J.}~\bibnamefont {Žitka}}, \bibinfo {author} {\bibfnamefont {J.}~\bibnamefont {Hnát}},\ and\ \bibinfo {author} {\bibfnamefont {K.}~\bibnamefont {Bouzek}},\ }\bibfield  {title} {\bibinfo {title} {Overview: State-of-the art commercial membranes for anion exchange membrane water electrolysis},\ }\href {https://doi.org/10.1115/1.4047963} {\bibfield  {journal} {\bibinfo  {journal} {J. Electrochem. Energy Convers. Storage}\ }\textbf {\bibinfo {volume} {18}},\ \bibinfo {pages} {024001} (\bibinfo {year} {2020})}\BibitemShut {NoStop}%
\bibitem [{\citenamefont {Hren}\ \emph {et~al.}(2021)\citenamefont {Hren}, \citenamefont {Božič}, \citenamefont {Fakin}, \citenamefont {Kleinschek},\ and\ \citenamefont {Gorgieva}}]{hren2021}%
  \BibitemOpen
  \bibfield  {author} {\bibinfo {author} {\bibfnamefont {M.}~\bibnamefont {Hren}}, \bibinfo {author} {\bibfnamefont {M.}~\bibnamefont {Božič}}, \bibinfo {author} {\bibfnamefont {D.}~\bibnamefont {Fakin}}, \bibinfo {author} {\bibfnamefont {K.~S.}\ \bibnamefont {Kleinschek}},\ and\ \bibinfo {author} {\bibfnamefont {S.}~\bibnamefont {Gorgieva}},\ }\bibfield  {title} {\bibinfo {title} {Alkaline membrane fuel cells: anion exchange membranes and fuels},\ }\href {https://doi.org/10.1039/D0SE01373K} {\bibfield  {journal} {\bibinfo  {journal} {Sustain. Energy Fuels}\ }\textbf {\bibinfo {volume} {5}},\ \bibinfo {pages} {604} (\bibinfo {year} {2021})}\BibitemShut {NoStop}%
\bibitem [{\citenamefont {Muhyuddin}\ \emph {et~al.}(2025)\citenamefont {Muhyuddin}, \citenamefont {Santoro}, \citenamefont {Osmieri}, \citenamefont {Ficca}, \citenamefont {Friedman}, \citenamefont {Yassin}, \citenamefont {Pagot}, \citenamefont {Negro}, \citenamefont {Konovalova}, \citenamefont {Lindquist} \emph {et~al.}}]{muhyuddin2025anion}%
  \BibitemOpen
  \bibfield  {author} {\bibinfo {author} {\bibfnamefont {M.}~\bibnamefont {Muhyuddin}}, \bibinfo {author} {\bibfnamefont {C.}~\bibnamefont {Santoro}}, \bibinfo {author} {\bibfnamefont {L.}~\bibnamefont {Osmieri}}, \bibinfo {author} {\bibfnamefont {V.~C.}\ \bibnamefont {Ficca}}, \bibinfo {author} {\bibfnamefont {A.}~\bibnamefont {Friedman}}, \bibinfo {author} {\bibfnamefont {K.}~\bibnamefont {Yassin}}, \bibinfo {author} {\bibfnamefont {G.}~\bibnamefont {Pagot}}, \bibinfo {author} {\bibfnamefont {E.}~\bibnamefont {Negro}}, \bibinfo {author} {\bibfnamefont {A.}~\bibnamefont {Konovalova}}, \bibinfo {author} {\bibfnamefont {G.}~\bibnamefont {Lindquist}}, \emph {et~al.},\ }\bibfield  {title} {\bibinfo {title} {Anion-exchange-membrane electrolysis with alkali-free water feed},\ }\href {https://doi.org/10.1021/acs.chemrev.4c00466} {\bibfield  {journal} {\bibinfo  {journal} {Chem. Rev.}\ }\textbf {\bibinfo {volume} {125}},\ \bibinfo {pages} {6906} (\bibinfo {year} {2025})}\BibitemShut {NoStop}%
\bibitem [{\citenamefont {Zou}\ and\ \citenamefont {Zhang}(2015)}]{zou2015}%
  \BibitemOpen
  \bibfield  {author} {\bibinfo {author} {\bibfnamefont {X.}~\bibnamefont {Zou}}\ and\ \bibinfo {author} {\bibfnamefont {Y.}~\bibnamefont {Zhang}},\ }\bibfield  {title} {\bibinfo {title} {Noble metal-free hydrogen evolution catalysts for water splitting},\ }\href {https://doi.org/10.1039/C4CS00448E} {\bibfield  {journal} {\bibinfo  {journal} {Chem. Soc. Rev.}\ }\textbf {\bibinfo {volume} {44}},\ \bibinfo {pages} {5148} (\bibinfo {year} {2015})}\BibitemShut {NoStop}%
\bibitem [{\citenamefont {Schalenbach}\ \emph {et~al.}(2018)\citenamefont {Schalenbach}, \citenamefont {Zeradjanin}, \citenamefont {Kasian}, \citenamefont {Cherevko},\ and\ \citenamefont {Mayrhofer}}]{schalenbach2018}%
  \BibitemOpen
  \bibfield  {author} {\bibinfo {author} {\bibfnamefont {M.}~\bibnamefont {Schalenbach}}, \bibinfo {author} {\bibfnamefont {A.~R.}\ \bibnamefont {Zeradjanin}}, \bibinfo {author} {\bibfnamefont {O.}~\bibnamefont {Kasian}}, \bibinfo {author} {\bibfnamefont {S.}~\bibnamefont {Cherevko}},\ and\ \bibinfo {author} {\bibfnamefont {K.~J.~J.}\ \bibnamefont {Mayrhofer}},\ }\bibfield  {title} {\bibinfo {title} {A perspective on low-temperature water electrolysis – challenges in alkaline and acidic technology},\ }\href {https://doi.org/10.20964/2018.02.26} {\bibfield  {journal} {\bibinfo  {journal} {Int. J. Electrochem. Sci.}\ }\textbf {\bibinfo {volume} {13}},\ \bibinfo {pages} {1173} (\bibinfo {year} {2018})}\BibitemShut {NoStop}%
\bibitem [{\citenamefont {Alia}\ \emph {et~al.}(2021)\citenamefont {Alia}, \citenamefont {Ding}, \citenamefont {McDaniel}, \citenamefont {Toma},\ and\ \citenamefont {Dinh}}]{alia2021}%
  \BibitemOpen
  \bibfield  {author} {\bibinfo {author} {\bibfnamefont {S.}~\bibnamefont {Alia}}, \bibinfo {author} {\bibfnamefont {D.}~\bibnamefont {Ding}}, \bibinfo {author} {\bibfnamefont {A.}~\bibnamefont {McDaniel}}, \bibinfo {author} {\bibfnamefont {F.}~\bibnamefont {Toma}},\ and\ \bibinfo {author} {\bibfnamefont {H.}~\bibnamefont {Dinh}},\ }\bibfield  {title} {\bibinfo {title} {Chalkboard 2 - how to make clean hydrogen},\ }\bibfield  {journal} {\bibinfo  {journal} {Interface}\ }\textbf {\bibinfo {volume} {30}},\ \href {https://doi.org/10.1149/2.f13214if} {10.1149/2.f13214if} (\bibinfo {year} {2021})\BibitemShut {NoStop}%
\bibitem [{\citenamefont {Douglin}\ \emph {et~al.}(2020)\citenamefont {Douglin}, \citenamefont {Varcoe},\ and\ \citenamefont {Dekel}}]{douglin2020high}%
  \BibitemOpen
  \bibfield  {author} {\bibinfo {author} {\bibfnamefont {J.~C.}\ \bibnamefont {Douglin}}, \bibinfo {author} {\bibfnamefont {J.~R.}\ \bibnamefont {Varcoe}},\ and\ \bibinfo {author} {\bibfnamefont {D.~R.}\ \bibnamefont {Dekel}},\ }\bibfield  {title} {\bibinfo {title} {A high-temperature anion-exchange membrane fuel cell},\ }\href {https://doi.org/10.1016/j.powera.2020.100023} {\bibfield  {journal} {\bibinfo  {journal} {J. Power Sources Advances}\ }\textbf {\bibinfo {volume} {5}},\ \bibinfo {pages} {100023} (\bibinfo {year} {2020})}\BibitemShut {NoStop}%
\bibitem [{\citenamefont {Douglin}\ \emph {et~al.}(2021)\citenamefont {Douglin}, \citenamefont {Singh}, \citenamefont {Haj-Bsoul}, \citenamefont {Li}, \citenamefont {Biemolt}, \citenamefont {Yan}, \citenamefont {Varcoe}, \citenamefont {Rothenberg},\ and\ \citenamefont {Dekel}}]{douglin2021high}%
  \BibitemOpen
  \bibfield  {author} {\bibinfo {author} {\bibfnamefont {J.~C.}\ \bibnamefont {Douglin}}, \bibinfo {author} {\bibfnamefont {R.~K.}\ \bibnamefont {Singh}}, \bibinfo {author} {\bibfnamefont {S.}~\bibnamefont {Haj-Bsoul}}, \bibinfo {author} {\bibfnamefont {S.}~\bibnamefont {Li}}, \bibinfo {author} {\bibfnamefont {J.}~\bibnamefont {Biemolt}}, \bibinfo {author} {\bibfnamefont {N.}~\bibnamefont {Yan}}, \bibinfo {author} {\bibfnamefont {J.~R.}\ \bibnamefont {Varcoe}}, \bibinfo {author} {\bibfnamefont {G.}~\bibnamefont {Rothenberg}},\ and\ \bibinfo {author} {\bibfnamefont {D.~R.}\ \bibnamefont {Dekel}},\ }\bibfield  {title} {\bibinfo {title} {A high-temperature anion-exchange membrane fuel cell with a critical raw material-free cathode},\ }\href {https://doi.org/10.1016/j.ceja.2021.100153} {\bibfield  {journal} {\bibinfo  {journal} {Chem. Eng. J. Adv.}\ }\textbf {\bibinfo {volume} {8}},\ \bibinfo {pages} {100153} (\bibinfo {year} {2021})}\BibitemShut {NoStop}%
\bibitem [{\citenamefont {Wijaya}\ \emph {et~al.}(2024)\citenamefont {Wijaya}, \citenamefont {Im},\ and\ \citenamefont {Nam}}]{wijaya2024}%
  \BibitemOpen
  \bibfield  {author} {\bibinfo {author} {\bibfnamefont {G.~H.~A.}\ \bibnamefont {Wijaya}}, \bibinfo {author} {\bibfnamefont {K.~S.}\ \bibnamefont {Im}},\ and\ \bibinfo {author} {\bibfnamefont {S.~Y.}\ \bibnamefont {Nam}},\ }\bibfield  {title} {\bibinfo {title} {Advancements in commercial anion exchange membranes: A review of membrane properties in water electrolysis applications},\ }\href {https://doi.org/10.1016/j.dwt.2024.100605} {\bibfield  {journal} {\bibinfo  {journal} {Desalin. Water Treat.}\ }\textbf {\bibinfo {volume} {320}},\ \bibinfo {pages} {100605} (\bibinfo {year} {2024})}\BibitemShut {NoStop}%
\bibitem [{\citenamefont {Han}\ \emph {et~al.}(2014)\citenamefont {Han}, \citenamefont {Ko}, \citenamefont {Abu-Hakmeh}, \citenamefont {Bae}, \citenamefont {Sohn},\ and\ \citenamefont {Jang}}]{han2014}%
  \BibitemOpen
  \bibfield  {author} {\bibinfo {author} {\bibfnamefont {K.~W.}\ \bibnamefont {Han}}, \bibinfo {author} {\bibfnamefont {K.~H.}\ \bibnamefont {Ko}}, \bibinfo {author} {\bibfnamefont {K.}~\bibnamefont {Abu-Hakmeh}}, \bibinfo {author} {\bibfnamefont {C.}~\bibnamefont {Bae}}, \bibinfo {author} {\bibfnamefont {Y.~J.}\ \bibnamefont {Sohn}},\ and\ \bibinfo {author} {\bibfnamefont {S.~S.}\ \bibnamefont {Jang}},\ }\bibfield  {title} {\bibinfo {title} {Molecular dynamics simulation study of a polysulfone-based anion exchange membrane in comparison with the proton exchange membrane},\ }\href {https://doi.org/10.1021/jp412473j} {\bibfield  {journal} {\bibinfo  {journal} {J. Phys. Chem. C}\ }\textbf {\bibinfo {volume} {118}},\ \bibinfo {pages} {12577} (\bibinfo {year} {2014})}\BibitemShut {NoStop}%
\bibitem [{\citenamefont {Dekel}\ \emph {et~al.}(2018)\citenamefont {Dekel}, \citenamefont {Rasin}, \citenamefont {Page},\ and\ \citenamefont {Brandon}}]{dekel2018_sim}%
  \BibitemOpen
  \bibfield  {author} {\bibinfo {author} {\bibfnamefont {D.~R.}\ \bibnamefont {Dekel}}, \bibinfo {author} {\bibfnamefont {I.~G.}\ \bibnamefont {Rasin}}, \bibinfo {author} {\bibfnamefont {M.}~\bibnamefont {Page}},\ and\ \bibinfo {author} {\bibfnamefont {S.}~\bibnamefont {Brandon}},\ }\bibfield  {title} {\bibinfo {title} {Steady state and transient simulation of anion exchange membrane fuel cells},\ }\href {https://doi.org/10.1016/j.jpowsour.2017.07.012} {\bibfield  {journal} {\bibinfo  {journal} {J. Power Sources}\ }\textbf {\bibinfo {volume} {375}},\ \bibinfo {pages} {191} (\bibinfo {year} {2018})}\BibitemShut {NoStop}%
\bibitem [{\citenamefont {Wang}\ \emph {et~al.}(2018)\citenamefont {Wang}, \citenamefont {Mo}, \citenamefont {He}, \citenamefont {Xie}, \citenamefont {Zhao}, \citenamefont {Zhang}, \citenamefont {Shao}, \citenamefont {Guo}, \citenamefont {Wujcik},\ and\ \citenamefont {Guo}}]{wang2018}%
  \BibitemOpen
  \bibfield  {author} {\bibinfo {author} {\bibfnamefont {C.}~\bibnamefont {Wang}}, \bibinfo {author} {\bibfnamefont {B.}~\bibnamefont {Mo}}, \bibinfo {author} {\bibfnamefont {Z.}~\bibnamefont {He}}, \bibinfo {author} {\bibfnamefont {X.}~\bibnamefont {Xie}}, \bibinfo {author} {\bibfnamefont {C.~X.}\ \bibnamefont {Zhao}}, \bibinfo {author} {\bibfnamefont {L.}~\bibnamefont {Zhang}}, \bibinfo {author} {\bibfnamefont {Q.}~\bibnamefont {Shao}}, \bibinfo {author} {\bibfnamefont {X.}~\bibnamefont {Guo}}, \bibinfo {author} {\bibfnamefont {E.~K.}\ \bibnamefont {Wujcik}},\ and\ \bibinfo {author} {\bibfnamefont {Z.}~\bibnamefont {Guo}},\ }\bibfield  {title} {\bibinfo {title} {Hydroxide ions transportation in polynorbornene anion exchange membrane},\ }\href {https://doi.org/10.1016/j.polymer.2018.01.079} {\bibfield  {journal} {\bibinfo  {journal} {Polymer}\ }\textbf {\bibinfo {volume} {138}},\ \bibinfo {pages} {363} (\bibinfo {year} {2018})}\BibitemShut {NoStop}%
\bibitem [{\citenamefont {Karibayev}\ \emph {et~al.}(2022)\citenamefont {Karibayev}, \citenamefont {Kalybekkyzy}, \citenamefont {Wang},\ and\ \citenamefont {Mentbayeva}}]{karibayev2022}%
  \BibitemOpen
  \bibfield  {author} {\bibinfo {author} {\bibfnamefont {M.}~\bibnamefont {Karibayev}}, \bibinfo {author} {\bibfnamefont {S.}~\bibnamefont {Kalybekkyzy}}, \bibinfo {author} {\bibfnamefont {Y.}~\bibnamefont {Wang}},\ and\ \bibinfo {author} {\bibfnamefont {A.}~\bibnamefont {Mentbayeva}},\ }\bibfield  {title} {\bibinfo {title} {Molecular modeling in anion exchange membrane research: A brief review of recent applications},\ }\bibfield  {journal} {\bibinfo  {journal} {Molecules}\ }\textbf {\bibinfo {volume} {27}},\ \href {https://doi.org/10.3390/molecules27113574} {10.3390/molecules27113574} (\bibinfo {year} {2022})\BibitemShut {NoStop}%
\bibitem [{\citenamefont {Park}\ \emph {et~al.}(2017)\citenamefont {Park}, \citenamefont {Kim}, \citenamefont {Kim},\ and\ \citenamefont {Nam}}]{park2017}%
  \BibitemOpen
  \bibfield  {author} {\bibinfo {author} {\bibfnamefont {C.~H.}\ \bibnamefont {Park}}, \bibinfo {author} {\bibfnamefont {T.-H.}\ \bibnamefont {Kim}}, \bibinfo {author} {\bibfnamefont {D.~J.}\ \bibnamefont {Kim}},\ and\ \bibinfo {author} {\bibfnamefont {S.~Y.}\ \bibnamefont {Nam}},\ }\bibfield  {title} {\bibinfo {title} {Molecular dynamics simulation of the functional group effect in hydrocarbon anionic exchange membranes},\ }\href {https://doi.org/10.1016/j.ijhydene.2017.05.146} {\bibfield  {journal} {\bibinfo  {journal} {Int. J. Hydrog. Energy}\ }\textbf {\bibinfo {volume} {42}},\ \bibinfo {pages} {20895} (\bibinfo {year} {2017})}\BibitemShut {NoStop}%
\bibitem [{\citenamefont {Takaba}\ \emph {et~al.}(2017)\citenamefont {Takaba}, \citenamefont {Hisabe}, \citenamefont {Shimizu},\ and\ \citenamefont {Alam}}]{takaba2017}%
  \BibitemOpen
  \bibfield  {author} {\bibinfo {author} {\bibfnamefont {H.}~\bibnamefont {Takaba}}, \bibinfo {author} {\bibfnamefont {T.}~\bibnamefont {Hisabe}}, \bibinfo {author} {\bibfnamefont {T.}~\bibnamefont {Shimizu}},\ and\ \bibinfo {author} {\bibfnamefont {M.~K.}\ \bibnamefont {Alam}},\ }\bibfield  {title} {\bibinfo {title} {Molecular modeling of oh- transport in poly(arylene ether sulfone ketone)s containing quaternized ammonio-substituted fluorenyl groups as anion exchange membranes},\ }\href {https://doi.org/10.1016/j.memsci.2016.09.019} {\bibfield  {journal} {\bibinfo  {journal} {J. Membr. Sci.}\ }\textbf {\bibinfo {volume} {522}},\ \bibinfo {pages} {237} (\bibinfo {year} {2017})}\BibitemShut {NoStop}%
\bibitem [{\citenamefont {Zelovich}\ \emph {et~al.}(2019{\natexlab{a}})\citenamefont {Zelovich}, \citenamefont {Vogt-Maranto}, \citenamefont {Hickner}, \citenamefont {Paddison}, \citenamefont {Bae}, \citenamefont {Dekel},\ and\ \citenamefont {Tuckerman}}]{zelovich2019hydration}%
  \BibitemOpen
  \bibfield  {author} {\bibinfo {author} {\bibfnamefont {T.}~\bibnamefont {Zelovich}}, \bibinfo {author} {\bibfnamefont {L.}~\bibnamefont {Vogt-Maranto}}, \bibinfo {author} {\bibfnamefont {M.~A.}\ \bibnamefont {Hickner}}, \bibinfo {author} {\bibfnamefont {S.~J.}\ \bibnamefont {Paddison}}, \bibinfo {author} {\bibfnamefont {C.}~\bibnamefont {Bae}}, \bibinfo {author} {\bibfnamefont {D.~R.}\ \bibnamefont {Dekel}},\ and\ \bibinfo {author} {\bibfnamefont {M.~E.}\ \bibnamefont {Tuckerman}},\ }\bibfield  {title} {\bibinfo {title} {Hydroxide ion diffusion in anion-exchange membranes at low hydration: Insights from ab initio molecular dynamics},\ }\href {https://doi.org/10.1021/acs.chemmater.9b01824} {\bibfield  {journal} {\bibinfo  {journal} {Chem. Mater.}\ }\textbf {\bibinfo {volume} {31}},\ \bibinfo {pages} {5778} (\bibinfo {year} {2019}{\natexlab{a}})}\BibitemShut {NoStop}%
\bibitem [{\citenamefont {Zelovich}\ \emph {et~al.}(2019{\natexlab{b}})\citenamefont {Zelovich}, \citenamefont {Long}, \citenamefont {Hickner}, \citenamefont {Paddison}, \citenamefont {Bae},\ and\ \citenamefont {Tuckerman}}]{zelovich2019mimics}%
  \BibitemOpen
  \bibfield  {author} {\bibinfo {author} {\bibfnamefont {T.}~\bibnamefont {Zelovich}}, \bibinfo {author} {\bibfnamefont {Z.}~\bibnamefont {Long}}, \bibinfo {author} {\bibfnamefont {M.}~\bibnamefont {Hickner}}, \bibinfo {author} {\bibfnamefont {S.~J.}\ \bibnamefont {Paddison}}, \bibinfo {author} {\bibfnamefont {C.}~\bibnamefont {Bae}},\ and\ \bibinfo {author} {\bibfnamefont {M.~E.}\ \bibnamefont {Tuckerman}},\ }\bibfield  {title} {\bibinfo {title} {Ab initio molecular dynamics study of hydroxide diffusion mechanisms in nanoconfined structural mimics of anion exchange membranes},\ }\href {https://doi.org/10.1021/acs.jpcc.8b10298} {\bibfield  {journal} {\bibinfo  {journal} {J. Phys. Chem. C}\ }\textbf {\bibinfo {volume} {123}},\ \bibinfo {pages} {4638} (\bibinfo {year} {2019}{\natexlab{b}})}\BibitemShut {NoStop}%
\bibitem [{\citenamefont {Zelovich}\ and\ \citenamefont {Tuckerman}(2020)}]{zelovich2020}%
  \BibitemOpen
  \bibfield  {author} {\bibinfo {author} {\bibfnamefont {T.}~\bibnamefont {Zelovich}}\ and\ \bibinfo {author} {\bibfnamefont {M.~E.}\ \bibnamefont {Tuckerman}},\ }\bibfield  {title} {\bibinfo {title} {Water layering affects hydroxide diffusion in functionalized nanoconfined environments},\ }\href {https://doi.org/10.1021/acs.jpclett.0c01141} {\bibfield  {journal} {\bibinfo  {journal} {J. Phys. Chem. Lett.}\ }\textbf {\bibinfo {volume} {11}},\ \bibinfo {pages} {5087} (\bibinfo {year} {2020})}\BibitemShut {NoStop}%
\bibitem [{\citenamefont {Zelovich}\ and\ \citenamefont {Tuckerman}(2021)}]{zelovich21_ohvsh3o}%
  \BibitemOpen
  \bibfield  {author} {\bibinfo {author} {\bibfnamefont {T.}~\bibnamefont {Zelovich}}\ and\ \bibinfo {author} {\bibfnamefont {M.}~\bibnamefont {Tuckerman}},\ }\bibfield  {title} {\bibinfo {title} {Oh- and h3o+ diffusion in model aems and pems at low hydration: Insights from ab initio molecular dynamics},\ }\bibfield  {journal} {\bibinfo  {journal} {Membranes}\ }\textbf {\bibinfo {volume} {11}},\ \href {https://doi.org/10.3390/membranes11050355} {10.3390/membranes11050355} (\bibinfo {year} {2021})\BibitemShut {NoStop}%
\bibitem [{\citenamefont {Zelovich}\ \emph {et~al.}(2024)\citenamefont {Zelovich}, \citenamefont {Dekel},\ and\ \citenamefont {Tuckerman}}]{zelovich2024}%
  \BibitemOpen
  \bibfield  {author} {\bibinfo {author} {\bibfnamefont {T.}~\bibnamefont {Zelovich}}, \bibinfo {author} {\bibfnamefont {D.}~\bibnamefont {Dekel}},\ and\ \bibinfo {author} {\bibfnamefont {M.}~\bibnamefont {Tuckerman}},\ }\bibfield  {title} {\bibinfo {title} {Electrostatic potential of functional cations as a predictor of hydroxide diffusion pathways in nanoconfined environments of anion exchange membranes},\ }\href {https://doi.org/10.1021/acs.jpclett.3c02800} {\bibfield  {journal} {\bibinfo  {journal} {J. Phys. Chem. Lett.}\ }\textbf {\bibinfo {volume} {15}},\ \bibinfo {pages} {408} (\bibinfo {year} {2024})}\BibitemShut {NoStop}%
\bibitem [{\citenamefont {Zelovich}\ \emph {et~al.}(2023)\citenamefont {Zelovich}, \citenamefont {Dekel},\ and\ \citenamefont {Tuckerman}}]{zelovich23}%
  \BibitemOpen
  \bibfield  {author} {\bibinfo {author} {\bibfnamefont {T.}~\bibnamefont {Zelovich}}, \bibinfo {author} {\bibfnamefont {D.~R.}\ \bibnamefont {Dekel}},\ and\ \bibinfo {author} {\bibfnamefont {M.~E.}\ \bibnamefont {Tuckerman}},\ }\bibfield  {title} {\bibinfo {title} {Functional groups in anion exchange membranes: Insights from ab initio molecular dynamics},\ }\href {https://doi.org/10.1016/j.memsci.2023.121638} {\bibfield  {journal} {\bibinfo  {journal} {J. Membr. Sci.}\ }\textbf {\bibinfo {volume} {678}},\ \bibinfo {pages} {121638} (\bibinfo {year} {2023})}\BibitemShut {NoStop}%
\bibitem [{\citenamefont {Chakraborti}\ \emph {et~al.}(2024)\citenamefont {Chakraborti}, \citenamefont {Sharma}, \citenamefont {Krishnamoorthy}, \citenamefont {Chaudhari}, \citenamefont {Mamtani},\ and\ \citenamefont {Singh}}]{chakraborti2024unravelling}%
  \BibitemOpen
  \bibfield  {author} {\bibinfo {author} {\bibfnamefont {T.}~\bibnamefont {Chakraborti}}, \bibinfo {author} {\bibfnamefont {R.}~\bibnamefont {Sharma}}, \bibinfo {author} {\bibfnamefont {A.~N.}\ \bibnamefont {Krishnamoorthy}}, \bibinfo {author} {\bibfnamefont {H.}~\bibnamefont {Chaudhari}}, \bibinfo {author} {\bibfnamefont {K.}~\bibnamefont {Mamtani}},\ and\ \bibinfo {author} {\bibfnamefont {J.~K.}\ \bibnamefont {Singh}},\ }\bibfield  {title} {\bibinfo {title} {Unravelling the effect of molecular interactions on macroscale properties in sustainion anion exchange membrane (aem) under hydrated conditions using md simulations},\ }\href {https://doi.org/10.1016/j.memsci.2024.122887} {\bibfield  {journal} {\bibinfo  {journal} {J. Membr. Sci.}\ }\textbf {\bibinfo {volume} {705}},\ \bibinfo {pages} {122887} (\bibinfo {year} {2024})}\BibitemShut {NoStop}%
\bibitem [{\citenamefont {de~Lucas}\ \emph {et~al.}(2024)\citenamefont {de~Lucas}, \citenamefont {Blazquez}, \citenamefont {Troncoso}, \citenamefont {Vega},\ and\ \citenamefont {Gámez}}]{delucas2024}%
  \BibitemOpen
  \bibfield  {author} {\bibinfo {author} {\bibfnamefont {M.}~\bibnamefont {de~Lucas}}, \bibinfo {author} {\bibfnamefont {S.}~\bibnamefont {Blazquez}}, \bibinfo {author} {\bibfnamefont {J.}~\bibnamefont {Troncoso}}, \bibinfo {author} {\bibfnamefont {C.}~\bibnamefont {Vega}},\ and\ \bibinfo {author} {\bibfnamefont {F.}~\bibnamefont {Gámez}},\ }\bibfield  {title} {\bibinfo {title} {Dressing a nonpolarizable force field for oh– in tip4p/2005 aqueous solutions with corrected hirshfeld charges},\ }\href {https://doi.org/10.1021/acs.jpclett.4c02261} {\bibfield  {journal} {\bibinfo  {journal} {J. Phys. Chem. Lett.}\ }\textbf {\bibinfo {volume} {15}},\ \bibinfo {pages} {9411} (\bibinfo {year} {2024})}\BibitemShut {NoStop}%
\bibitem [{\citenamefont {Sharma}\ \emph {et~al.}(2025)\citenamefont {Sharma}, \citenamefont {Chakraborti},\ and\ \citenamefont {Singh}}]{sharma2025understanding}%
  \BibitemOpen
  \bibfield  {author} {\bibinfo {author} {\bibfnamefont {R.}~\bibnamefont {Sharma}}, \bibinfo {author} {\bibfnamefont {T.}~\bibnamefont {Chakraborti}},\ and\ \bibinfo {author} {\bibfnamefont {J.~K.}\ \bibnamefont {Singh}},\ }\bibfield  {title} {\bibinfo {title} {Understanding the role of alkyl chain architecture and functional group chemistry in governing ion transport in peek based anion exchange membranes: A molecular dynamics study},\ }\href {https://doi.org/10.1016/j.memsci.2025.124598} {\bibfield  {journal} {\bibinfo  {journal} {J. Membr. Sci.}\ ,\ \bibinfo {pages} {124598}} (\bibinfo {year} {2025})}\BibitemShut {NoStop}%
\bibitem [{\citenamefont {Lagerweij}\ \emph {et~al.}(2026)\citenamefont {Lagerweij}, \citenamefont {Moultos},\ and\ \citenamefont {Vlugt}}]{lagerweij2026}%
  \BibitemOpen
  \bibfield  {author} {\bibinfo {author} {\bibfnamefont {V.~J.}\ \bibnamefont {Lagerweij}}, \bibinfo {author} {\bibfnamefont {O.~A.}\ \bibnamefont {Moultos}},\ and\ \bibinfo {author} {\bibfnamefont {T.~J.}\ \bibnamefont {Vlugt}},\ }\bibfield  {title} {\bibinfo {title} {Electrical conductivity differences between experiments and classical simulations reveal self-diffusion coefficients and ion lifetimes of hydroxide and hydronium in aqueous solutions},\ }\href {https://doi.org/10.1021/acs.jpcb.5c07132} {\bibfield  {journal} {\bibinfo  {journal} {J. Phys. Chem. B}\ }\textbf {\bibinfo {volume} {130}},\ \bibinfo {pages} {1332} (\bibinfo {year} {2026})}\BibitemShut {NoStop}%
\bibitem [{\citenamefont {Tuckerman}\ \emph {et~al.}(2006)\citenamefont {Tuckerman}, \citenamefont {Chandra},\ and\ \citenamefont {Marx}}]{tuckerman2006acs}%
  \BibitemOpen
  \bibfield  {author} {\bibinfo {author} {\bibfnamefont {M.~E.}\ \bibnamefont {Tuckerman}}, \bibinfo {author} {\bibfnamefont {A.}~\bibnamefont {Chandra}},\ and\ \bibinfo {author} {\bibfnamefont {D.}~\bibnamefont {Marx}},\ }\bibfield  {title} {\bibinfo {title} {{Structure and Dynamics of OH-(aq)}},\ }\href {https://doi.org/10.1021/ar040207n} {\bibfield  {journal} {\bibinfo  {journal} {Acc. Chem. Res.}\ }\textbf {\bibinfo {volume} {39}},\ \bibinfo {pages} {151} (\bibinfo {year} {2006})}\BibitemShut {NoStop}%
\bibitem [{\citenamefont {Ouma}\ \emph {et~al.}(2022)\citenamefont {Ouma}, \citenamefont {Obodo},\ and\ \citenamefont {Bessarabov}}]{ouma2022}%
  \BibitemOpen
  \bibfield  {author} {\bibinfo {author} {\bibfnamefont {C.~N.~M.}\ \bibnamefont {Ouma}}, \bibinfo {author} {\bibfnamefont {K.~O.}\ \bibnamefont {Obodo}},\ and\ \bibinfo {author} {\bibfnamefont {D.}~\bibnamefont {Bessarabov}},\ }\bibfield  {title} {\bibinfo {title} {Computational approaches to alkaline anion-exchange membranes for fuel cell applications},\ }\bibfield  {journal} {\bibinfo  {journal} {Membranes}\ }\textbf {\bibinfo {volume} {12}},\ \href {https://doi.org/10.3390/membranes12111051} {10.3390/membranes12111051} (\bibinfo {year} {2022})\BibitemShut {NoStop}%
\bibitem [{\citenamefont {Dreßler}\ \emph {et~al.}(2016)\citenamefont {Dreßler}, \citenamefont {Kabbe},\ and\ \citenamefont {Sebastiani}}]{dressler2016}%
  \BibitemOpen
  \bibfield  {author} {\bibinfo {author} {\bibfnamefont {C.}~\bibnamefont {Dreßler}}, \bibinfo {author} {\bibfnamefont {G.}~\bibnamefont {Kabbe}},\ and\ \bibinfo {author} {\bibfnamefont {D.}~\bibnamefont {Sebastiani}},\ }\bibfield  {title} {\bibinfo {title} {{Proton conductivity in hydrogen phosphate/sulfates from a coupled molecular dynamics/lattice Monte Carlo (cMD/LMC) approach}},\ }\href {https://doi.org/10.1021/acs.jpcc.6b05822} {\bibfield  {journal} {\bibinfo  {journal} {J. Phys. Chem. C}\ }\textbf {\bibinfo {volume} {120}},\ \bibinfo {pages} {19913} (\bibinfo {year} {2016})}\BibitemShut {NoStop}%
\bibitem [{\citenamefont {Kabbe}\ \emph {et~al.}(2016)\citenamefont {Kabbe}, \citenamefont {Dreßler},\ and\ \citenamefont {Sebastiani}}]{kabbe2016}%
  \BibitemOpen
  \bibfield  {author} {\bibinfo {author} {\bibfnamefont {G.}~\bibnamefont {Kabbe}}, \bibinfo {author} {\bibfnamefont {C.}~\bibnamefont {Dreßler}},\ and\ \bibinfo {author} {\bibfnamefont {D.}~\bibnamefont {Sebastiani}},\ }\bibfield  {title} {\bibinfo {title} {Toward realistic transfer rates within the coupled molecular dynamics/lattice monte carlo approach},\ }\href {https://doi.org/10.1021/acs.jpcc.6b05821} {\bibfield  {journal} {\bibinfo  {journal} {J. Phys. Chem. C}\ }\textbf {\bibinfo {volume} {120}},\ \bibinfo {pages} {19905} (\bibinfo {year} {2016})}\BibitemShut {NoStop}%
\bibitem [{\citenamefont {Kabbe}\ \emph {et~al.}(2017)\citenamefont {Kabbe}, \citenamefont {Dreßler},\ and\ \citenamefont {Sebastiani}}]{kabbe2017}%
  \BibitemOpen
  \bibfield  {author} {\bibinfo {author} {\bibfnamefont {G.}~\bibnamefont {Kabbe}}, \bibinfo {author} {\bibfnamefont {C.}~\bibnamefont {Dreßler}},\ and\ \bibinfo {author} {\bibfnamefont {D.}~\bibnamefont {Sebastiani}},\ }\bibfield  {title} {\bibinfo {title} {Proton mobility in aqueous systems: combining ab initio accuracy with millisecond timescales},\ }\href {https://doi.org/10.1039/C7CP05632J} {\bibfield  {journal} {\bibinfo  {journal} {Phys. Chem. Chem. Phys.}\ }\textbf {\bibinfo {volume} {19}},\ \bibinfo {pages} {28604} (\bibinfo {year} {2017})}\BibitemShut {NoStop}%
\bibitem [{\citenamefont {Dutta}\ and\ \citenamefont {Lazaridis}(2024)}]{dutta2024}%
  \BibitemOpen
  \bibfield  {author} {\bibinfo {author} {\bibfnamefont {A.}~\bibnamefont {Dutta}}\ and\ \bibinfo {author} {\bibfnamefont {T.}~\bibnamefont {Lazaridis}},\ }\bibfield  {title} {\bibinfo {title} {Classical models of hydroxide for proton hopping simulations},\ }\href {https://doi.org/10.1021/acs.jpcb.4c05499} {\bibfield  {journal} {\bibinfo  {journal} {J. Phys. Chem. B}\ }\textbf {\bibinfo {volume} {128}},\ \bibinfo {pages} {12161} (\bibinfo {year} {2024})}\BibitemShut {NoStop}%
\bibitem [{\citenamefont {Qaisrani}\ \emph {et~al.}(2025)\citenamefont {Qaisrani}, \citenamefont {Kirsch}, \citenamefont {Fl{\"o}totto}, \citenamefont {H{\"a}nseroth}, \citenamefont {Oumard}, \citenamefont {Sebastiani},\ and\ \citenamefont {Dre{\ss}ler}}]{qaisrani2025bridging}%
  \BibitemOpen
  \bibfield  {author} {\bibinfo {author} {\bibfnamefont {M.~N.}\ \bibnamefont {Qaisrani}}, \bibinfo {author} {\bibfnamefont {C.}~\bibnamefont {Kirsch}}, \bibinfo {author} {\bibfnamefont {A.}~\bibnamefont {Fl{\"o}totto}}, \bibinfo {author} {\bibfnamefont {J.}~\bibnamefont {H{\"a}nseroth}}, \bibinfo {author} {\bibfnamefont {J.}~\bibnamefont {Oumard}}, \bibinfo {author} {\bibfnamefont {D.}~\bibnamefont {Sebastiani}},\ and\ \bibinfo {author} {\bibfnamefont {C.}~\bibnamefont {Dre{\ss}ler}},\ }\bibfield  {title} {\bibinfo {title} {Bridging atomistic and mesoscale lithium transport via machine-learned force fields and markov state models},\ }\bibfield  {journal} {\bibinfo  {journal} {arXiv preprint arXiv:2511.20863}\ }\href {https://doi.org/10.48550/arXiv.2511.20863} {10.48550/arXiv.2511.20863} (\bibinfo {year} {2025})\BibitemShut {NoStop}%
\bibitem [{\citenamefont {Hänseroth}\ \emph {et~al.}(2025{\natexlab{a}})\citenamefont {Hänseroth}, \citenamefont {Sebastiani}, \citenamefont {Jimenez~Siegert}, \citenamefont {Scholl}, \citenamefont {Skadell},\ and\ \citenamefont {Dreßler}}]{haenseroth2025lmc}%
  \BibitemOpen
  \bibfield  {author} {\bibinfo {author} {\bibfnamefont {J.}~\bibnamefont {Hänseroth}}, \bibinfo {author} {\bibfnamefont {D.}~\bibnamefont {Sebastiani}}, \bibinfo {author} {\bibfnamefont {J.~A.}\ \bibnamefont {Jimenez~Siegert}}, \bibinfo {author} {\bibfnamefont {J.}~\bibnamefont {Scholl}}, \bibinfo {author} {\bibfnamefont {K.}~\bibnamefont {Skadell}},\ and\ \bibinfo {author} {\bibfnamefont {C.}~\bibnamefont {Dreßler}},\ }\bibfield  {title} {\bibinfo {title} {Hydroxide mobility in aqueous systems: Combining ab initio accuracy with millisecond timescales},\ }\href {https://doi.org/10.1002/smll.202500931} {\bibfield  {journal} {\bibinfo  {journal} {Small}\ ,\ \bibinfo {pages} {2500931}} (\bibinfo {year} {2025}{\natexlab{a}})}\BibitemShut {NoStop}%
\bibitem [{\citenamefont {Hellström}\ \emph {et~al.}(2018)\citenamefont {Hellström}, \citenamefont {Ceriotti},\ and\ \citenamefont {Behler}}]{hellstrom2018}%
  \BibitemOpen
  \bibfield  {author} {\bibinfo {author} {\bibfnamefont {M.}~\bibnamefont {Hellström}}, \bibinfo {author} {\bibfnamefont {M.}~\bibnamefont {Ceriotti}},\ and\ \bibinfo {author} {\bibfnamefont {J.}~\bibnamefont {Behler}},\ }\bibfield  {title} {\bibinfo {title} {Nuclear quantum effects in sodium hydroxide solutions from neural network molecular dynamics simulations},\ }\href {https://doi.org/10.1021/acs.jpcb.8b06433} {\bibfield  {journal} {\bibinfo  {journal} {J. Phys. Chem. B}\ }\textbf {\bibinfo {volume} {122}},\ \bibinfo {pages} {10158} (\bibinfo {year} {2018})}\BibitemShut {NoStop}%
\bibitem [{\citenamefont {Jinnouchi}\ \emph {et~al.}(2023)\citenamefont {Jinnouchi}, \citenamefont {Minami}, \citenamefont {Karsai}, \citenamefont {Verdi},\ and\ \citenamefont {Kresse}}]{jinnouchi23_proton}%
  \BibitemOpen
  \bibfield  {author} {\bibinfo {author} {\bibfnamefont {R.}~\bibnamefont {Jinnouchi}}, \bibinfo {author} {\bibfnamefont {S.}~\bibnamefont {Minami}}, \bibinfo {author} {\bibfnamefont {F.}~\bibnamefont {Karsai}}, \bibinfo {author} {\bibfnamefont {C.}~\bibnamefont {Verdi}},\ and\ \bibinfo {author} {\bibfnamefont {G.}~\bibnamefont {Kresse}},\ }\bibfield  {title} {\bibinfo {title} {Proton transport in perfluorinated ionomer simulated by machine-learned interatomic potential},\ }\href {https://doi.org/10.1021/acs.jpclett.3c00293} {\bibfield  {journal} {\bibinfo  {journal} {J. Phys. Chem. Lett.}\ }\textbf {\bibinfo {volume} {14}},\ \bibinfo {pages} {3581} (\bibinfo {year} {2023})}\BibitemShut {NoStop}%
\bibitem [{\citenamefont {Grunert}\ \emph {et~al.}(2025)\citenamefont {Grunert}, \citenamefont {Großmann}, \citenamefont {Hänseroth}, \citenamefont {Fl{\"o}totto}, \citenamefont {Oumard}, \citenamefont {Wolf}, \citenamefont {Runge},\ and\ \citenamefont {Dreßler}}]{grunert2025}%
  \BibitemOpen
  \bibfield  {author} {\bibinfo {author} {\bibfnamefont {M.}~\bibnamefont {Grunert}}, \bibinfo {author} {\bibfnamefont {M.}~\bibnamefont {Großmann}}, \bibinfo {author} {\bibfnamefont {J.}~\bibnamefont {Hänseroth}}, \bibinfo {author} {\bibfnamefont {A.}~\bibnamefont {Fl{\"o}totto}}, \bibinfo {author} {\bibfnamefont {J.}~\bibnamefont {Oumard}}, \bibinfo {author} {\bibfnamefont {J.~L.}\ \bibnamefont {Wolf}}, \bibinfo {author} {\bibfnamefont {E.}~\bibnamefont {Runge}},\ and\ \bibinfo {author} {\bibfnamefont {C.}~\bibnamefont {Dreßler}},\ }\bibfield  {title} {\bibinfo {title} {Modeling complex proton transport phenomena - exploring the limits of fine-tuning and transferability of foundational machine-learned force fields},\ }\href {https://doi.org/10.1021/acs.jpcc.5c02064} {\bibfield  {journal} {\bibinfo  {journal} {J. Phys. Chem. C}\ }\textbf {\bibinfo {volume} {129}},\ \bibinfo {pages} {9662} (\bibinfo {year} {2025})}\BibitemShut {NoStop}%
\bibitem [{\citenamefont {H{\"a}nseroth}\ and\ \citenamefont {Dre{\ss}ler}(2025)}]{haenseroth2025mace}%
  \BibitemOpen
  \bibfield  {author} {\bibinfo {author} {\bibfnamefont {J.}~\bibnamefont {H{\"a}nseroth}}\ and\ \bibinfo {author} {\bibfnamefont {C.}~\bibnamefont {Dre{\ss}ler}},\ }\bibfield  {title} {\bibinfo {title} {Optimizing machine learning interatomic potentials for hydroxide transport: Surprising efficiency of single-concentration training},\ }\bibfield  {journal} {\bibinfo  {journal} {J. Chem. Phys.}\ }\textbf {\bibinfo {volume} {163}},\ \href {https://doi.org/10.1063/5.0284063} {10.1063/5.0284063} (\bibinfo {year} {2025})\BibitemShut {NoStop}%
\bibitem [{\citenamefont {Fl{\"o}totto}\ \emph {et~al.}(2026)\citenamefont {Fl{\"o}totto}, \citenamefont {Spetzler}, \citenamefont {von Stackelberg}, \citenamefont {Ziegler}, \citenamefont {Runge},\ and\ \citenamefont {Dre{\ss}ler}}]{flototto2026large}%
  \BibitemOpen
  \bibfield  {author} {\bibinfo {author} {\bibfnamefont {A.}~\bibnamefont {Fl{\"o}totto}}, \bibinfo {author} {\bibfnamefont {B.}~\bibnamefont {Spetzler}}, \bibinfo {author} {\bibfnamefont {R.}~\bibnamefont {von Stackelberg}}, \bibinfo {author} {\bibfnamefont {M.}~\bibnamefont {Ziegler}}, \bibinfo {author} {\bibfnamefont {E.}~\bibnamefont {Runge}},\ and\ \bibinfo {author} {\bibfnamefont {C.}~\bibnamefont {Dre{\ss}ler}},\ }\bibfield  {title} {\bibinfo {title} {Large‐scale cooperative sulfur vacancy dynamics in two‐dimensional {MoS2} from machine learning interatomic potentials},\ }\bibfield  {journal} {\bibinfo  {journal} {Small}\ }\href {https://doi.org/10.1002/smll.202510679} {10.1002/smll.202510679} (\bibinfo {year} {2026})\BibitemShut {NoStop}%
\bibitem [{\citenamefont {H{\"a}nseroth}\ \emph {et~al.}(2026)\citenamefont {H{\"a}nseroth}, \citenamefont {Gro{\ss}mann}, \citenamefont {Grunert}, \citenamefont {Runge},\ and\ \citenamefont {Dre{\ss}ler}}]{haenseroth2026htscreening}%
  \BibitemOpen
  \bibfield  {author} {\bibinfo {author} {\bibfnamefont {J.}~\bibnamefont {H{\"a}nseroth}}, \bibinfo {author} {\bibfnamefont {M.}~\bibnamefont {Gro{\ss}mann}}, \bibinfo {author} {\bibfnamefont {M.}~\bibnamefont {Grunert}}, \bibinfo {author} {\bibfnamefont {E.}~\bibnamefont {Runge}},\ and\ \bibinfo {author} {\bibfnamefont {C.}~\bibnamefont {Dre{\ss}ler}},\ }\bibfield  {title} {\bibinfo {title} {High-throughput screening and mechanistic insights into solid acid proton conductors},\ }\bibfield  {journal} {\bibinfo  {journal} {arXiv preprint arXiv:2602.15268}\ }\href {https://doi.org/10.48550/arXiv.2602.15268} {10.48550/arXiv.2602.15268} (\bibinfo {year} {2026})\BibitemShut {NoStop}%
\bibitem [{\citenamefont {Bartók}\ \emph {et~al.}(2010)\citenamefont {Bartók}, \citenamefont {Payne}, \citenamefont {Kondor},\ and\ \citenamefont {Csányi}}]{bartok2010}%
  \BibitemOpen
  \bibfield  {author} {\bibinfo {author} {\bibfnamefont {A.~P.}\ \bibnamefont {Bartók}}, \bibinfo {author} {\bibfnamefont {M.~C.}\ \bibnamefont {Payne}}, \bibinfo {author} {\bibfnamefont {R.}~\bibnamefont {Kondor}},\ and\ \bibinfo {author} {\bibfnamefont {G.}~\bibnamefont {Csányi}},\ }\bibfield  {title} {\bibinfo {title} {Gaussian approximation potentials: {T}he accuracy of quantum mechanics, without the electrons},\ }\href {https://doi.org/10.1103/PhysRevLett.104.136403} {\bibfield  {journal} {\bibinfo  {journal} {Phys. Rev. Lett.}\ }\textbf {\bibinfo {volume} {104}},\ \bibinfo {pages} {136403} (\bibinfo {year} {2010})}\BibitemShut {NoStop}%
\bibitem [{\citenamefont {Unke}\ \emph {et~al.}(2021)\citenamefont {Unke}, \citenamefont {Chmiela}, \citenamefont {Sauceda}, \citenamefont {Gastegger}, \citenamefont {Poltavsky}, \citenamefont {Schütt}, \citenamefont {Tkatchenko},\ and\ \citenamefont {Müller}}]{unke2021}%
  \BibitemOpen
  \bibfield  {author} {\bibinfo {author} {\bibfnamefont {O.~T.}\ \bibnamefont {Unke}}, \bibinfo {author} {\bibfnamefont {S.}~\bibnamefont {Chmiela}}, \bibinfo {author} {\bibfnamefont {H.~E.}\ \bibnamefont {Sauceda}}, \bibinfo {author} {\bibfnamefont {M.}~\bibnamefont {Gastegger}}, \bibinfo {author} {\bibfnamefont {I.}~\bibnamefont {Poltavsky}}, \bibinfo {author} {\bibfnamefont {K.~T.}\ \bibnamefont {Schütt}}, \bibinfo {author} {\bibfnamefont {A.}~\bibnamefont {Tkatchenko}},\ and\ \bibinfo {author} {\bibfnamefont {K.~R.}\ \bibnamefont {Müller}},\ }\bibfield  {title} {\bibinfo {title} {Machine {L}earning {F}orce {F}ields},\ }\href {https://doi.org/10.1021/acs.chemrev.0c01111} {\bibfield  {journal} {\bibinfo  {journal} {Chem. Rev.}\ }\textbf {\bibinfo {volume} {121}},\ \bibinfo {pages} {10142} (\bibinfo {year} {2021})}\BibitemShut {NoStop}%
\bibitem [{\citenamefont {Friederich}\ \emph {et~al.}(2021)\citenamefont {Friederich}, \citenamefont {H{\"a}se}, \citenamefont {Proppe},\ and\ \citenamefont {Aspuru-Guzik}}]{friederich2021}%
  \BibitemOpen
  \bibfield  {author} {\bibinfo {author} {\bibfnamefont {P.}~\bibnamefont {Friederich}}, \bibinfo {author} {\bibfnamefont {F.}~\bibnamefont {H{\"a}se}}, \bibinfo {author} {\bibfnamefont {J.}~\bibnamefont {Proppe}},\ and\ \bibinfo {author} {\bibfnamefont {A.}~\bibnamefont {Aspuru-Guzik}},\ }\bibfield  {title} {\bibinfo {title} {Machine-learned potentials for next-generation matter simulations},\ }\href {https://doi.org/10.1038/s41563-020-0777-6} {\bibfield  {journal} {\bibinfo  {journal} {Nat. Mater.}\ }\textbf {\bibinfo {volume} {20}},\ \bibinfo {pages} {750} (\bibinfo {year} {2021})}\BibitemShut {NoStop}%
\bibitem [{\citenamefont {Reiser}\ \emph {et~al.}(2022)\citenamefont {Reiser}, \citenamefont {Neubert}, \citenamefont {Eberhard}, \citenamefont {Torresi}, \citenamefont {Zhou}, \citenamefont {Shao}, \citenamefont {Metni}, \citenamefont {van Hoesel}, \citenamefont {Schopmans}, \citenamefont {Sommer},\ and\ \citenamefont {Friederich}}]{reiser2022}%
  \BibitemOpen
  \bibfield  {author} {\bibinfo {author} {\bibfnamefont {P.}~\bibnamefont {Reiser}}, \bibinfo {author} {\bibfnamefont {M.}~\bibnamefont {Neubert}}, \bibinfo {author} {\bibfnamefont {A.}~\bibnamefont {Eberhard}}, \bibinfo {author} {\bibfnamefont {L.}~\bibnamefont {Torresi}}, \bibinfo {author} {\bibfnamefont {C.}~\bibnamefont {Zhou}}, \bibinfo {author} {\bibfnamefont {C.}~\bibnamefont {Shao}}, \bibinfo {author} {\bibfnamefont {H.}~\bibnamefont {Metni}}, \bibinfo {author} {\bibfnamefont {C.}~\bibnamefont {van Hoesel}}, \bibinfo {author} {\bibfnamefont {H.}~\bibnamefont {Schopmans}}, \bibinfo {author} {\bibfnamefont {T.}~\bibnamefont {Sommer}},\ and\ \bibinfo {author} {\bibfnamefont {P.}~\bibnamefont {Friederich}},\ }\bibfield  {title} {\bibinfo {title} {Graph neural networks for materials science and chemistry},\ }\bibfield  {journal} {\bibinfo  {journal} {Commun. Mater.}\ }\textbf {\bibinfo {volume} {3}},\ \href {https://doi.org/10.1038/s43246-022-00315-6} {10.1038/s43246-022-00315-6} (\bibinfo {year}
  {2022})\BibitemShut {NoStop}%
\bibitem [{\citenamefont {Batzner}\ \emph {et~al.}(2022)\citenamefont {Batzner}, \citenamefont {Musaelian}, \citenamefont {Sun}, \citenamefont {Geiger}, \citenamefont {Mailoa}, \citenamefont {Kornbluth}, \citenamefont {Molinari}, \citenamefont {Smidt},\ and\ \citenamefont {Kozinsky}}]{batzner2022}%
  \BibitemOpen
  \bibfield  {author} {\bibinfo {author} {\bibfnamefont {S.}~\bibnamefont {Batzner}}, \bibinfo {author} {\bibfnamefont {A.}~\bibnamefont {Musaelian}}, \bibinfo {author} {\bibfnamefont {L.}~\bibnamefont {Sun}}, \bibinfo {author} {\bibfnamefont {M.}~\bibnamefont {Geiger}}, \bibinfo {author} {\bibfnamefont {J.~P.}\ \bibnamefont {Mailoa}}, \bibinfo {author} {\bibfnamefont {M.}~\bibnamefont {Kornbluth}}, \bibinfo {author} {\bibfnamefont {N.}~\bibnamefont {Molinari}}, \bibinfo {author} {\bibfnamefont {T.~E.}\ \bibnamefont {Smidt}},\ and\ \bibinfo {author} {\bibfnamefont {B.}~\bibnamefont {Kozinsky}},\ }\bibfield  {title} {\bibinfo {title} {E(3)-equivariant graph neural networks for data-efficient and accurate interatomic potentials},\ }\href {https://doi.org/10.1038/s41467-022-29939-5} {\bibfield  {journal} {\bibinfo  {journal} {Nat. Commun.}\ }\textbf {\bibinfo {volume} {13}},\ \bibinfo {pages} {2453} (\bibinfo {year} {2022})}\BibitemShut {NoStop}%
\bibitem [{\citenamefont {Behler}\ and\ \citenamefont {Parrinello}(2007)}]{behler2007}%
  \BibitemOpen
  \bibfield  {author} {\bibinfo {author} {\bibfnamefont {J.}~\bibnamefont {Behler}}\ and\ \bibinfo {author} {\bibfnamefont {M.}~\bibnamefont {Parrinello}},\ }\bibfield  {title} {\bibinfo {title} {Generalized neural-network representation of high-dimensional potential-energy surfaces},\ }\href {https://doi.org/10.1103/PhysRevLett.98.146401} {\bibfield  {journal} {\bibinfo  {journal} {Phys. Rev. Lett.}\ }\textbf {\bibinfo {volume} {98}},\ \bibinfo {pages} {146401} (\bibinfo {year} {2007})}\BibitemShut {NoStop}%
\bibitem [{\citenamefont {Bart{\'o}k}\ \emph {et~al.}(2013)\citenamefont {Bart{\'o}k}, \citenamefont {Kondor},\ and\ \citenamefont {Cs{\'a}nyi}}]{bartok2013representing}%
  \BibitemOpen
  \bibfield  {author} {\bibinfo {author} {\bibfnamefont {A.~P.}\ \bibnamefont {Bart{\'o}k}}, \bibinfo {author} {\bibfnamefont {R.}~\bibnamefont {Kondor}},\ and\ \bibinfo {author} {\bibfnamefont {G.}~\bibnamefont {Cs{\'a}nyi}},\ }\bibfield  {title} {\bibinfo {title} {On representing chemical environments},\ }\href {https://doi.org/10.1103/PhysRevB.87.184115} {\bibfield  {journal} {\bibinfo  {journal} {Phys. Rev. B}\ }\textbf {\bibinfo {volume} {87}},\ \bibinfo {pages} {184115} (\bibinfo {year} {2013})}\BibitemShut {NoStop}%
\bibitem [{\citenamefont {Drautz}(2019)}]{drautz2019}%
  \BibitemOpen
  \bibfield  {author} {\bibinfo {author} {\bibfnamefont {R.}~\bibnamefont {Drautz}},\ }\bibfield  {title} {\bibinfo {title} {Atomic cluster expansion for accurate and transferable interatomic potentials},\ }\href {https://doi.org/10.1103/PhysRevB.99.014104} {\bibfield  {journal} {\bibinfo  {journal} {Phys. Rev. B}\ }\textbf {\bibinfo {volume} {99}},\ \bibinfo {pages} {014104} (\bibinfo {year} {2019})}\BibitemShut {NoStop}%
\bibitem [{\citenamefont {Jain}\ \emph {et~al.}(2013)\citenamefont {Jain}, \citenamefont {Ong}, \citenamefont {Hautier}, \citenamefont {Chen}, \citenamefont {Richards}, \citenamefont {Dacek}, \citenamefont {Cholia}, \citenamefont {Gunter}, \citenamefont {Skinner}, \citenamefont {Ceder},\ and\ \citenamefont {Persson}}]{mp_1}%
  \BibitemOpen
  \bibfield  {author} {\bibinfo {author} {\bibfnamefont {A.}~\bibnamefont {Jain}}, \bibinfo {author} {\bibfnamefont {S.~P.}\ \bibnamefont {Ong}}, \bibinfo {author} {\bibfnamefont {G.}~\bibnamefont {Hautier}}, \bibinfo {author} {\bibfnamefont {W.}~\bibnamefont {Chen}}, \bibinfo {author} {\bibfnamefont {W.~D.}\ \bibnamefont {Richards}}, \bibinfo {author} {\bibfnamefont {S.}~\bibnamefont {Dacek}}, \bibinfo {author} {\bibfnamefont {S.}~\bibnamefont {Cholia}}, \bibinfo {author} {\bibfnamefont {D.}~\bibnamefont {Gunter}}, \bibinfo {author} {\bibfnamefont {D.}~\bibnamefont {Skinner}}, \bibinfo {author} {\bibfnamefont {G.}~\bibnamefont {Ceder}},\ and\ \bibinfo {author} {\bibfnamefont {K.~A.}\ \bibnamefont {Persson}},\ }\bibfield  {title} {\bibinfo {title} {Commentary: {T}he {M}aterials {P}roject: {A} materials genome approach to accelerating materials innovation},\ }\href {https://doi.org/10.1063/1.4812323} {\bibfield  {journal} {\bibinfo  {journal} {APL Mater.}\ }\textbf {\bibinfo {volume} {1}},\ \bibinfo {pages}
  {011002} (\bibinfo {year} {2013})}\BibitemShut {NoStop}%
\bibitem [{\citenamefont {Ong}\ \emph {et~al.}(2015)\citenamefont {Ong}, \citenamefont {Cholia}, \citenamefont {Jain}, \citenamefont {Brafman}, \citenamefont {Gunter}, \citenamefont {Ceder},\ and\ \citenamefont {Persson}}]{mp_2}%
  \BibitemOpen
  \bibfield  {author} {\bibinfo {author} {\bibfnamefont {S.~P.}\ \bibnamefont {Ong}}, \bibinfo {author} {\bibfnamefont {S.}~\bibnamefont {Cholia}}, \bibinfo {author} {\bibfnamefont {A.}~\bibnamefont {Jain}}, \bibinfo {author} {\bibfnamefont {M.}~\bibnamefont {Brafman}}, \bibinfo {author} {\bibfnamefont {D.}~\bibnamefont {Gunter}}, \bibinfo {author} {\bibfnamefont {G.}~\bibnamefont {Ceder}},\ and\ \bibinfo {author} {\bibfnamefont {K.~A.}\ \bibnamefont {Persson}},\ }\bibfield  {title} {\bibinfo {title} {The {Materials Application Programming Interface (API)}: A simple, flexible and efficient {API} for materials data based on {REpresentational State Transfer (REST)} principles},\ }\href {https://doi.org/10.1016/j.commatsci.2014.10.037} {\bibfield  {journal} {\bibinfo  {journal} {Comput. Mater. Sci.}\ }\textbf {\bibinfo {volume} {97}},\ \bibinfo {pages} {209–215} (\bibinfo {year} {2015})}\BibitemShut {NoStop}%
\bibitem [{\citenamefont {Ong}\ \emph {et~al.}(2013)\citenamefont {Ong}, \citenamefont {Richards}, \citenamefont {Jain}, \citenamefont {Hautier}, \citenamefont {Kocher}, \citenamefont {Cholia}, \citenamefont {Gunter}, \citenamefont {Chevrier}, \citenamefont {Persson},\ and\ \citenamefont {Ceder}}]{pymatgen}%
  \BibitemOpen
  \bibfield  {author} {\bibinfo {author} {\bibfnamefont {S.~P.}\ \bibnamefont {Ong}}, \bibinfo {author} {\bibfnamefont {W.~D.}\ \bibnamefont {Richards}}, \bibinfo {author} {\bibfnamefont {A.}~\bibnamefont {Jain}}, \bibinfo {author} {\bibfnamefont {G.}~\bibnamefont {Hautier}}, \bibinfo {author} {\bibfnamefont {M.}~\bibnamefont {Kocher}}, \bibinfo {author} {\bibfnamefont {S.}~\bibnamefont {Cholia}}, \bibinfo {author} {\bibfnamefont {D.}~\bibnamefont {Gunter}}, \bibinfo {author} {\bibfnamefont {V.~L.}\ \bibnamefont {Chevrier}}, \bibinfo {author} {\bibfnamefont {K.~A.}\ \bibnamefont {Persson}},\ and\ \bibinfo {author} {\bibfnamefont {G.}~\bibnamefont {Ceder}},\ }\bibfield  {title} {\bibinfo {title} {Python materials genomics (pymatgen): A robust, open-source python library for materials analysis},\ }\href {https://doi.org/10.1016/j.commatsci.2012.10.028} {\bibfield  {journal} {\bibinfo  {journal} {Comput. Mater. Sci.}\ }\textbf {\bibinfo {volume} {68}},\ \bibinfo {pages} {314} (\bibinfo {year}
  {2013})}\BibitemShut {NoStop}%
\bibitem [{\citenamefont {Deng}\ \emph {et~al.}(2023)\citenamefont {Deng}, \citenamefont {Zhong}, \citenamefont {Jun}, \citenamefont {Riebesell}, \citenamefont {Han}, \citenamefont {Bartel},\ and\ \citenamefont {Ceder}}]{mptrj}%
  \BibitemOpen
  \bibfield  {author} {\bibinfo {author} {\bibfnamefont {B.}~\bibnamefont {Deng}}, \bibinfo {author} {\bibfnamefont {P.}~\bibnamefont {Zhong}}, \bibinfo {author} {\bibfnamefont {K.}~\bibnamefont {Jun}}, \bibinfo {author} {\bibfnamefont {J.}~\bibnamefont {Riebesell}}, \bibinfo {author} {\bibfnamefont {K.}~\bibnamefont {Han}}, \bibinfo {author} {\bibfnamefont {C.~J.}\ \bibnamefont {Bartel}},\ and\ \bibinfo {author} {\bibfnamefont {G.}~\bibnamefont {Ceder}},\ }\bibfield  {title} {\bibinfo {title} {Chgnet as a pretrained universal neural network potential for charge-informed atomistic modelling},\ }\href {https://doi.org/10.1038/s42256-023-00716-3} {\bibfield  {journal} {\bibinfo  {journal} {Nat. Mach. Intell.}\ }\textbf {\bibinfo {volume} {5}},\ \bibinfo {pages} {1031} (\bibinfo {year} {2023})}\BibitemShut {NoStop}%
\bibitem [{\citenamefont {Batatia}\ \emph {et~al.}(2025{\natexlab{a}})\citenamefont {Batatia}, \citenamefont {Benner}, \citenamefont {Chiang}, \citenamefont {Elena}, \citenamefont {Kovács}, \citenamefont {Riebesell}, \citenamefont {Advincula}, \citenamefont {Asta}, \citenamefont {Avaylon}, \citenamefont {Baldwin}, \citenamefont {Berger}, \citenamefont {Bernstein}, \citenamefont {Bhowmik}, \citenamefont {Bigi}, \citenamefont {Blau}, \citenamefont {Cărare}, \citenamefont {Ceriotti}, \citenamefont {Chong}, \citenamefont {Darby}, \citenamefont {De}, \citenamefont {Della~Pia}, \citenamefont {Deringer}, \citenamefont {Elijošius}, \citenamefont {El-Machachi}, \citenamefont {Fako}, \citenamefont {Falcioni}, \citenamefont {Ferrari}, \citenamefont {Gardner}, \citenamefont {Gawkowski}, \citenamefont {Genreith-Schriever}, \citenamefont {George}, \citenamefont {Goodall}, \citenamefont {Grandel}, \citenamefont {Grey}, \citenamefont {Grigorev}, \citenamefont {Han}, \citenamefont {Handley}, \citenamefont {Heenen},
  \citenamefont {Hermansson}, \citenamefont {Ho}, \citenamefont {Hofmann}, \citenamefont {Holm}, \citenamefont {Jaafar}, \citenamefont {Jakob}, \citenamefont {Jung}, \citenamefont {Kapil}, \citenamefont {Kaplan}, \citenamefont {Karimitari}, \citenamefont {Kermode}, \citenamefont {Kourtis}, \citenamefont {Kroupa}, \citenamefont {Kullgren}, \citenamefont {Kuner}, \citenamefont {Kuryla}, \citenamefont {Liepuoniute}, \citenamefont {Lin}, \citenamefont {Margraf}, \citenamefont {Magdău}, \citenamefont {Michaelides}, \citenamefont {Moore}, \citenamefont {Naik}, \citenamefont {Niblett}, \citenamefont {Norwood}, \citenamefont {O’Neill}, \citenamefont {Ortner}, \citenamefont {Persson}, \citenamefont {Reuter}, \citenamefont {Rosen}, \citenamefont {Rosset}, \citenamefont {Schaaf}, \citenamefont {Schran}, \citenamefont {Shi}, \citenamefont {Sivonxay}, \citenamefont {Stenczel}, \citenamefont {Sutton}, \citenamefont {Svahn}, \citenamefont {Swinburne}, \citenamefont {Tilly}, \citenamefont {van~der Oord}, \citenamefont
  {Vargas}, \citenamefont {Varga-Umbrich}, \citenamefont {Vegge}, \citenamefont {Vondrák}, \citenamefont {Wang}, \citenamefont {Witt}, \citenamefont {Wolf}, \citenamefont {Zills},\ and\ \citenamefont {Csányi}}]{mace_mp}%
  \BibitemOpen
  \bibfield  {author} {\bibinfo {author} {\bibfnamefont {I.}~\bibnamefont {Batatia}}, \bibinfo {author} {\bibfnamefont {P.}~\bibnamefont {Benner}}, \bibinfo {author} {\bibfnamefont {Y.}~\bibnamefont {Chiang}}, \bibinfo {author} {\bibfnamefont {A.~M.}\ \bibnamefont {Elena}}, \bibinfo {author} {\bibfnamefont {D.~P.}\ \bibnamefont {Kovács}}, \bibinfo {author} {\bibfnamefont {J.}~\bibnamefont {Riebesell}}, \bibinfo {author} {\bibfnamefont {X.~R.}\ \bibnamefont {Advincula}}, \bibinfo {author} {\bibfnamefont {M.}~\bibnamefont {Asta}}, \bibinfo {author} {\bibfnamefont {M.}~\bibnamefont {Avaylon}}, \bibinfo {author} {\bibfnamefont {W.~J.}\ \bibnamefont {Baldwin}}, \bibinfo {author} {\bibfnamefont {F.}~\bibnamefont {Berger}}, \bibinfo {author} {\bibfnamefont {N.}~\bibnamefont {Bernstein}}, \bibinfo {author} {\bibfnamefont {A.}~\bibnamefont {Bhowmik}}, \bibinfo {author} {\bibfnamefont {F.}~\bibnamefont {Bigi}}, \bibinfo {author} {\bibfnamefont {S.~M.}\ \bibnamefont {Blau}}, \bibinfo {author} {\bibfnamefont
  {V.}~\bibnamefont {Cărare}}, \bibinfo {author} {\bibfnamefont {M.}~\bibnamefont {Ceriotti}}, \bibinfo {author} {\bibfnamefont {S.}~\bibnamefont {Chong}}, \bibinfo {author} {\bibfnamefont {J.~P.}\ \bibnamefont {Darby}}, \bibinfo {author} {\bibfnamefont {S.}~\bibnamefont {De}}, \bibinfo {author} {\bibfnamefont {F.}~\bibnamefont {Della~Pia}}, \bibinfo {author} {\bibfnamefont {V.~L.}\ \bibnamefont {Deringer}}, \bibinfo {author} {\bibfnamefont {R.}~\bibnamefont {Elijošius}}, \bibinfo {author} {\bibfnamefont {Z.}~\bibnamefont {El-Machachi}}, \bibinfo {author} {\bibfnamefont {E.}~\bibnamefont {Fako}}, \bibinfo {author} {\bibfnamefont {F.}~\bibnamefont {Falcioni}}, \bibinfo {author} {\bibfnamefont {A.~C.}\ \bibnamefont {Ferrari}}, \bibinfo {author} {\bibfnamefont {J.~L.~A.}\ \bibnamefont {Gardner}}, \bibinfo {author} {\bibfnamefont {M.~J.}\ \bibnamefont {Gawkowski}}, \bibinfo {author} {\bibfnamefont {A.}~\bibnamefont {Genreith-Schriever}}, \bibinfo {author} {\bibfnamefont {J.}~\bibnamefont {George}}, \bibinfo
  {author} {\bibfnamefont {R.~E.~A.}\ \bibnamefont {Goodall}}, \bibinfo {author} {\bibfnamefont {J.}~\bibnamefont {Grandel}}, \bibinfo {author} {\bibfnamefont {C.~P.}\ \bibnamefont {Grey}}, \bibinfo {author} {\bibfnamefont {P.}~\bibnamefont {Grigorev}}, \bibinfo {author} {\bibfnamefont {S.}~\bibnamefont {Han}}, \bibinfo {author} {\bibfnamefont {W.}~\bibnamefont {Handley}}, \bibinfo {author} {\bibfnamefont {H.~H.}\ \bibnamefont {Heenen}}, \bibinfo {author} {\bibfnamefont {K.}~\bibnamefont {Hermansson}}, \bibinfo {author} {\bibfnamefont {C.~H.}\ \bibnamefont {Ho}}, \bibinfo {author} {\bibfnamefont {S.}~\bibnamefont {Hofmann}}, \bibinfo {author} {\bibfnamefont {C.}~\bibnamefont {Holm}}, \bibinfo {author} {\bibfnamefont {J.}~\bibnamefont {Jaafar}}, \bibinfo {author} {\bibfnamefont {K.~S.}\ \bibnamefont {Jakob}}, \bibinfo {author} {\bibfnamefont {H.}~\bibnamefont {Jung}}, \bibinfo {author} {\bibfnamefont {V.}~\bibnamefont {Kapil}}, \bibinfo {author} {\bibfnamefont {A.~D.}\ \bibnamefont {Kaplan}}, \bibinfo {author}
  {\bibfnamefont {N.}~\bibnamefont {Karimitari}}, \bibinfo {author} {\bibfnamefont {J.~R.}\ \bibnamefont {Kermode}}, \bibinfo {author} {\bibfnamefont {P.}~\bibnamefont {Kourtis}}, \bibinfo {author} {\bibfnamefont {N.}~\bibnamefont {Kroupa}}, \bibinfo {author} {\bibfnamefont {J.}~\bibnamefont {Kullgren}}, \bibinfo {author} {\bibfnamefont {M.~C.}\ \bibnamefont {Kuner}}, \bibinfo {author} {\bibfnamefont {D.}~\bibnamefont {Kuryla}}, \bibinfo {author} {\bibfnamefont {G.}~\bibnamefont {Liepuoniute}}, \bibinfo {author} {\bibfnamefont {C.}~\bibnamefont {Lin}}, \bibinfo {author} {\bibfnamefont {J.~T.}\ \bibnamefont {Margraf}}, \bibinfo {author} {\bibfnamefont {I.-B.}\ \bibnamefont {Magdău}}, \bibinfo {author} {\bibfnamefont {A.}~\bibnamefont {Michaelides}}, \bibinfo {author} {\bibfnamefont {J.~H.}\ \bibnamefont {Moore}}, \bibinfo {author} {\bibfnamefont {A.~A.}\ \bibnamefont {Naik}}, \bibinfo {author} {\bibfnamefont {S.~P.}\ \bibnamefont {Niblett}}, \bibinfo {author} {\bibfnamefont {S.~W.}\ \bibnamefont {Norwood}},
  \bibinfo {author} {\bibfnamefont {N.}~\bibnamefont {O’Neill}}, \bibinfo {author} {\bibfnamefont {C.}~\bibnamefont {Ortner}}, \bibinfo {author} {\bibfnamefont {K.~A.}\ \bibnamefont {Persson}}, \bibinfo {author} {\bibfnamefont {K.}~\bibnamefont {Reuter}}, \bibinfo {author} {\bibfnamefont {A.~S.}\ \bibnamefont {Rosen}}, \bibinfo {author} {\bibfnamefont {L.~A.~M.}\ \bibnamefont {Rosset}}, \bibinfo {author} {\bibfnamefont {L.~L.}\ \bibnamefont {Schaaf}}, \bibinfo {author} {\bibfnamefont {C.}~\bibnamefont {Schran}}, \bibinfo {author} {\bibfnamefont {B.~X.}\ \bibnamefont {Shi}}, \bibinfo {author} {\bibfnamefont {E.}~\bibnamefont {Sivonxay}}, \bibinfo {author} {\bibfnamefont {T.~K.}\ \bibnamefont {Stenczel}}, \bibinfo {author} {\bibfnamefont {C.}~\bibnamefont {Sutton}}, \bibinfo {author} {\bibfnamefont {V.}~\bibnamefont {Svahn}}, \bibinfo {author} {\bibfnamefont {T.~D.}\ \bibnamefont {Swinburne}}, \bibinfo {author} {\bibfnamefont {J.}~\bibnamefont {Tilly}}, \bibinfo {author} {\bibfnamefont {C.}~\bibnamefont
  {van~der Oord}}, \bibinfo {author} {\bibfnamefont {S.}~\bibnamefont {Vargas}}, \bibinfo {author} {\bibfnamefont {E.}~\bibnamefont {Varga-Umbrich}}, \bibinfo {author} {\bibfnamefont {T.}~\bibnamefont {Vegge}}, \bibinfo {author} {\bibfnamefont {M.}~\bibnamefont {Vondrák}}, \bibinfo {author} {\bibfnamefont {Y.}~\bibnamefont {Wang}}, \bibinfo {author} {\bibfnamefont {W.~C.}\ \bibnamefont {Witt}}, \bibinfo {author} {\bibfnamefont {T.}~\bibnamefont {Wolf}}, \bibinfo {author} {\bibfnamefont {F.}~\bibnamefont {Zills}},\ and\ \bibinfo {author} {\bibfnamefont {G.}~\bibnamefont {Csányi}},\ }\bibfield  {title} {\bibinfo {title} {A foundation model for atomistic materials chemistry},\ }\href {https://doi.org/10.1063/5.0297006} {\bibfield  {journal} {\bibinfo  {journal} {J. Chem. Phys.}\ }\textbf {\bibinfo {volume} {163}},\ \bibinfo {pages} {184110} (\bibinfo {year} {2025}{\natexlab{a}})}\BibitemShut {NoStop}%
\bibitem [{\citenamefont {Lysogorskiy}\ \emph {et~al.}(2025)\citenamefont {Lysogorskiy}, \citenamefont {Bochkarev},\ and\ \citenamefont {Drautz}}]{grace_2}%
  \BibitemOpen
  \bibfield  {author} {\bibinfo {author} {\bibfnamefont {Y.}~\bibnamefont {Lysogorskiy}}, \bibinfo {author} {\bibfnamefont {A.}~\bibnamefont {Bochkarev}},\ and\ \bibinfo {author} {\bibfnamefont {R.}~\bibnamefont {Drautz}},\ }\bibfield  {title} {\bibinfo {title} {Graph atomic cluster expansion for foundational machine learning interatomic potentials},\ }\bibfield  {journal} {\bibinfo  {journal} {arXiv preprint arXiv:2508.17936}\ }\href {https://doi.org/10.48550/arXiv.2508.17936} {10.48550/arXiv.2508.17936} (\bibinfo {year} {2025})\BibitemShut {NoStop}%
\bibitem [{\citenamefont {Batatia}\ \emph {et~al.}(2022)\citenamefont {Batatia}, \citenamefont {Kovacs}, \citenamefont {Simm}, \citenamefont {Ortner},\ and\ \citenamefont {Cs{\'a}nyi}}]{mace_1}%
  \BibitemOpen
  \bibfield  {author} {\bibinfo {author} {\bibfnamefont {I.}~\bibnamefont {Batatia}}, \bibinfo {author} {\bibfnamefont {D.~P.}\ \bibnamefont {Kovacs}}, \bibinfo {author} {\bibfnamefont {G.}~\bibnamefont {Simm}}, \bibinfo {author} {\bibfnamefont {C.}~\bibnamefont {Ortner}},\ and\ \bibinfo {author} {\bibfnamefont {G.}~\bibnamefont {Cs{\'a}nyi}},\ }\bibfield  {title} {\bibinfo {title} {Mace: Higher order equivariant message passing neural networks for fast and accurate force fields},\ }\href {https://doi.org/10.48550/arXiv.2206.07697} {\bibfield  {journal} {\bibinfo  {journal} {Adv. Neural Inf. Process. Syst.}\ }\textbf {\bibinfo {volume} {35}},\ \bibinfo {pages} {11423} (\bibinfo {year} {2022})}\BibitemShut {NoStop}%
\bibitem [{\citenamefont {Batatia}\ \emph {et~al.}(2025{\natexlab{b}})\citenamefont {Batatia}, \citenamefont {Batzner}, \citenamefont {Kov{\'a}cs}, \citenamefont {Musaelian}, \citenamefont {Simm}, \citenamefont {Drautz}, \citenamefont {Ortner}, \citenamefont {Kozinsky},\ and\ \citenamefont {Cs{\'a}nyi}}]{mace_2}%
  \BibitemOpen
  \bibfield  {author} {\bibinfo {author} {\bibfnamefont {I.}~\bibnamefont {Batatia}}, \bibinfo {author} {\bibfnamefont {S.}~\bibnamefont {Batzner}}, \bibinfo {author} {\bibfnamefont {D.~P.}\ \bibnamefont {Kov{\'a}cs}}, \bibinfo {author} {\bibfnamefont {A.}~\bibnamefont {Musaelian}}, \bibinfo {author} {\bibfnamefont {G.~N.}\ \bibnamefont {Simm}}, \bibinfo {author} {\bibfnamefont {R.}~\bibnamefont {Drautz}}, \bibinfo {author} {\bibfnamefont {C.}~\bibnamefont {Ortner}}, \bibinfo {author} {\bibfnamefont {B.}~\bibnamefont {Kozinsky}},\ and\ \bibinfo {author} {\bibfnamefont {G.}~\bibnamefont {Cs{\'a}nyi}},\ }\bibfield  {title} {\bibinfo {title} {The design space of e (3)-equivariant atom-centred interatomic potentials},\ }\href {https://doi.org/10.1038/s42256-024-00956-x} {\bibfield  {journal} {\bibinfo  {journal} {Nat. Mach. Intell.}\ }\textbf {\bibinfo {volume} {7}},\ \bibinfo {pages} {56} (\bibinfo {year} {2025}{\natexlab{b}})}\BibitemShut {NoStop}%
\bibitem [{\citenamefont {de~Grotthuss}(1805)}]{grotthuss}%
  \BibitemOpen
  \bibfield  {author} {\bibinfo {author} {\bibfnamefont {C.}~\bibnamefont {de~Grotthuss}},\ }\href@noop {} {\emph {\bibinfo {title} {M{\'e}moire sur la d{\'e}composition de l'eau: et des corps qu' elle tient en dissolution {\`a} l'aide de l'{\'e}lectricit{\'e} galvanique}}}\ (\bibinfo {year} {1805})\BibitemShut {NoStop}%
\bibitem [{\citenamefont {Tuckerman}\ \emph {et~al.}(1995)\citenamefont {Tuckerman}, \citenamefont {Laasonen}, \citenamefont {Sprik},\ and\ \citenamefont {Parrinello}}]{tuckerman1995jpc}%
  \BibitemOpen
  \bibfield  {author} {\bibinfo {author} {\bibfnamefont {M.}~\bibnamefont {Tuckerman}}, \bibinfo {author} {\bibfnamefont {K.}~\bibnamefont {Laasonen}}, \bibinfo {author} {\bibfnamefont {M.}~\bibnamefont {Sprik}},\ and\ \bibinfo {author} {\bibfnamefont {M.}~\bibnamefont {Parrinello}},\ }\bibfield  {title} {\bibinfo {title} {{Ab Initio Molecular Dynamics Simulation of the Solvation and Transport of H3O+ and OH- Ions in Water}},\ }\href {https://doi.org/10.1021/j100016a003} {\bibfield  {journal} {\bibinfo  {journal} {J. Phys. Chem.}\ }\textbf {\bibinfo {volume} {99}},\ \bibinfo {pages} {5749} (\bibinfo {year} {1995})}\BibitemShut {NoStop}%
\bibitem [{\citenamefont {Marx}(2006)}]{marx2006}%
  \BibitemOpen
  \bibfield  {author} {\bibinfo {author} {\bibfnamefont {D.}~\bibnamefont {Marx}},\ }\bibfield  {title} {\bibinfo {title} {Proton transfer 200 years after von grotthuss: Insights from ab initio simulations},\ }\href {https://doi.org/10.1002/cphc.200600128} {\bibfield  {journal} {\bibinfo  {journal} {ChemPhysChem}\ }\textbf {\bibinfo {volume} {7}},\ \bibinfo {pages} {1848} (\bibinfo {year} {2006})}\BibitemShut {NoStop}%
\bibitem [{\citenamefont {Marx}\ \emph {et~al.}(2010)\citenamefont {Marx}, \citenamefont {Chandra},\ and\ \citenamefont {Tuckerman}}]{tuckerman2010rev}%
  \BibitemOpen
  \bibfield  {author} {\bibinfo {author} {\bibfnamefont {D.}~\bibnamefont {Marx}}, \bibinfo {author} {\bibfnamefont {A.}~\bibnamefont {Chandra}},\ and\ \bibinfo {author} {\bibfnamefont {M.}~\bibnamefont {Tuckerman}},\ }\bibfield  {title} {\bibinfo {title} {Aqueous basic solutions: Hydroxide solvation, structural diffusion, and comparison to the hydrated proton},\ }\href {https://doi.org/10.1021/cr900233f} {\bibfield  {journal} {\bibinfo  {journal} {Chem. Rev.}\ }\textbf {\bibinfo {volume} {110}},\ \bibinfo {pages} {2174} (\bibinfo {year} {2010})}\BibitemShut {NoStop}%
\bibitem [{\citenamefont {Franck}\ \emph {et~al.}(1965)\citenamefont {Franck}, \citenamefont {Hartmann},\ and\ \citenamefont {Hensel}}]{franck1965}%
  \BibitemOpen
  \bibfield  {author} {\bibinfo {author} {\bibfnamefont {E.~U.}\ \bibnamefont {Franck}}, \bibinfo {author} {\bibfnamefont {D.}~\bibnamefont {Hartmann}},\ and\ \bibinfo {author} {\bibfnamefont {F.}~\bibnamefont {Hensel}},\ }\bibfield  {title} {\bibinfo {title} {Proton mobility in water at high temperatures and pressures},\ }\href {https://doi.org/10.1039/DF9653900200} {\bibfield  {journal} {\bibinfo  {journal} {Faraday Discuss.}\ }\textbf {\bibinfo {volume} {39}},\ \bibinfo {pages} {200} (\bibinfo {year} {1965})}\BibitemShut {NoStop}%
\bibitem [{\citenamefont {Marino}\ \emph {et~al.}(2014)\citenamefont {Marino}, \citenamefont {Melchior}, \citenamefont {Wohlfarth},\ and\ \citenamefont {Kreuer}}]{marino2014hydroxide}%
  \BibitemOpen
  \bibfield  {author} {\bibinfo {author} {\bibfnamefont {M.}~\bibnamefont {Marino}}, \bibinfo {author} {\bibfnamefont {J.}~\bibnamefont {Melchior}}, \bibinfo {author} {\bibfnamefont {A.}~\bibnamefont {Wohlfarth}},\ and\ \bibinfo {author} {\bibfnamefont {K.}~\bibnamefont {Kreuer}},\ }\bibfield  {title} {\bibinfo {title} {Hydroxide, halide and water transport in a model anion exchange membrane},\ }\href {https://doi.org/10.1016/j.memsci.2014.04.003} {\bibfield  {journal} {\bibinfo  {journal} {J. Membr. Sci.}\ }\textbf {\bibinfo {volume} {464}},\ \bibinfo {pages} {61} (\bibinfo {year} {2014})}\BibitemShut {NoStop}%
\bibitem [{\citenamefont {Miller}\ \emph {et~al.}(2020)\citenamefont {Miller}, \citenamefont {Bouzek}, \citenamefont {Hnat}, \citenamefont {Loos}, \citenamefont {Bern{\"a}cker}, \citenamefont {Wei{\ss}g{\"a}rber}, \citenamefont {R{\"o}ntzsch},\ and\ \citenamefont {Meier-Haack}}]{miller2020}%
  \BibitemOpen
  \bibfield  {author} {\bibinfo {author} {\bibfnamefont {H.~A.}\ \bibnamefont {Miller}}, \bibinfo {author} {\bibfnamefont {K.}~\bibnamefont {Bouzek}}, \bibinfo {author} {\bibfnamefont {J.}~\bibnamefont {Hnat}}, \bibinfo {author} {\bibfnamefont {S.}~\bibnamefont {Loos}}, \bibinfo {author} {\bibfnamefont {C.~I.}\ \bibnamefont {Bern{\"a}cker}}, \bibinfo {author} {\bibfnamefont {T.}~\bibnamefont {Wei{\ss}g{\"a}rber}}, \bibinfo {author} {\bibfnamefont {L.}~\bibnamefont {R{\"o}ntzsch}},\ and\ \bibinfo {author} {\bibfnamefont {J.}~\bibnamefont {Meier-Haack}},\ }\bibfield  {title} {\bibinfo {title} {Green hydrogen from anion exchange membrane water electrolysis: a review of recent developments in critical materials and operating conditions},\ }\href {https://doi.org/10.1039/C9SE01240K} {\bibfield  {journal} {\bibinfo  {journal} {Sustain. Energy Fuels}\ }\textbf {\bibinfo {volume} {4}},\ \bibinfo {pages} {2114} (\bibinfo {year} {2020})}\BibitemShut {NoStop}%
\bibitem [{\citenamefont {Busacca}\ \emph {et~al.}(2019)\citenamefont {Busacca}, \citenamefont {Zignani}, \citenamefont {{Di Blasi}}, \citenamefont {{Di Blasi}}, \citenamefont {{Lo Faro}}, \citenamefont {Antonucci},\ and\ \citenamefont {Aricò}}]{busacca22019}%
  \BibitemOpen
  \bibfield  {author} {\bibinfo {author} {\bibfnamefont {C.}~\bibnamefont {Busacca}}, \bibinfo {author} {\bibfnamefont {S.}~\bibnamefont {Zignani}}, \bibinfo {author} {\bibfnamefont {A.}~\bibnamefont {{Di Blasi}}}, \bibinfo {author} {\bibfnamefont {O.}~\bibnamefont {{Di Blasi}}}, \bibinfo {author} {\bibfnamefont {M.}~\bibnamefont {{Lo Faro}}}, \bibinfo {author} {\bibfnamefont {V.}~\bibnamefont {Antonucci}},\ and\ \bibinfo {author} {\bibfnamefont {A.}~\bibnamefont {Aricò}},\ }\bibfield  {title} {\bibinfo {title} {{Electrospun NiMn2O4 and NiCo2O4 spinel oxides supported on carbon nanofibers as electrocatalysts for the oxygen evolution reaction in an anion exchange membrane-based electrolysis cell}},\ }\href {https://doi.org/10.1016/j.ijhydene.2019.02.214} {\bibfield  {journal} {\bibinfo  {journal} {Int. J. Hydrog. Energy}\ }\textbf {\bibinfo {volume} {44}},\ \bibinfo {pages} {20987} (\bibinfo {year} {2019})},\ \bibinfo {note} {special Issue on HYPOTHESIS XIII}\BibitemShut {NoStop}%
\bibitem [{\citenamefont {Giovanelli}\ \emph {et~al.}(2024)\citenamefont {Giovanelli}, \citenamefont {Pozio}, \citenamefont {Pucci}, \citenamefont {Geppi},\ and\ \citenamefont {Martini}}]{giovanelli24_faa3}%
  \BibitemOpen
  \bibfield  {author} {\bibinfo {author} {\bibfnamefont {A.}~\bibnamefont {Giovanelli}}, \bibinfo {author} {\bibfnamefont {A.}~\bibnamefont {Pozio}}, \bibinfo {author} {\bibfnamefont {A.}~\bibnamefont {Pucci}}, \bibinfo {author} {\bibfnamefont {M.}~\bibnamefont {Geppi}},\ and\ \bibinfo {author} {\bibfnamefont {F.}~\bibnamefont {Martini}},\ }\bibfield  {title} {\bibinfo {title} {{Fumasep FAA-3-PK-130: Exploiting Multinuclear Solid-State NMR to Shed Light on Undisclosed Structural Properties}},\ }\href {https://doi.org/10.1016/j.polymer.2024.127536} {\bibfield  {journal} {\bibinfo  {journal} {Polymer}\ ,\ \bibinfo {pages} {127536}} (\bibinfo {year} {2024})}\BibitemShut {NoStop}%
\bibitem [{\citenamefont {Favero}\ \emph {et~al.}(2024)\citenamefont {Favero}, \citenamefont {Stephens},\ and\ \citenamefont {Titirci}}]{favero2024}%
  \BibitemOpen
  \bibfield  {author} {\bibinfo {author} {\bibfnamefont {S.}~\bibnamefont {Favero}}, \bibinfo {author} {\bibfnamefont {I.~E.~L.}\ \bibnamefont {Stephens}},\ and\ \bibinfo {author} {\bibfnamefont {M.-M.}\ \bibnamefont {Titirci}},\ }\bibfield  {title} {\bibinfo {title} {Anion exchange ionomers: Design considerations and recent advances - an electrochemical perspective},\ }\href {https://doi.org/10.1002/adma.202308238} {\bibfield  {journal} {\bibinfo  {journal} {Adv. Mater.}\ }\textbf {\bibinfo {volume} {36}},\ \bibinfo {pages} {2308238} (\bibinfo {year} {2024})}\BibitemShut {NoStop}%
\bibitem [{\citenamefont {Zhegur-Khais}\ \emph {et~al.}(2020)\citenamefont {Zhegur-Khais}, \citenamefont {Kubannek}, \citenamefont {Krewer},\ and\ \citenamefont {Dekel}}]{zhegur2020measuring}%
  \BibitemOpen
  \bibfield  {author} {\bibinfo {author} {\bibfnamefont {A.}~\bibnamefont {Zhegur-Khais}}, \bibinfo {author} {\bibfnamefont {F.}~\bibnamefont {Kubannek}}, \bibinfo {author} {\bibfnamefont {U.}~\bibnamefont {Krewer}},\ and\ \bibinfo {author} {\bibfnamefont {D.~R.}\ \bibnamefont {Dekel}},\ }\bibfield  {title} {\bibinfo {title} {Measuring the true hydroxide conductivity of anion exchange membranes},\ }\href {https://doi.org/10.1016/j.memsci.2020.118461} {\bibfield  {journal} {\bibinfo  {journal} {J. Membr. Sci.}\ }\textbf {\bibinfo {volume} {612}},\ \bibinfo {pages} {118461} (\bibinfo {year} {2020})}\BibitemShut {NoStop}%
\bibitem [{\citenamefont {Tuckerman}\ \emph {et~al.}(2002)\citenamefont {Tuckerman}, \citenamefont {Marx},\ and\ \citenamefont {Parrinello}}]{tuckerman2002nature}%
  \BibitemOpen
  \bibfield  {author} {\bibinfo {author} {\bibfnamefont {M.}~\bibnamefont {Tuckerman}}, \bibinfo {author} {\bibfnamefont {D.}~\bibnamefont {Marx}},\ and\ \bibinfo {author} {\bibfnamefont {M.}~\bibnamefont {Parrinello}},\ }\bibfield  {title} {\bibinfo {title} {The nature and transport mechanism of hydrated hydroxide ions in aqueous solution},\ }\href@noop {} {\bibfield  {journal} {\bibinfo  {journal} {Nature}\ }\textbf {\bibinfo {volume} {417}},\ \bibinfo {pages} {925} (\bibinfo {year} {2002})}\BibitemShut {NoStop}%
\bibitem [{\citenamefont {Long}\ and\ \citenamefont {Tuckerman}(2023)}]{zhuoran23cylindernano}%
  \BibitemOpen
  \bibfield  {author} {\bibinfo {author} {\bibfnamefont {Z.}~\bibnamefont {Long}}\ and\ \bibinfo {author} {\bibfnamefont {M.~E.}\ \bibnamefont {Tuckerman}},\ }\bibfield  {title} {\bibinfo {title} {Hydroxide diffusion in functionalized cylindrical nanopores as idealized models of anion exchange membrane environments: An ab initio molecular dynamics study},\ }\href {https://doi.org/10.1021/acs.jpcc.2c05747} {\bibfield  {journal} {\bibinfo  {journal} {J. Phys. Chem. C}\ }\textbf {\bibinfo {volume} {127}},\ \bibinfo {pages} {2792} (\bibinfo {year} {2023})}\BibitemShut {NoStop}%
\bibitem [{\citenamefont {Lin}\ \emph {et~al.}(2012)\citenamefont {Lin}, \citenamefont {Seitsonen}, \citenamefont {Tavernelli},\ and\ \citenamefont {Rothlisberger}}]{blyp_overstructure1}%
  \BibitemOpen
  \bibfield  {author} {\bibinfo {author} {\bibfnamefont {I.-C.}\ \bibnamefont {Lin}}, \bibinfo {author} {\bibfnamefont {A.~P.}\ \bibnamefont {Seitsonen}}, \bibinfo {author} {\bibfnamefont {I.}~\bibnamefont {Tavernelli}},\ and\ \bibinfo {author} {\bibfnamefont {U.}~\bibnamefont {Rothlisberger}},\ }\bibfield  {title} {\bibinfo {title} {Structure and dynamics of liquid water from ab initio molecular dynamics--comparison of blyp, pbe, and revpbe density functionals with and without van der waals corrections},\ }\href {https://doi.org/10.1021/ct3001848} {\bibfield  {journal} {\bibinfo  {journal} {J. Chem. Theory Comput.}\ }\textbf {\bibinfo {volume} {8}},\ \bibinfo {pages} {3902} (\bibinfo {year} {2012})}\BibitemShut {NoStop}%
\bibitem [{\citenamefont {Mu\~{n}oz Santiburcio}(2022)}]{blyp_overstructure2}%
  \BibitemOpen
  \bibfield  {author} {\bibinfo {author} {\bibfnamefont {D.}~\bibnamefont {Mu\~{n}oz Santiburcio}},\ }\bibfield  {title} {\bibinfo {title} {Accurate diffusion coefficients of the excess proton and hydroxide in water via extensive ab initio simulations with different schemes},\ }\href {https://doi.org/10.1063/5.0093958} {\bibfield  {journal} {\bibinfo  {journal} {J. Chem. Phys.}\ }\textbf {\bibinfo {volume} {157}},\ \bibinfo {pages} {024504} (\bibinfo {year} {2022})}\BibitemShut {NoStop}%
\bibitem [{\citenamefont {Hänseroth}\ \emph {et~al.}(2025{\natexlab{b}})\citenamefont {Hänseroth}, \citenamefont {Flötotto}, \citenamefont {Qaisrani},\ and\ \citenamefont {Dreßler}}]{haenseroth2025atk}%
  \BibitemOpen
  \bibfield  {author} {\bibinfo {author} {\bibfnamefont {J.}~\bibnamefont {Hänseroth}}, \bibinfo {author} {\bibfnamefont {A.}~\bibnamefont {Flötotto}}, \bibinfo {author} {\bibfnamefont {M.~N.}\ \bibnamefont {Qaisrani}},\ and\ \bibinfo {author} {\bibfnamefont {C.}~\bibnamefont {Dreßler}},\ }\bibfield  {title} {\bibinfo {title} {Fine-tuning unifies foundational machine-learned interatomic potential architectures at ab initio accuracy},\ }\bibfield  {journal} {\bibinfo  {journal} {arXiv preprint arXiv:2511.05337}\ }\href {https://doi.org/10.48550/arXiv.2511.05337} {10.48550/arXiv.2511.05337} (\bibinfo {year} {2025}{\natexlab{b}})\BibitemShut {NoStop}%
\bibitem [{\citenamefont {Lippert}\ and\ \citenamefont {Parrinello}(1997)}]{cp2k_1}%
  \BibitemOpen
  \bibfield  {author} {\bibinfo {author} {\bibfnamefont {J.}~\bibnamefont {Lippert}, \bibfnamefont {G.~Hutter}}\ and\ \bibinfo {author} {\bibfnamefont {M.}~\bibnamefont {Parrinello}},\ }\bibfield  {title} {\bibinfo {title} {A hybrid gaussian and plane wave density functional scheme},\ }\href {https://doi.org/10.1080/002689797170220} {\bibfield  {journal} {\bibinfo  {journal} {Mol. Phys.}\ }\textbf {\bibinfo {volume} {92}},\ \bibinfo {pages} {477} (\bibinfo {year} {1997})}\BibitemShut {NoStop}%
\bibitem [{\citenamefont {Hutter}\ \emph {et~al.}(2014)\citenamefont {Hutter}, \citenamefont {Iannuzzi}, \citenamefont {Schiffmann},\ and\ \citenamefont {VandeVondele}}]{cp2k_2}%
  \BibitemOpen
  \bibfield  {author} {\bibinfo {author} {\bibfnamefont {J.}~\bibnamefont {Hutter}}, \bibinfo {author} {\bibfnamefont {M.}~\bibnamefont {Iannuzzi}}, \bibinfo {author} {\bibfnamefont {F.}~\bibnamefont {Schiffmann}},\ and\ \bibinfo {author} {\bibfnamefont {J.}~\bibnamefont {VandeVondele}},\ }\bibfield  {title} {\bibinfo {title} {cp2k: atomistic simulations of condensed matter systems},\ }\href {https://doi.org/10.1002/wcms.1159} {\bibfield  {journal} {\bibinfo  {journal} {Wiley Interdiscip. Rev. Comput. Mol. Sci.}\ }\textbf {\bibinfo {volume} {4}},\ \bibinfo {pages} {15} (\bibinfo {year} {2014})}\BibitemShut {NoStop}%
\bibitem [{\citenamefont {Borštnik}\ \emph {et~al.}(2014)\citenamefont {Borštnik}, \citenamefont {VandeVondele}, \citenamefont {Weber},\ and\ \citenamefont {Hutter}}]{cp2k_3}%
  \BibitemOpen
  \bibfield  {author} {\bibinfo {author} {\bibfnamefont {U.}~\bibnamefont {Borštnik}}, \bibinfo {author} {\bibfnamefont {J.}~\bibnamefont {VandeVondele}}, \bibinfo {author} {\bibfnamefont {V.}~\bibnamefont {Weber}},\ and\ \bibinfo {author} {\bibfnamefont {J.}~\bibnamefont {Hutter}},\ }\bibfield  {title} {\bibinfo {title} {Sparse matrix multiplication: The distributed block-compressed sparse row library},\ }\href {https://doi.org/10.1016/j.parco.2014.03.012} {\bibfield  {journal} {\bibinfo  {journal} {Parallel Comput.}\ }\textbf {\bibinfo {volume} {40}},\ \bibinfo {pages} {47} (\bibinfo {year} {2014})}\BibitemShut {NoStop}%
\bibitem [{\citenamefont {K{\"u}hne}\ \emph {et~al.}(2020)\citenamefont {K{\"u}hne}, \citenamefont {Iannuzzi}, \citenamefont {Del~Ben}, \citenamefont {Rybkin}, \citenamefont {Seewald}, \citenamefont {Stein}, \citenamefont {Laino}, \citenamefont {Khaliullin}, \citenamefont {Sch{\"u}tt}, \citenamefont {Schiffmann} \emph {et~al.}}]{cp2k_4}%
  \BibitemOpen
  \bibfield  {author} {\bibinfo {author} {\bibfnamefont {T.~D.}\ \bibnamefont {K{\"u}hne}}, \bibinfo {author} {\bibfnamefont {M.}~\bibnamefont {Iannuzzi}}, \bibinfo {author} {\bibfnamefont {M.}~\bibnamefont {Del~Ben}}, \bibinfo {author} {\bibfnamefont {V.~V.}\ \bibnamefont {Rybkin}}, \bibinfo {author} {\bibfnamefont {P.}~\bibnamefont {Seewald}}, \bibinfo {author} {\bibfnamefont {F.}~\bibnamefont {Stein}}, \bibinfo {author} {\bibfnamefont {T.}~\bibnamefont {Laino}}, \bibinfo {author} {\bibfnamefont {R.~Z.}\ \bibnamefont {Khaliullin}}, \bibinfo {author} {\bibfnamefont {O.}~\bibnamefont {Sch{\"u}tt}}, \bibinfo {author} {\bibfnamefont {F.}~\bibnamefont {Schiffmann}}, \emph {et~al.},\ }\bibfield  {title} {\bibinfo {title} {{CP2K}: An electronic structure and molecular dynamics software package-{Q}uickstep: {E}fficient and accurate electronic structure calculations},\ }\bibfield  {journal} {\bibinfo  {journal} {J. Chem. Phys.}\ }\textbf {\bibinfo {volume} {152}},\ \href {https://doi.org/10.1063/5.0007045}
  {10.1063/5.0007045} (\bibinfo {year} {2020})\BibitemShut {NoStop}%
\bibitem [{\citenamefont {Iannuzzi}\ \emph {et~al.}(2026)\citenamefont {Iannuzzi}, \citenamefont {Wilhelm}, \citenamefont {Stein}, \citenamefont {Bussy}, \citenamefont {Elgabarty}, \citenamefont {Golze}, \citenamefont {Hehn}, \citenamefont {Graml}, \citenamefont {Marek}, \citenamefont {G{\"o}kmen}, \citenamefont {Schran}, \citenamefont {Forbert}, \citenamefont {Khaliullin}, \citenamefont {Kozhevnikov}, \citenamefont {Taillefumier}, \citenamefont {Meli}, \citenamefont {Rybkin}, \citenamefont {Brehm}, \citenamefont {Schade}, \citenamefont {Sch{\"u}tt}, \citenamefont {Pototschnig}, \citenamefont {Mirhosseini}, \citenamefont {Kn{\"u}pfer}, \citenamefont {Marx}, \citenamefont {Krack}, \citenamefont {Hutter},\ and\ \citenamefont {K{\"u}hne}}]{cp2k_5}%
  \BibitemOpen
  \bibfield  {author} {\bibinfo {author} {\bibfnamefont {M.}~\bibnamefont {Iannuzzi}}, \bibinfo {author} {\bibfnamefont {J.}~\bibnamefont {Wilhelm}}, \bibinfo {author} {\bibfnamefont {F.}~\bibnamefont {Stein}}, \bibinfo {author} {\bibfnamefont {A.}~\bibnamefont {Bussy}}, \bibinfo {author} {\bibfnamefont {H.}~\bibnamefont {Elgabarty}}, \bibinfo {author} {\bibfnamefont {D.}~\bibnamefont {Golze}}, \bibinfo {author} {\bibfnamefont {A.-S.}\ \bibnamefont {Hehn}}, \bibinfo {author} {\bibfnamefont {M.}~\bibnamefont {Graml}}, \bibinfo {author} {\bibfnamefont {S.}~\bibnamefont {Marek}}, \bibinfo {author} {\bibfnamefont {B.~S.}\ \bibnamefont {G{\"o}kmen}}, \bibinfo {author} {\bibfnamefont {C.}~\bibnamefont {Schran}}, \bibinfo {author} {\bibfnamefont {H.}~\bibnamefont {Forbert}}, \bibinfo {author} {\bibfnamefont {R.~Z.}\ \bibnamefont {Khaliullin}}, \bibinfo {author} {\bibfnamefont {A.}~\bibnamefont {Kozhevnikov}}, \bibinfo {author} {\bibfnamefont {M.}~\bibnamefont {Taillefumier}}, \bibinfo {author} {\bibfnamefont
  {R.}~\bibnamefont {Meli}}, \bibinfo {author} {\bibfnamefont {V.~V.}\ \bibnamefont {Rybkin}}, \bibinfo {author} {\bibfnamefont {M.}~\bibnamefont {Brehm}}, \bibinfo {author} {\bibfnamefont {R.}~\bibnamefont {Schade}}, \bibinfo {author} {\bibfnamefont {O.}~\bibnamefont {Sch{\"u}tt}}, \bibinfo {author} {\bibfnamefont {J.~V.}\ \bibnamefont {Pototschnig}}, \bibinfo {author} {\bibfnamefont {H.}~\bibnamefont {Mirhosseini}}, \bibinfo {author} {\bibfnamefont {A.}~\bibnamefont {Kn{\"u}pfer}}, \bibinfo {author} {\bibfnamefont {D.}~\bibnamefont {Marx}}, \bibinfo {author} {\bibfnamefont {M.}~\bibnamefont {Krack}}, \bibinfo {author} {\bibfnamefont {J.}~\bibnamefont {Hutter}},\ and\ \bibinfo {author} {\bibfnamefont {T.~D.}\ \bibnamefont {K{\"u}hne}},\ }\bibfield  {title} {\bibinfo {title} {{The CP2K Program Package Made Simple}},\ }\href {https://doi.org/10.1021/acs.jpcb.5c05851} {\bibfield  {journal} {\bibinfo  {journal} {J. Phys. Chem. B}\ }\textbf {\bibinfo {volume} {130}},\ \bibinfo {pages} {1237} (\bibinfo {year}
  {2026})}\BibitemShut {NoStop}%
\bibitem [{\citenamefont {VandeVondele}\ \emph {et~al.}(2005)\citenamefont {VandeVondele}, \citenamefont {Krack}, \citenamefont {Mohamed}, \citenamefont {Parrinello}, \citenamefont {Chassaing},\ and\ \citenamefont {Hutter}}]{cp2k_quickstep}%
  \BibitemOpen
  \bibfield  {author} {\bibinfo {author} {\bibfnamefont {J.}~\bibnamefont {VandeVondele}}, \bibinfo {author} {\bibfnamefont {M.}~\bibnamefont {Krack}}, \bibinfo {author} {\bibfnamefont {F.}~\bibnamefont {Mohamed}}, \bibinfo {author} {\bibfnamefont {M.}~\bibnamefont {Parrinello}}, \bibinfo {author} {\bibfnamefont {T.}~\bibnamefont {Chassaing}},\ and\ \bibinfo {author} {\bibfnamefont {J.}~\bibnamefont {Hutter}},\ }\bibfield  {title} {\bibinfo {title} {Quickstep: Fast and accurate density functional calculations using a mixed gaussian and plane waves approach},\ }\href {https://doi.org/10.1016/j.cpc.2004.12.014} {\bibfield  {journal} {\bibinfo  {journal} {Comput. Phys. Commun.}\ }\textbf {\bibinfo {volume} {167}},\ \bibinfo {pages} {103} (\bibinfo {year} {2005})}\BibitemShut {NoStop}%
\bibitem [{\citenamefont {VandeVondele}\ and\ \citenamefont {Hutter}(2007)}]{cp2k_basis-set}%
  \BibitemOpen
  \bibfield  {author} {\bibinfo {author} {\bibfnamefont {J.}~\bibnamefont {VandeVondele}}\ and\ \bibinfo {author} {\bibfnamefont {J.}~\bibnamefont {Hutter}},\ }\bibfield  {title} {\bibinfo {title} {Gaussian basis sets for accurate calculations on molecular systems in gas and condensed phases},\ }\href {https://doi.org/10.1063/1.2770708} {\bibfield  {journal} {\bibinfo  {journal} {J. Chem. Phys.}\ }\textbf {\bibinfo {volume} {127}},\ \bibinfo {pages} {114105} (\bibinfo {year} {2007})}\BibitemShut {NoStop}%
\bibitem [{\citenamefont {VandeVondele}\ and\ \citenamefont {Hutter}(2003)}]{cp2k_orb_trans}%
  \BibitemOpen
  \bibfield  {author} {\bibinfo {author} {\bibfnamefont {J.}~\bibnamefont {VandeVondele}}\ and\ \bibinfo {author} {\bibfnamefont {J.}~\bibnamefont {Hutter}},\ }\bibfield  {title} {\bibinfo {title} {An efficient orbital transformation method for electronic structure calculations},\ }\href {https://doi.org/10.1063/1.1543154} {\bibfield  {journal} {\bibinfo  {journal} {J. Chem. Phys.}\ }\textbf {\bibinfo {volume} {118}},\ \bibinfo {pages} {4365} (\bibinfo {year} {2003})}\BibitemShut {NoStop}%
\bibitem [{\citenamefont {Hartwigsen}\ \emph {et~al.}(1998)\citenamefont {Hartwigsen}, \citenamefont {Goedecker},\ and\ \citenamefont {Hutter}}]{cp2k_gth-pseudopot1}%
  \BibitemOpen
  \bibfield  {author} {\bibinfo {author} {\bibfnamefont {C.}~\bibnamefont {Hartwigsen}}, \bibinfo {author} {\bibfnamefont {S.}~\bibnamefont {Goedecker}},\ and\ \bibinfo {author} {\bibfnamefont {J.}~\bibnamefont {Hutter}},\ }\bibfield  {title} {\bibinfo {title} {{Relativistic separable dual-space Gaussian pseudopotentials from H to Rn}},\ }\href {https://doi.org/10.1103/PhysRevB.58.3641} {\bibfield  {journal} {\bibinfo  {journal} {Phys. Rev. B}\ }\textbf {\bibinfo {volume} {58}},\ \bibinfo {pages} {3641} (\bibinfo {year} {1998})}\BibitemShut {NoStop}%
\bibitem [{\citenamefont {Krack}(2005)}]{cp2k_gth-pseudopot2}%
  \BibitemOpen
  \bibfield  {author} {\bibinfo {author} {\bibfnamefont {M.}~\bibnamefont {Krack}},\ }\bibfield  {title} {\bibinfo {title} {{Pseudopotentials for H to Kr optimized for gradient-corrected exchange-correlation functionals}},\ }\href {https://doi.org/10.1007/s00214-005-0655-y} {\bibfield  {journal} {\bibinfo  {journal} {Theor. Chem. Acc.}\ }\textbf {\bibinfo {volume} {114}},\ \bibinfo {pages} {145} (\bibinfo {year} {2005})}\BibitemShut {NoStop}%
\bibitem [{\citenamefont {Goedecker}\ \emph {et~al.}(1996)\citenamefont {Goedecker}, \citenamefont {Teter},\ and\ \citenamefont {Hutter}}]{cp2k_gth-pseudopot3}%
  \BibitemOpen
  \bibfield  {author} {\bibinfo {author} {\bibfnamefont {S.}~\bibnamefont {Goedecker}}, \bibinfo {author} {\bibfnamefont {M.}~\bibnamefont {Teter}},\ and\ \bibinfo {author} {\bibfnamefont {J.}~\bibnamefont {Hutter}},\ }\bibfield  {title} {\bibinfo {title} {Separable dual-space gaussian pseudopotentials},\ }\href {https://doi.org/10.1103/PhysRevB.54.1703} {\bibfield  {journal} {\bibinfo  {journal} {Phys. Rev. B}\ }\textbf {\bibinfo {volume} {54}},\ \bibinfo {pages} {1703} (\bibinfo {year} {1996})}\BibitemShut {NoStop}%
\bibitem [{\citenamefont {Becke}(1988)}]{blyp1}%
  \BibitemOpen
  \bibfield  {author} {\bibinfo {author} {\bibfnamefont {A.}~\bibnamefont {Becke}},\ }\bibfield  {title} {\bibinfo {title} {Density-functional exchange-energy approximation with correct asymptotic behavior},\ }\href {https://doi.org/10.1103/PhysRevA.38.3098} {\bibfield  {journal} {\bibinfo  {journal} {Phys. Rev. A}\ }\textbf {\bibinfo {volume} {38}},\ \bibinfo {pages} {3098} (\bibinfo {year} {1988})}\BibitemShut {NoStop}%
\bibitem [{\citenamefont {Lee}\ \emph {et~al.}(1988)\citenamefont {Lee}, \citenamefont {Yang},\ and\ \citenamefont {Parr}}]{blyp2}%
  \BibitemOpen
  \bibfield  {author} {\bibinfo {author} {\bibfnamefont {C.}~\bibnamefont {Lee}}, \bibinfo {author} {\bibfnamefont {W.}~\bibnamefont {Yang}},\ and\ \bibinfo {author} {\bibfnamefont {R.}~\bibnamefont {Parr}},\ }\bibfield  {title} {\bibinfo {title} {Development of the colle-salvetti correlation-energy formula into a functional of the electron density},\ }\href {https://doi.org/10.1103/PhysRevB.37.785} {\bibfield  {journal} {\bibinfo  {journal} {Phys. Rev. A}\ }\textbf {\bibinfo {volume} {37}},\ \bibinfo {pages} {785} (\bibinfo {year} {1988})}\BibitemShut {NoStop}%
\bibitem [{\citenamefont {Nos\'{e}}(1984)}]{nose1}%
  \BibitemOpen
  \bibfield  {author} {\bibinfo {author} {\bibfnamefont {S.}~\bibnamefont {Nos\'{e}}},\ }\bibfield  {title} {\bibinfo {title} {A {U}nified {F}ormulation of the {C}onstant {T}emperature {M}olecular {D}ynamics {M}ethods},\ }\href {https://doi.org/10.1063/1.447334} {\bibfield  {journal} {\bibinfo  {journal} {J. Chem. Phys.}\ }\textbf {\bibinfo {volume} {81}},\ \bibinfo {pages} {511} (\bibinfo {year} {1984})}\BibitemShut {NoStop}%
\bibitem [{\citenamefont {Nos\'{e}}(1970)}]{nose2}%
  \BibitemOpen
  \bibfield  {author} {\bibinfo {author} {\bibfnamefont {S.}~\bibnamefont {Nos\'{e}}},\ }\bibfield  {title} {\bibinfo {title} {A {M}olecular {D}ynamics {M}ethod for {S}imulations in the {C}anonical {E}nsemble},\ }\href {https://doi.org/10.1080/00268978400101201} {\bibfield  {journal} {\bibinfo  {journal} {Mol. Phys.}\ }\textbf {\bibinfo {volume} {52}},\ \bibinfo {pages} {255} (\bibinfo {year} {1970})}\BibitemShut {NoStop}%
\bibitem [{\citenamefont {Martyna}\ \emph {et~al.}(1992)\citenamefont {Martyna}, \citenamefont {Klein},\ and\ \citenamefont {Tuckerman}}]{nose3}%
  \BibitemOpen
  \bibfield  {author} {\bibinfo {author} {\bibfnamefont {G.~J.}\ \bibnamefont {Martyna}}, \bibinfo {author} {\bibfnamefont {M.~L.}\ \bibnamefont {Klein}},\ and\ \bibinfo {author} {\bibfnamefont {M.}~\bibnamefont {Tuckerman}},\ }\bibfield  {title} {\bibinfo {title} {Nos\'{e}-{H}oover chains: {T}he {C}anonical {E}nsemble via {C}ontinuous {D}ynamics},\ }\href {https://doi.org/10.1063/1.463940} {\bibfield  {journal} {\bibinfo  {journal} {J. Chem. Phys.}\ }\textbf {\bibinfo {volume} {97}},\ \bibinfo {pages} {2635} (\bibinfo {year} {1992})}\BibitemShut {NoStop}%
\bibitem [{\citenamefont {Thompson}\ \emph {et~al.}(2022)\citenamefont {Thompson}, \citenamefont {Aktulga}, \citenamefont {Berger}, \citenamefont {Bolintineanu}, \citenamefont {Brown}, \citenamefont {Crozier}, \citenamefont {in~'t Veld}, \citenamefont {Kohlmeyer}, \citenamefont {Moore}, \citenamefont {Nguyen}, \citenamefont {Shan}, \citenamefont {Stevens}, \citenamefont {Tranchida}, \citenamefont {Trott},\ and\ \citenamefont {Plimpton}}]{lammps}%
  \BibitemOpen
  \bibfield  {author} {\bibinfo {author} {\bibfnamefont {A.~P.}\ \bibnamefont {Thompson}}, \bibinfo {author} {\bibfnamefont {H.~M.}\ \bibnamefont {Aktulga}}, \bibinfo {author} {\bibfnamefont {R.}~\bibnamefont {Berger}}, \bibinfo {author} {\bibfnamefont {D.~S.}\ \bibnamefont {Bolintineanu}}, \bibinfo {author} {\bibfnamefont {W.~M.}\ \bibnamefont {Brown}}, \bibinfo {author} {\bibfnamefont {P.~S.}\ \bibnamefont {Crozier}}, \bibinfo {author} {\bibfnamefont {P.~J.}\ \bibnamefont {in~'t Veld}}, \bibinfo {author} {\bibfnamefont {A.}~\bibnamefont {Kohlmeyer}}, \bibinfo {author} {\bibfnamefont {S.~G.}\ \bibnamefont {Moore}}, \bibinfo {author} {\bibfnamefont {T.~D.}\ \bibnamefont {Nguyen}}, \bibinfo {author} {\bibfnamefont {R.}~\bibnamefont {Shan}}, \bibinfo {author} {\bibfnamefont {M.~J.}\ \bibnamefont {Stevens}}, \bibinfo {author} {\bibfnamefont {J.}~\bibnamefont {Tranchida}}, \bibinfo {author} {\bibfnamefont {C.}~\bibnamefont {Trott}},\ and\ \bibinfo {author} {\bibfnamefont {S.~J.}\ \bibnamefont {Plimpton}},\
  }\bibfield  {title} {\bibinfo {title} {{LAMMPS - a flexible simulation tool for particle-based materials modeling at the atomic, meso, and continuum scales}},\ }\href {https://doi.org/10.1016/j.cpc.2021.108171} {\bibfield  {journal} {\bibinfo  {journal} {Comput. Phys. Commun.}\ }\textbf {\bibinfo {volume} {271}},\ \bibinfo {pages} {108171} (\bibinfo {year} {2022})}\BibitemShut {NoStop}%
\bibitem [{\citenamefont {Larsen}\ \emph {et~al.}(2017)\citenamefont {Larsen}, \citenamefont {Mortensen}, \citenamefont {Blomqvist}, \citenamefont {Castelli}, \citenamefont {Christensen}, \citenamefont {Dułak}, \citenamefont {Friis}, \citenamefont {Groves}, \citenamefont {Hammer}, \citenamefont {Hargus}, \citenamefont {Hermes}, \citenamefont {Jennings}, \citenamefont {Jensen}, \citenamefont {Kermode}, \citenamefont {Kitchin}, \citenamefont {Kolsbjerg}, \citenamefont {Kubal}, \citenamefont {Kaasbjerg}, \citenamefont {Lysgaard}, \citenamefont {Maronsson}, \citenamefont {Maxson}, \citenamefont {Olsen}, \citenamefont {Pastewka}, \citenamefont {Peterson}, \citenamefont {Rostgaard}, \citenamefont {Schiøtz}, \citenamefont {Schütt}, \citenamefont {Strange}, \citenamefont {Thygesen}, \citenamefont {Vegge}, \citenamefont {Vilhelmsen}, \citenamefont {Walter}, \citenamefont {Zeng},\ and\ \citenamefont {Jacobsen}}]{ase}%
  \BibitemOpen
  \bibfield  {author} {\bibinfo {author} {\bibfnamefont {A.~H.}\ \bibnamefont {Larsen}}, \bibinfo {author} {\bibfnamefont {J.~J.}\ \bibnamefont {Mortensen}}, \bibinfo {author} {\bibfnamefont {J.}~\bibnamefont {Blomqvist}}, \bibinfo {author} {\bibfnamefont {I.~E.}\ \bibnamefont {Castelli}}, \bibinfo {author} {\bibfnamefont {R.}~\bibnamefont {Christensen}}, \bibinfo {author} {\bibfnamefont {M.}~\bibnamefont {Dułak}}, \bibinfo {author} {\bibfnamefont {J.}~\bibnamefont {Friis}}, \bibinfo {author} {\bibfnamefont {M.~N.}\ \bibnamefont {Groves}}, \bibinfo {author} {\bibfnamefont {B.}~\bibnamefont {Hammer}}, \bibinfo {author} {\bibfnamefont {C.}~\bibnamefont {Hargus}}, \bibinfo {author} {\bibfnamefont {E.~D.}\ \bibnamefont {Hermes}}, \bibinfo {author} {\bibfnamefont {P.~C.}\ \bibnamefont {Jennings}}, \bibinfo {author} {\bibfnamefont {P.~B.}\ \bibnamefont {Jensen}}, \bibinfo {author} {\bibfnamefont {J.}~\bibnamefont {Kermode}}, \bibinfo {author} {\bibfnamefont {J.~R.}\ \bibnamefont {Kitchin}}, \bibinfo {author}
  {\bibfnamefont {E.~L.}\ \bibnamefont {Kolsbjerg}}, \bibinfo {author} {\bibfnamefont {J.}~\bibnamefont {Kubal}}, \bibinfo {author} {\bibfnamefont {K.}~\bibnamefont {Kaasbjerg}}, \bibinfo {author} {\bibfnamefont {S.}~\bibnamefont {Lysgaard}}, \bibinfo {author} {\bibfnamefont {J.~B.}\ \bibnamefont {Maronsson}}, \bibinfo {author} {\bibfnamefont {T.}~\bibnamefont {Maxson}}, \bibinfo {author} {\bibfnamefont {T.}~\bibnamefont {Olsen}}, \bibinfo {author} {\bibfnamefont {L.}~\bibnamefont {Pastewka}}, \bibinfo {author} {\bibfnamefont {A.}~\bibnamefont {Peterson}}, \bibinfo {author} {\bibfnamefont {C.}~\bibnamefont {Rostgaard}}, \bibinfo {author} {\bibfnamefont {J.}~\bibnamefont {Schiøtz}}, \bibinfo {author} {\bibfnamefont {O.}~\bibnamefont {Schütt}}, \bibinfo {author} {\bibfnamefont {M.}~\bibnamefont {Strange}}, \bibinfo {author} {\bibfnamefont {K.~S.}\ \bibnamefont {Thygesen}}, \bibinfo {author} {\bibfnamefont {T.}~\bibnamefont {Vegge}}, \bibinfo {author} {\bibfnamefont {L.}~\bibnamefont {Vilhelmsen}}, \bibinfo
  {author} {\bibfnamefont {M.}~\bibnamefont {Walter}}, \bibinfo {author} {\bibfnamefont {Z.}~\bibnamefont {Zeng}},\ and\ \bibinfo {author} {\bibfnamefont {K.~W.}\ \bibnamefont {Jacobsen}},\ }\bibfield  {title} {\bibinfo {title} {The atomic simulation environment—a python library for working with atoms},\ }\href {https://doi.org/10.1088/1361-648X/aa680e} {\bibfield  {journal} {\bibinfo  {journal} {J. Phys. Condens. Matter}\ }\textbf {\bibinfo {volume} {29}},\ \bibinfo {pages} {273002} (\bibinfo {year} {2017})}\BibitemShut {NoStop}%
\bibitem [{\citenamefont {Mart{\'\i}nez}\ \emph {et~al.}(2009)\citenamefont {Mart{\'\i}nez}, \citenamefont {Andrade}, \citenamefont {Birgin},\ and\ \citenamefont {Mart{\'\i}nez}}]{martinez2009packmol}%
  \BibitemOpen
  \bibfield  {author} {\bibinfo {author} {\bibfnamefont {L.}~\bibnamefont {Mart{\'\i}nez}}, \bibinfo {author} {\bibfnamefont {R.}~\bibnamefont {Andrade}}, \bibinfo {author} {\bibfnamefont {E.~G.}\ \bibnamefont {Birgin}},\ and\ \bibinfo {author} {\bibfnamefont {J.~M.}\ \bibnamefont {Mart{\'\i}nez}},\ }\bibfield  {title} {\bibinfo {title} {{PACKMOL: A package for building initial configurations for molecular dynamics simulations}},\ }\href {https://doi.org/10.1002/jcc.21224} {\bibfield  {journal} {\bibinfo  {journal} {J. Comput. Chem.}\ }\textbf {\bibinfo {volume} {30}},\ \bibinfo {pages} {2157} (\bibinfo {year} {2009})}\BibitemShut {NoStop}%
\bibitem [{\citenamefont {Hanwell}\ \emph {et~al.}(2012)\citenamefont {Hanwell}, \citenamefont {Curtis}, \citenamefont {Lonie}, \citenamefont {Vandermeersch}, \citenamefont {Zurek},\ and\ \citenamefont {Hutchison}}]{hanwell2012avogadro}%
  \BibitemOpen
  \bibfield  {author} {\bibinfo {author} {\bibfnamefont {M.~D.}\ \bibnamefont {Hanwell}}, \bibinfo {author} {\bibfnamefont {D.~E.}\ \bibnamefont {Curtis}}, \bibinfo {author} {\bibfnamefont {D.~C.}\ \bibnamefont {Lonie}}, \bibinfo {author} {\bibfnamefont {T.}~\bibnamefont {Vandermeersch}}, \bibinfo {author} {\bibfnamefont {E.}~\bibnamefont {Zurek}},\ and\ \bibinfo {author} {\bibfnamefont {G.~R.}\ \bibnamefont {Hutchison}},\ }\bibfield  {title} {\bibinfo {title} {Avogadro: an advanced semantic chemical editor, visualization, and analysis platform},\ }\href {https://doi.org/10.1186/1758-2946-4-17} {\bibfield  {journal} {\bibinfo  {journal} {J. Cheminform.}\ }\textbf {\bibinfo {volume} {4}},\ \bibinfo {pages} {17} (\bibinfo {year} {2012})}\BibitemShut {NoStop}%
\bibitem [{\citenamefont {Hanwell}\ \emph {et~al.}(2022)\citenamefont {Hanwell}, \citenamefont {Curtis}, \citenamefont {Lonie}, \citenamefont {Vandermeersch}, \citenamefont {Zurek},\ and\ \citenamefont {Hutchison}}]{avogadro2}%
  \BibitemOpen
  \bibfield  {author} {\bibinfo {author} {\bibfnamefont {M.~D.}\ \bibnamefont {Hanwell}}, \bibinfo {author} {\bibfnamefont {D.~E.}\ \bibnamefont {Curtis}}, \bibinfo {author} {\bibfnamefont {D.~C.}\ \bibnamefont {Lonie}}, \bibinfo {author} {\bibfnamefont {T.}~\bibnamefont {Vandermeersch}}, \bibinfo {author} {\bibfnamefont {E.}~\bibnamefont {Zurek}},\ and\ \bibinfo {author} {\bibfnamefont {G.~R.}\ \bibnamefont {Hutchison}},\ }\href@noop {} {\bibinfo {title} {Avogadro 2: An open-source molecular builder and visualization tool.}} (\bibinfo {year} {2022})\BibitemShut {NoStop}%
\end{thebibliography}%


\begin{thebibliography}{2}%
\makeatletter
\providecommand \@ifxundefined [1]{%
 \@ifx{#1\undefined}
}%
\providecommand \@ifnum [1]{%
 \ifnum #1\expandafter \@firstoftwo
 \else \expandafter \@secondoftwo
 \fi
}%
\providecommand \@ifx [1]{%
 \ifx #1\expandafter \@firstoftwo
 \else \expandafter \@secondoftwo
 \fi
}%
\providecommand \natexlab [1]{#1}%
\providecommand \enquote  [1]{``#1''}%
\providecommand \bibnamefont  [1]{#1}%
\providecommand \bibfnamefont [1]{#1}%
\providecommand \citenamefont [1]{#1}%
\providecommand \href@noop [0]{\@secondoftwo}%
\providecommand \href [0]{\begingroup \@sanitize@url \@href}%
\providecommand \@href[1]{\@@startlink{#1}\@@href}%
\providecommand \@@href[1]{\endgroup#1\@@endlink}%
\providecommand \@sanitize@url [0]{\catcode `\\12\catcode `\$12\catcode `\&12\catcode `\#12\catcode `\^12\catcode `\_12\catcode `\%12\relax}%
\providecommand \@@startlink[1]{}%
\providecommand \@@endlink[0]{}%
\providecommand \url  [0]{\begingroup\@sanitize@url \@url }%
\providecommand \@url [1]{\endgroup\@href {#1}{\urlprefix }}%
\providecommand \urlprefix  [0]{URL }%
\providecommand \Eprint [0]{\href }%
\providecommand \doibase [0]{https://doi.org/}%
\providecommand \selectlanguage [0]{\@gobble}%
\providecommand \bibinfo  [0]{\@secondoftwo}%
\providecommand \bibfield  [0]{\@secondoftwo}%
\providecommand \translation [1]{[#1]}%
\providecommand \BibitemOpen [0]{}%
\providecommand \bibitemStop [0]{}%
\providecommand \bibitemNoStop [0]{.\EOS\space}%
\providecommand \EOS [0]{\spacefactor3000\relax}%
\providecommand \BibitemShut  [1]{\csname bibitem#1\endcsname}%
\let\auto@bib@innerbib\@empty
\bibitem [{\citenamefont {Marino}\ \emph {et~al.}(2014)\citenamefont {Marino}, \citenamefont {Melchior}, \citenamefont {Wohlfarth},\ and\ \citenamefont {Kreuer}}]{marino2014hydroxide}%
  \BibitemOpen
  \bibfield  {author} {\bibinfo {author} {\bibfnamefont {M.}~\bibnamefont {Marino}}, \bibinfo {author} {\bibfnamefont {J.}~\bibnamefont {Melchior}}, \bibinfo {author} {\bibfnamefont {A.}~\bibnamefont {Wohlfarth}},\ and\ \bibinfo {author} {\bibfnamefont {K.}~\bibnamefont {Kreuer}},\ }\bibfield  {title} {\bibinfo {title} {Hydroxide, halide and water transport in a model anion exchange membrane},\ }\href {https://doi.org/10.1016/j.memsci.2014.04.003} {\bibfield  {journal} {\bibinfo  {journal} {J. Membr. Sci.}\ }\textbf {\bibinfo {volume} {464}},\ \bibinfo {pages} {61} (\bibinfo {year} {2014})}\BibitemShut {NoStop}%
\bibitem [{\citenamefont {H{\"a}nseroth}\ and\ \citenamefont {Dre{\ss}ler}(2025)}]{haenseroth2025mace}%
  \BibitemOpen
  \bibfield  {author} {\bibinfo {author} {\bibfnamefont {J.}~\bibnamefont {H{\"a}nseroth}}\ and\ \bibinfo {author} {\bibfnamefont {C.}~\bibnamefont {Dre{\ss}ler}},\ }\bibfield  {title} {\bibinfo {title} {Optimizing machine learning interatomic potentials for hydroxide transport: Surprising efficiency of single-concentration training},\ }\bibfield  {journal} {\bibinfo  {journal} {J. Chem. Phys.}\ }\textbf {\bibinfo {volume} {163}},\ \href {https://doi.org/10.1063/5.0284063} {10.1063/5.0284063} (\bibinfo {year} {2025})\BibitemShut {NoStop}%
\end{thebibliography}%

\end{document}


\title{Supplementary Information for "Revealing Hydroxide Ion Transport Mechanisms in Commercial Anion-Exchange Membranes at Nano-Scale"}

\author{Jonas H{\"a}nseroth}
\affiliation{Theoretical Solid State Physics, Institute of Physics, Technische Universität Ilmenau, 98693 Ilmenau, Germany}

\author{Muhammad Nawaz Qaisrani}
\affiliation{Theoretical Solid State Physics, Institute of Physics, Technische Universität Ilmenau, 98693 Ilmenau, Germany}

\author{Mostafa Moradi}
\affiliation{Fraunhofer-Institut für Keramische Technologien und Systeme IKTS, 
 Hydrogen Technologies, 99310 Arnstadt, Germany}

\author{Karl Skadell}
\affiliation{Fraunhofer-Institut für Keramische Technologien und Systeme IKTS, 
 Hydrogen Technologies, 99310 Arnstadt, Germany}

\author{Christian Dre{\ss}ler}
\email{christian.dressler@tu-ilmenau.de}
\affiliation{Theoretical Solid State Physics, Institute of Physics, Technische Universität Ilmenau, 98693 Ilmenau, Germany}

\date{\today}

\maketitle

For more information on abbreviations, please refer to the main text, where all abbreviations are defined in detail.
Abbreviations not introduced in the main text are defined here.

\tableofcontents
\newpage
\clearpage

\section*{Supplementary Note 1: Overview of fine-tuned MLIP MD simulations}

\begin{table}[h!]
    \centering
    \caption{\textbf{Overview of MLIP MD simulations.}
    Summary of $34$ simulations totaling $385$~ns near-ab initio accuracy, spanning hydration ratios $\lambda = N(\text{H}_2\text{O})/N(\text{functional group})$ from $3$--$50$ and temperatures $298$--$370$~K.}
    
    \begin{tabular}{l@{\hspace{2.0em}}c@{\hspace{1.0em}}c@{\hspace{1.0em}}c@{\hspace{1.0em}}c@{\hspace{1.0em}}c@{\hspace{1.0em}}c@{\hspace{1.0em}}c@{\hspace{1.0em}}c@{\hspace{1.0em}}c@{\hspace{1.0em}}c@{\hspace{1.0em}}c@{\hspace{1.0em}}}
        \toprule
        $\lambda=N(\text{H}_2\text{O})/N(\text{Func.-Group})$ & $3$ & $5$ & $8$ & $10$ & $12$ & $15$ & $20$ & $30$ & $40$ & $50$ \\
        \midrule
        NPT-MD @ $350$~K: T$_\mathrm{MD}$ [ns] & 10 & 10 & 10 & 10 & 10 & 10 & 10 & 10 & 10 & 10 \\
        NVT-MD @ $298$~K: T$_\mathrm{MD}$ [ns] & -- & 10 & -- & 10 & -- & -- & -- & -- & -- & -- \\
        NVT-MD @ $310$~K: T$_\mathrm{MD}$ [ns] & 10 & 10 & -- & 10 & -- & -- & 5 & -- & -- & -- \\
        NVT-MD @ $330$~K: T$_\mathrm{MD}$ [ns] & 10 & 10 & -- & 10 & -- & -- & 5 & -- & -- & -- \\
        NVT-MD @ $350$~K: T$_\mathrm{MD}$ [ns] & 20 & 20 & 20 & 20 & 25 & 15 & 15 & 10 & 10 & 5 \\
        NVT-MD @ $370$~K: T$_\mathrm{MD}$ [ns] & 10 & 10 & -- & 10 & -- & -- & 5 & -- & -- & -- \\
        \bottomrule
    \end{tabular}
    \label{stab:temp}
\end{table}

\section*{Supplementary Note 2: Hydration and temperature dependency of the diffusion coefficient}

\begin{table}[h!]
    \centering
    \caption{\textbf{Hydroxide diffusion coefficients $\mathbf{D}$(OH$^-$).}
    Values in {\AA}$^2$/ps from NVT-MD simulations across hydration ratios $\lambda = N(\text{H}_2\text{O})/N(\text{functional group})$ $3$--$50$ and temperatures $298$--$370$~K.}
    \begin{tabular}{l@{\hspace{2.0em}}c@{\hspace{1.0em}}c@{\hspace{1.0em}}c@{\hspace{1.0em}}c@{\hspace{1.0em}}c@{\hspace{1.0em}}c@{\hspace{1.0em}}c@{\hspace{1.0em}}c@{\hspace{1.0em}}c@{\hspace{1.0em}}c@{\hspace{1.0em}}c@{\hspace{1.0em}}}
        \toprule
        $\lambda=N(\text{H}_2\text{O})/N(\text{Func.-Group})$ & $3$ & $5$ & $8$ & $10$ & $12$ & $15$ & $20$ & $30$ & $40$ & $50$ \\
        \midrule
        D(OH$^-$) [{\AA}$^2$/ps] @ $298$~K & -- & 0.003 & -- & 0.009 & -- & -- & -- & -- & -- & -- \\
        D(OH$^-$) [{\AA}$^2$/ps] @ $310$~K &  0.002 & 0.004 & -- &0.019 & -- & -- & 0.042 & -- & -- & -- \\
        D(OH$^-$) [{\AA}$^2$/ps] @ $330$~K &  0.004 & 0.006 & -- & 0.042 & -- & -- & 0.085 & -- & -- & -- \\
        D(OH$^-$) [{\AA}$^2$/ps] @ $350$~K &  0.012 & 0.012 & 0.036 & 0.075 & 0.080 & 0.085 & 0.076 & 0.114 & 0.210 & 0.312 \\
        D(OH$^-$) [{\AA}$^2$/ps] @ $370$~K &  0.013 & 0.026 & -- & 0.061 & -- & -- & 0.195 & -- & -- & -- \\
        \bottomrule
    \end{tabular}
    \label{stab:oh_diff}
\end{table}

\section*{Supplementary Note 3: Low-hydration diffusion coefficients}

\begin{figure*}[h!]
    \centering
    \includegraphics[width=0.5\textwidth]{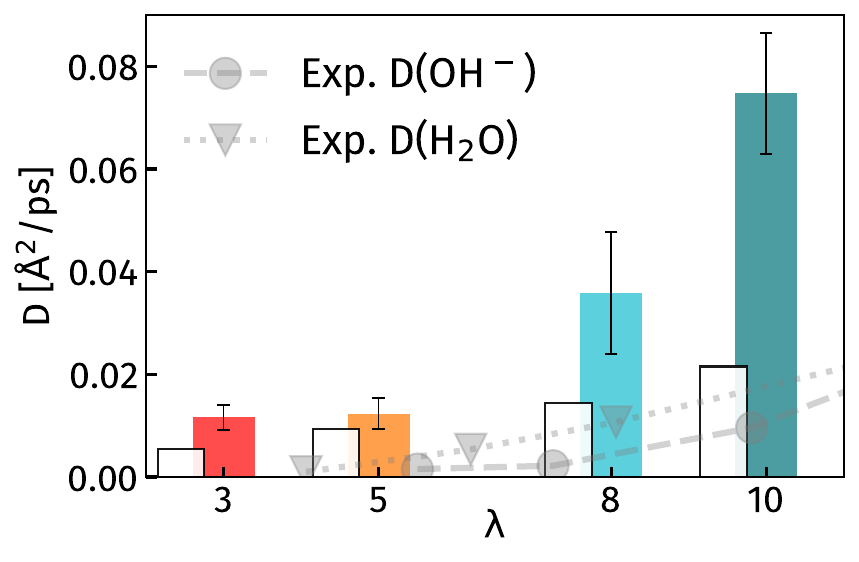}
    \caption{\textbf{Enlarged view of low-hydration diffusion coefficients at $350$~K.}
    Water ($D$(H$_2$O)) and hydroxide ($D$(OH$^-$)) diffusion coefficients from fine-tuned MLIP MD simulations across $\lambda=3$--$10$, expanding main text Figure 1.
    Experimental diffusion coefficients from Marino et al. \cite{marino2014hydroxide}, measured at $298$~K, are shown for comparison.}
    \label{fig:si1}
\end{figure*}

\newpage
\clearpage

\section*{Supplementary Note 4: Proton transfer rate functions}

\begin{figure*}[h!]
    \centering
    \includegraphics[width=0.75\textwidth]{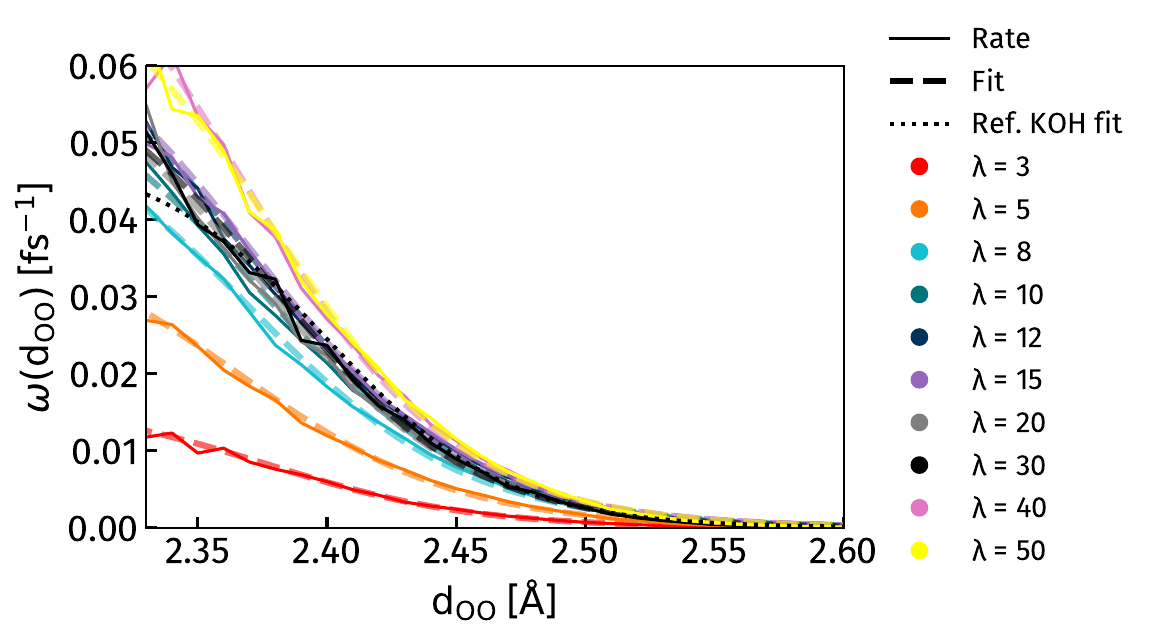}
    \caption{\textbf{Proton transfer probability $\omega(d_\text{OO})$ across hydration levels.}
    Jump rate functions for proton transfer events in membrane channels ($\lambda=3$--$50$), fitted to $\omega(d_\text{OO})=a/[1+\exp((d_\text{OO}-b)/c)]$ and obtained from the fine-tuned MLIP MD simulations at $350$~K. 
    KOH reference (2.3~M aqueous, AIMD) from \cite{haenseroth2025mace}.}
    \label{fig:si2}
\end{figure*}

\section*{Supplementary Note 5: Proton transfer event statistics}

\begin{table}[ht]
    \centering
    \caption{\textbf{Proton transfer rates per OH$^-$ and picosecond.}
    Proton transfer events counted from fine-tuned MLIP MD simulations at $350$~K across hydration conditions $\lambda=3$--$50$, normalized by number of hydroxide ions and trajectory length (ps). 
    Rattling events (proton wiggling between the same two O atoms) identified and excluded to yield the number of non-rattling PT events; non-rattling fraction = non-rattling PT events / total PT events.}
    \begin{tabular}{lc@{\hspace{1em}}c@{\hspace{1em}}c@{\hspace{1em}}}
        \toprule
                   &   PT events per    &   PT events per OH$^-$      &  Fraction of non-     \\
        $\lambda$  &   OH$^-$ and ps    &   and ps (excl. ratteling)  &  ratteling PT events  \\
        \midrule
            3 &  0.7380 &  0.0765 & 0.1037 \\
            5 &  2.8944 &  0.3143 & 0.1086 \\
            8 &  4.1811 &  0.5905 & 0.1412 \\
           10 &  4.1839 &  0.6733 & 0.1609 \\
           12 &  4.2504 &  0.7530 & 0.1772 \\
           15 &  5.2245 &  0.9282 & 0.1777 \\
           20 &  1.5698 &  0.3173 & 0.2021 \\
           30 &  1.0059 &  0.2211 & 0.2198 \\
           40 &  1.8356 &  0.4333 & 0.2361 \\
           50 &  2.6590 &  0.6077 & 0.2285 \\
        \bottomrule
    \end{tabular}
    \label{stab:pt_events}
\end{table}

\newpage
\clearpage

\section*{Supplementary Note 6: Free energy profiles for proton transfer}

\begin{figure*}[h!]
    \centering
    \includegraphics[width=0.55\textwidth]{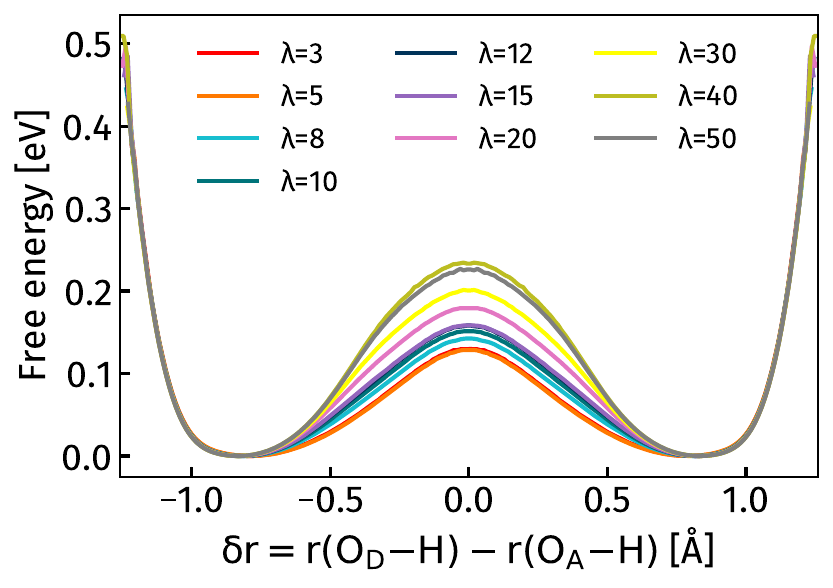}
    \caption{\textbf{Free energy profiles along the proton transfer coordinate.}
    Profiles as a function of $\delta r$, defined as the difference between the donor O--H and acceptor O$\cdots$H distances, for hydration levels $\lambda=3$--$50$, obtained from fine-tuned MLIP molecular dynamics simulations at $350$~K.
    The free energy is computed as $F(\delta r) = -k_{\mathrm{B}}T \ln P(\delta r)$, where $P(\delta r)$ denotes the normalized probability distribution of the proton transfer coordinate sampled along the trajectories.}
    \label{img:si3}
\end{figure*}

\section*{Supplementary Note 7: Multichannel AEM simulation setups}

\begin{figure*}[h!]
    \centering
    \includegraphics[width=0.65\textwidth]{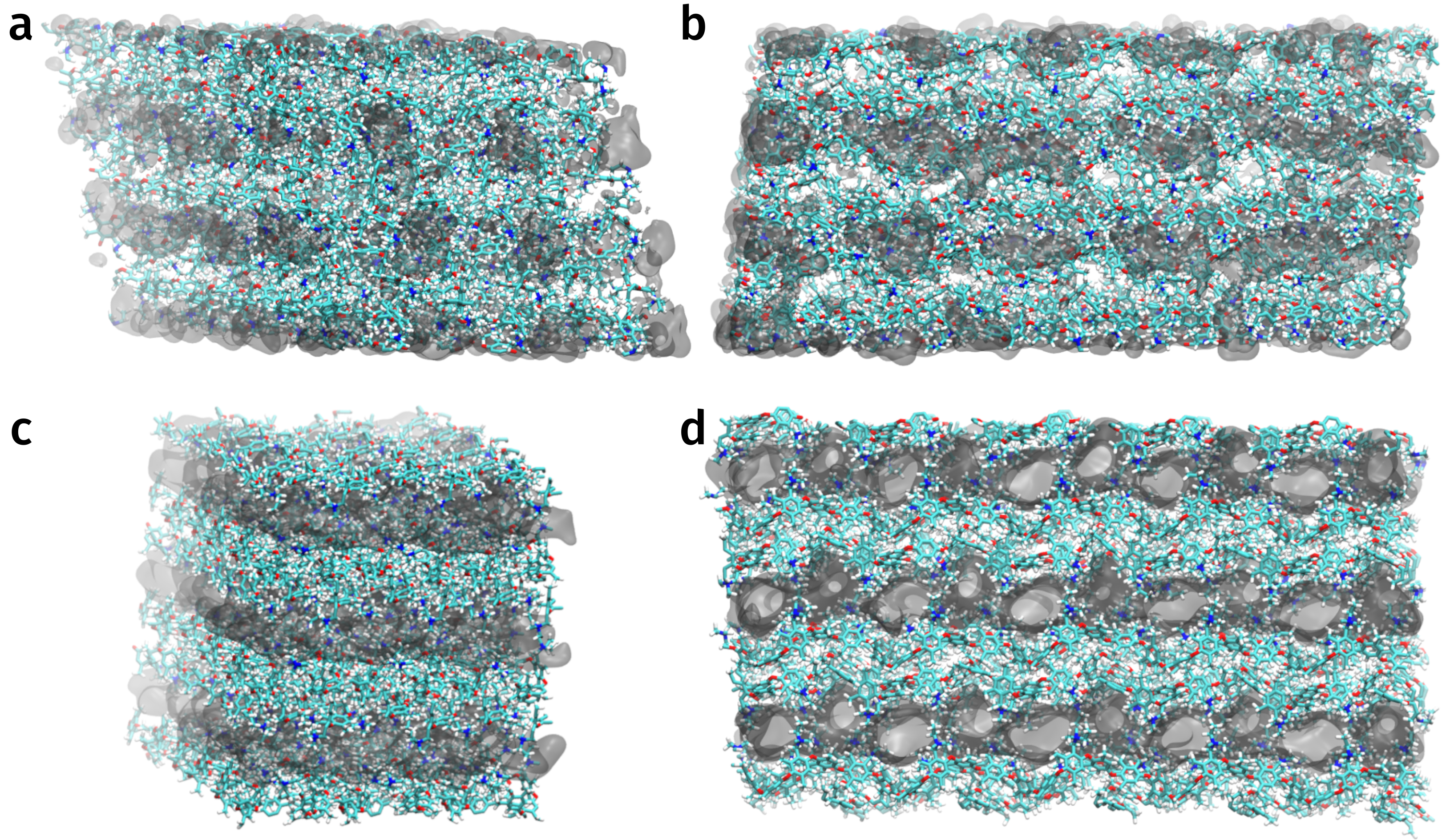}
    \caption{\textbf{$3\times3\times5$ multichannel anion exchange membrane systems.}
    Systems of anion exchange membranes consisting of $36$ polymer strands at hydration levels $\lambda=3$ (24,660 atoms) and $\lambda=10$ (32,220 atoms), elongated 100~{\AA} along the $z$-axis, with the aqueous phase rendered as a gray continuum. 
    Polymer atoms are colored as follows: C turquoise, O red, N blue. 
    Panels \textbf{a} and \textbf{b} show the $\lambda=3$ system in $z$- and $y$-views, respectively, while \textbf{c} and \textbf{d} depict the $\lambda=10$ system in the same orientations.}
    \label{fig:si4}
\end{figure*}

\newpage
\clearpage

\section*{Supplementary Note 8: Nyquist plots}

\begin{figure*}[h!]
    \centering
    \includegraphics[width=0.85\textwidth]{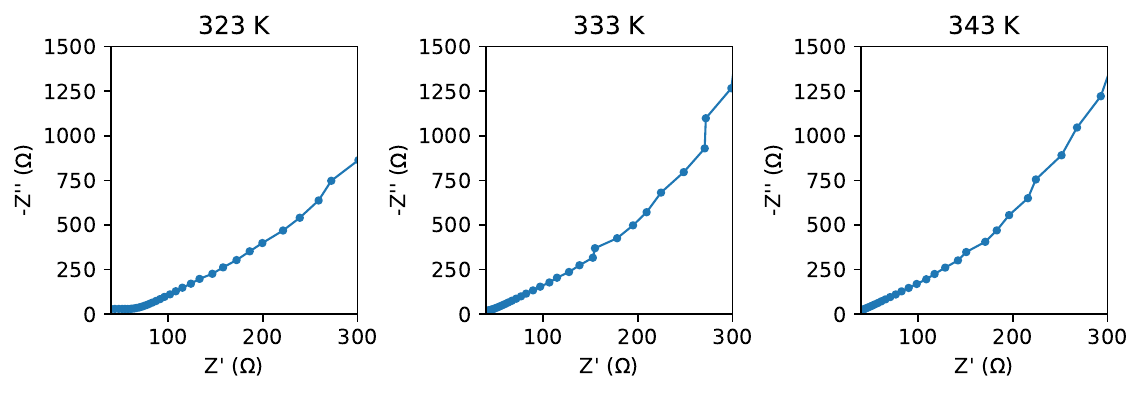}
    \caption{\textbf{Nyquist plots.}
    Nyquist plot for AEM-130-PK/KOH with $0.1$~M KOH at $323$, $333$ and $343$~K of the high frequency region used to determine the ohmic resistance.} 
    \label{img:si5}
\end{figure*}

\section*{Supplementary Note 9: Hydroxide environment at low hydration levels}

\begin{figure*}[h!]
    \centering
    \includegraphics[width=0.95\textwidth]{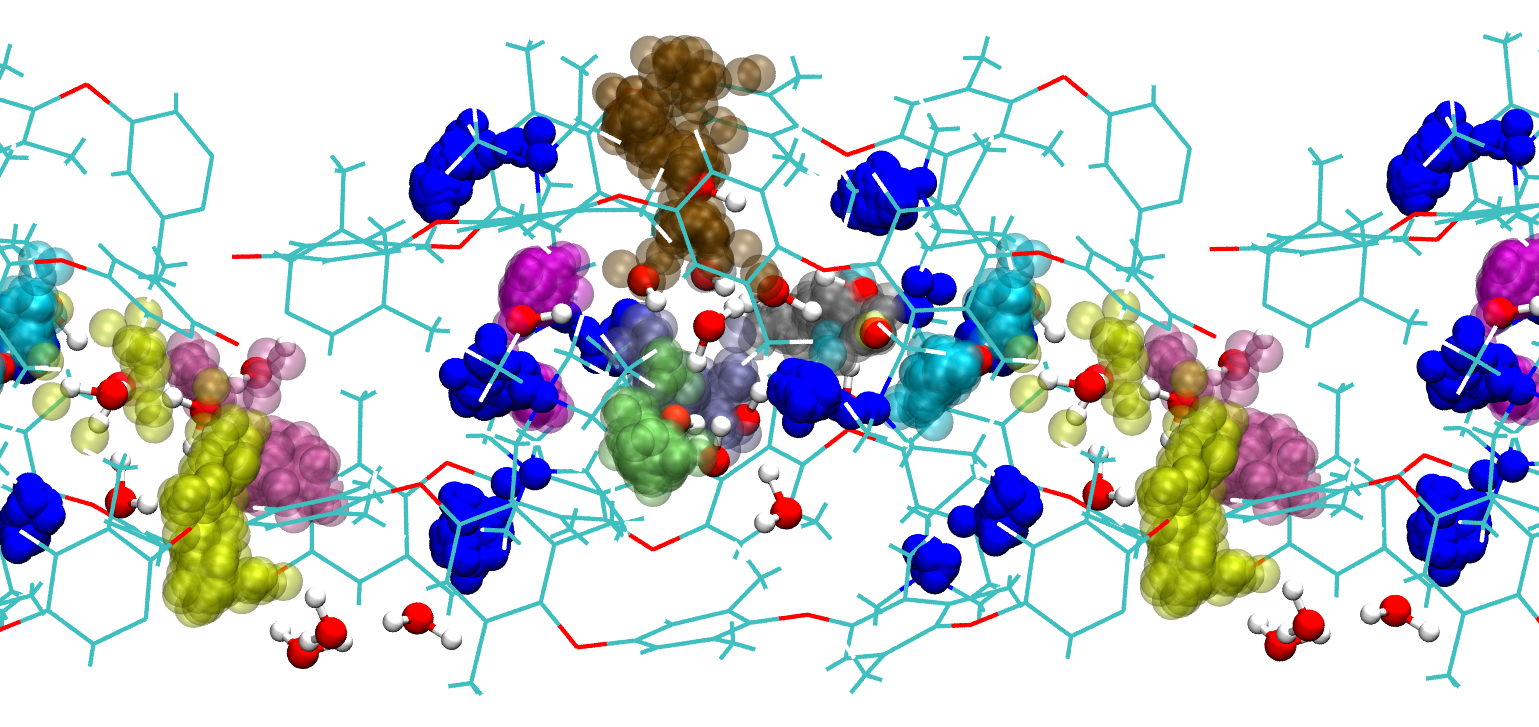}
    \caption{\textbf{Hydroxide diffusion pathways at $\lambda=3$.}
    Hydroxide moieties diffuse within the membrane channel at $350$~K and $\lambda=3$; pathways obtained from AIMD simulation. 
    The polymeric backbone appears as turquoise/red lines, hydroxide pathways over $75$~ps as transparent colored spheres (each ion differently colored), functional group N atoms as solid blue spheres (same time interval), and water molecules as red O (white H) spheres.}
    \label{img:si6}
\end{figure*}

\newpage
\clearpage

\section*{Supplementary Note 10: Overlap of nearest-neighbor distance distributions}

\begin{table}[ht]
    \centering
    \caption{\textbf{Quantitative overlap of hydroxide nearest-neighbor distance distributions.}
    Integrated overlap between the OH$^-$–water oxygen distance distribution and those to the nearest functional-group carbon (C\textsubscript{FG}) and backbone carbon (C\textsubscript{BB}) at different hydration levels $\lambda$. 
    Each individual distribution is normalized to unit area, and the overlap values correspond to the shared area between the respective pairs of curves.}
    \begin{tabular}{lc@{\hspace{1em}}c@{\hspace{1em}}c@{\hspace{1em}}}
        \toprule
        $\lambda$ & Total area per curve & Overlap O and C\textsubscript{FG} & Overlap O and C\textsubscript{BB} \\
        \midrule
        3  & 1.000 & 0.1070 & 0.0713 \\
        5  & 1.000 & 0.0647 & 0.0381 \\
        10 & 1.000 & 0.0179 & 0.099 \\
        20 & 1.000 & 0.0257 & 0.023 \\
        50 & 1.000 & 0.0086 & 0.007 \\
        \bottomrule
    \end{tabular}
    \label{stab:tab_overlap}
\end{table}

\section*{Supplementary Note 11: Fine-tuned MLIP accuracy on AIMD membrane data}

\begin{figure*}[h!]
    \centering
    \includegraphics[width=0.95\textwidth]{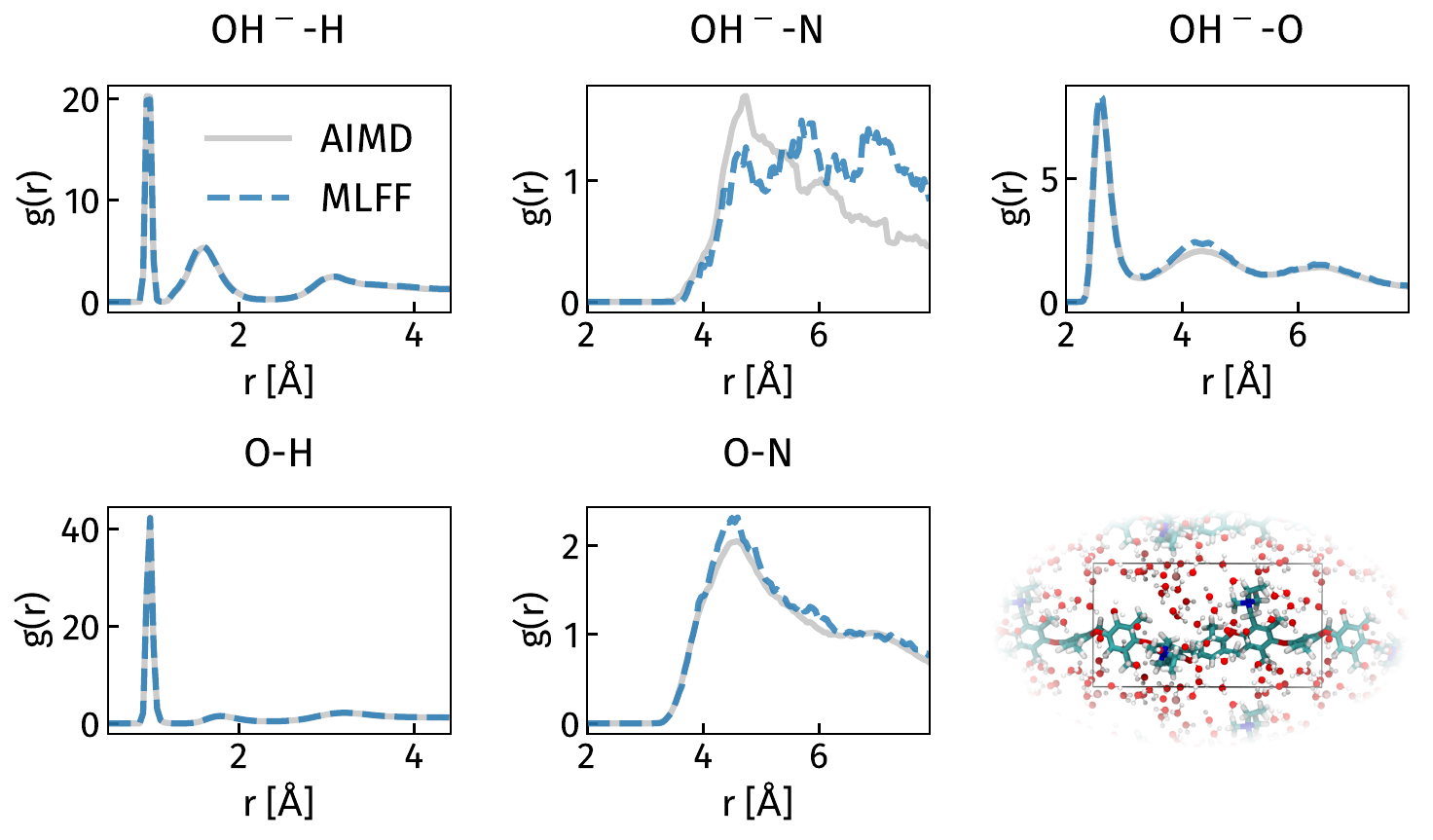}
    \caption{\textbf{Fine-tuned MLIP reproduces AIMD radial distribution functions.}
    Radial distribution functions ($g(r)$) for OH$^-$--H, OH$^-$--N, OH$^-$--O, H$_2$O--H, and H$_2$O--N pairs from AIMD reference (solid gray lines) and fine-tuned MLIP MD (dashed blue line) simulations at matching conditions. 
    Lower left: membrane subsystem used for AIMD training data generation (O: red, C: turquoise, N: blue, K: rose, H: white).}
    \label{fig:si7}
\end{figure*}

\newpage
\clearpage

\section*{Supplementary Note 12: Equivalent-circuit model for EIS analysis}

\begin{figure*}[h!]
    \centering
    \includegraphics[width=0.3\textwidth]{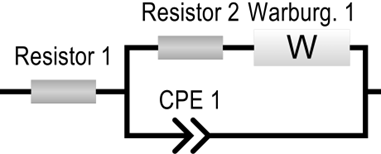}
    \caption{\textbf{Equivalent circuit model for EIS fitting.}
    Equivalent circuit model used for EIS analysis, consisting of series resistances, a constant phase element (CPE), and a Warburg diffusion element.} 
    \label{img:si8}
\end{figure*}

\section*{Supplementary Note 13: Experimental EIS cell setup}

\begin{figure*}[h!]
    \centering
    \includegraphics[width=0.5\textwidth]{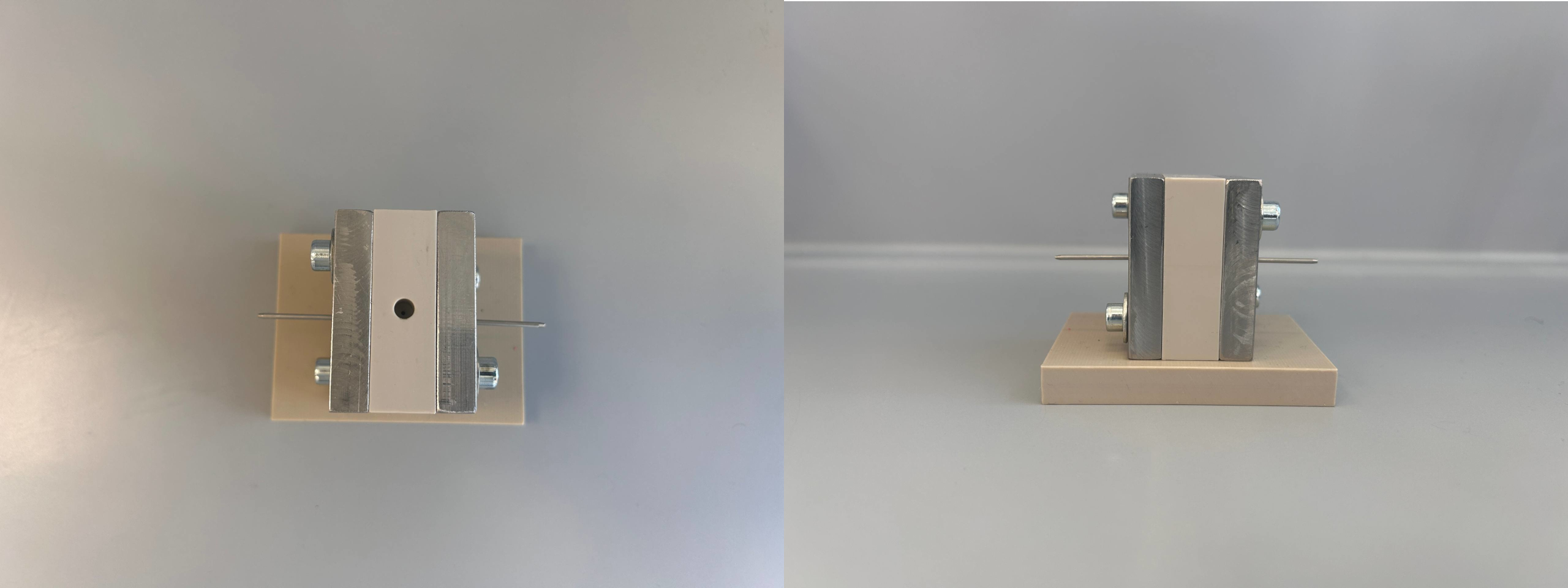}
    \caption{\textbf{In-house built EIS measurement cell.}
    Two-electrode cell for AEM-130-PK/KOH conductivity measurements using $0.1$~M KOH electrolyte at $323$--$343$~K. 
    The design features a single central electrolyte reservoir with two identical membranes in series, stainless steel electrodes (high Ni/Mo content), and active area $A=1.79$~cm$^2$.}
    \label{img:si9}
\end{figure*}

\bibliography{si_bibliography}